\documentclass[aps,
               pra,
               showpacs,
               amssymb,
               nofootinbib,
               superscriptaddress,
               twocolumn,
               longbibliography]{revtex4-2}
\usepackage{braket}
\usepackage[ansinew]{inputenc}
\usepackage{bbold}
\usepackage{bbm}
\usepackage{bm}
\usepackage{amsbsy}
\usepackage{amsthm}
\usepackage{amssymb}
\usepackage{amsfonts}
\usepackage{amsmath, mathtools}
\usepackage{dsfont}
\usepackage{graphicx}
\usepackage{epsfig}
\usepackage{epstopdf}
\usepackage{dsfont}
\usepackage{mathrsfs} 
\usepackage{multibib}
\usepackage[dvipsnames]{xcolor}
\usepackage[colorlinks]{hyperref}
\makeatletter
\newcommand\org@hypertarget{}
\let\org@hypertarget\hypertarget
\renewcommand\hypertarget[2]{%
  \Hy@raisedlink{\org@hypertarget{#1}{}}#2%
  }
\makeatother
\usepackage[figure,table]{hypcap}

\usepackage{MnSymbol}
\usepackage{enumerate}
\usepackage{float}
\usepackage{comment}
\usepackage{braket}
\usepackage{subfigure}
\usepackage{orcidlink}
\usepackage{tcolorbox}

\makeatletter
\newcommand{\providetcbcountername}[1]{%
  \@ifundefined{c@tcb@cnt@#1}{%
    --undefined--%
  }{%
    tcb@cnt@#1%
  }
}

\newcommand{\settcbcounter}[2]{%
  \@ifundefined{c@tcb@cnt@#1}{%
    \GenericError{Error}{counter name #1 is no tcb counter }{}{}%
  }{%
    \setcounter{tcb@cnt@#1}{#2}%
   }%
}%
\tcbuselibrary{theorems}
\newtcolorbox[blend into=tables]{smallboxtable}[3][]
{colback=mycolor!5, colframe=mycolor, float=ht, width=0.48\textwidth, lower separated=false, blend before title=colon hang,
title={#2}, label=#3 ,#1}
\newtcolorbox[blend into=tables]{medboxtable}[3][]
{colback=mycolor!5, colframe=mycolor, float=ht, width=0.99\textwidth, lower separated=false, blend before title=colon hang,
title={#2}, label=#3 ,#1}
\newtcbtheorem{Definitions}{Definition}%
{colback=mycolor!5,colframe=mycolor,fonttitle=\bfseries,
left=4pt,
right=4pt,
top=5pt,
bottom=5pt}{defi}
\newtcbtheorem{Theorems}{Theorem}%
{colback=mycolor!5,colframe=mycolor,fonttitle=\bfseries,
left=4pt,
right=4pt,
top=5pt,
bottom=5pt}{theorem}
\definecolor{mycolor}{rgb}{0.122, 0.435, 0.698}
\usepackage{subfigure}
\usepackage{makecell}

\hypersetup{
	bookmarksnumbered,
	pdfstartview={FitH},
	citecolor={darkgreen},
	linkcolor={darkred},
	urlcolor={darkblue},
	pdfpagemode={UseOutlines}}
\definecolor{darkgreen}{RGB}{50,190,50}
\definecolor{darkblue}{RGB}{0,0,190}
\definecolor{darkred}{RGB}{238,0,0}
\definecolor{mycolor}{rgb}{0.122, 0.435, 0.698}
\usepackage{soul}

\newcommand{\tr}{\textnormal{Tr}}

\renewcommand{\thesection}{\Roman{section}}
\renewcommand{\thesubsection}{\Roman{section}.\Alph{subsection}}
\renewcommand{\thesubsubsection}{\Roman{section}.\Alph{subsection}.\arabic{subsubsection}}
\makeatletter
\renewcommand{\p@subsection}{}
\renewcommand{\p@subsubsection}{}
\makeatother

\usepackage{siunitx}

\usepackage{array}
\usepackage{colortbl}

\begin{document}

\title{Generation of multipartite photonic entanglement using a trapped-ion quantum processing node}
\author{Marco Canteri\,\orcidlink{0000-0001-9726-2434}}
\affiliation{Institut f\"ur Experimentalphysik, Universit\"at Innsbruck, Technikerstrasse 25, 6020 Innsbruck, Austria}
\author{James Bate}
\affiliation{Institut f\"ur Experimentalphysik, Universit\"at Innsbruck, Technikerstrasse 25, 6020 Innsbruck, Austria}
\author{Ida Mishra}
\affiliation{Technische Universit{\"a}t Wien, Atominstitut \& Vienna Center for Quantum Science and Technology (VCQ),  Stadionallee 2, 1020 Vienna, Austria}
\author{Nicolai Friis\,\orcidlink{0000-0003-1950-8640}}
\email{nicolai.friis@tuwien.ac.at}
\affiliation{Technische Universit{\"a}t Wien, Atominstitut \& Vienna Center for Quantum Science and Technology (VCQ),  Stadionallee 2, 1020 Vienna, Austria}
\author{Victor Krutyanskiy\,\orcidlink{0000-0003-0620-4648}}
\affiliation{Institut f\"ur Experimentalphysik, Universit\"at Innsbruck, Technikerstrasse 25, 6020 Innsbruck, Austria}
\author{Benjamin P. Lanyon\,\orcidlink{0000-0002-7379-4572}}
\email[Correspondence should be sent to ]{ben.lanyon@uibk.ac.at}
\affiliation{Institut f\"ur Experimentalphysik, Universit\"at Innsbruck, Technikerstrasse 25, 6020 Innsbruck, Austria}

\date{\today}

\begin{abstract}
The ability to establish entanglement between the nodes of future quantum networks is essential for enabling a wide range of new applications in science and technology. A promising approach involves the use of a powerful central node capable of deterministically preparing arbitrary multipartite entangled states of its matter-based qubits and efficiently distributing these states to surrounding end nodes via flying photons. This central node, referred to as a ``factory node", serves as a hub for the production and distribution of multipartite entanglement. In this work, we demonstrate key functionalities of a factory node using a cavity-integrated trapped-ion quantum processor. Specifically, we program the system to generate genuinely multipartite entangled Greenberger-Horne-Zeilinger (GHZ) states of three path-switchable photons and verify them using custom-designed entanglement witnesses. These photons can, in the future, be used to establish stored multipartite entanglement between remote matter-based nodes. Our results demonstrate that the well-established techniques for the deterministic preparation of entangled states of co-trapped ion qubits can be used to prepare the same states of traveling photons, paving the way for multipartite entanglement distribution in quantum local area networks.

\end{abstract}

\date{\today}

\maketitle


\section{Introduction}\label{sec:intro}
\vspace*{-2mm}

{\noindent}There is a current research effort to develop the building blocks of future quantum networks: distributed matter-based nodes, for information storage and processing, that are connected with photonic links for the transfer of quantum information~\cite{Kimble2008, Duan_2010, Reiserer_2015,WehnerElkoussHanson2018, Wei_2022, Covey2023, Azuma2023, Tittel_2025}. 
Such networks are envisioned to span distances ranging from a single laboratory for, e.g., scalable quantum computing ~\cite{Monroe_2014, Brown2016}, to global scales, promising enhanced distributed timekeeping~\cite{Komar2014}, secure communication systems~\cite{Muralidharan2016, Zapatero2023}, and sensor networks~\cite{Zhang_2021,Proctor_2018, Zhang_2021}.

The ability of the nodes of a quantum network to become entangled is a key feature that makes such networks more powerful than their classical counterparts.  
While bipartite entanglement between pairs of nodes allows for a few applications, such as secure point-to-point communication~\cite{Ekert1991} and blind quantum computing~\cite{BroadbentFitzsimonsKashefi2009}, accessing the full power of quantum networks requires establishing multipartite entanglement between multiple nodes. 
Recently, experiments have established examples of multipartite entangled states between two~\cite{Ruskuc2025, shi2025} and three~\cite{Jing2019, PompiliEtAl2021} remote matter-based systems.

\begin{figure}[ht!]
\begin{center}
\includegraphics[angle = 0, width=0.35\textwidth,trim={0cm 0cm 0cm 0cm},clip]{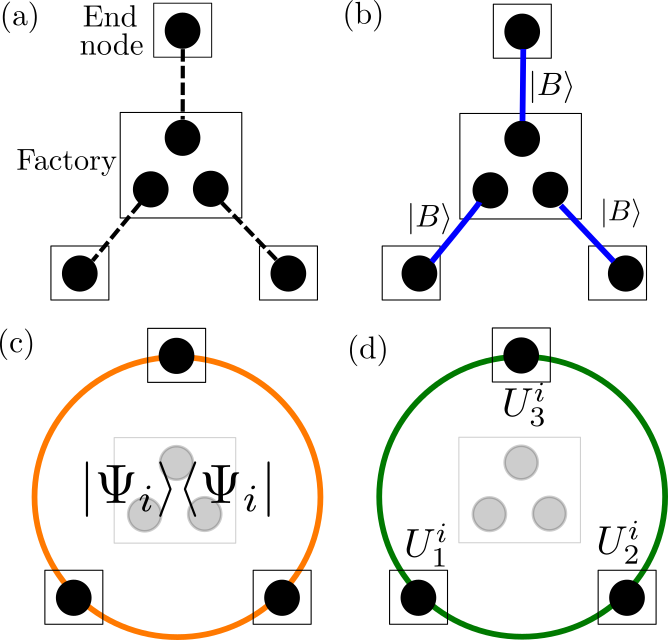}
\scriptsize{ \caption{\textbf{Multipartite entanglement distribution via a central factory node.} 
(a) The factory node contains as many qubits (black circles) as there are remote nodes (here, three). Dashed lines show photonic channels, e.g., optical fibers. 
(b) Bell states $\ket{B}$ (blue lines) are established between factory and remote nodes via the distribution of photons~\cite{Beukers_2024,Duan_2010}.
(c)-(d) Remote state preparation: 
(c) Projective measurement of the factory-node qubits in a basis of multipartite entangled states {$\ket{\Psi_i}\!\!\bra{\Psi_i}$ yields one of eight outcomes labelled by $i=1,\ldots,8$, resulting in the preparation of end node qubits in the state $\ket{\Psi_i}$ (orange ring); 
(d) Single-qubit rotations $U^i_k$ are performed at the remote nodes (labelled $k=1,2,3$), with parameters determined by the outcome in (c) to deterministically prepare any one chosen distributed state, e.g., $\ket{\Psi_1}$ (green ring).}
}
\label{fig:scheme}}
\vspace{-4mm}
\end{center}
\end{figure}

The question of how to best establish multipartite entanglement in future quantum networks has been extensively studied in theory~\cite{PhysRevA.94.052307, METER_2011, Pirker_2018, PhysRevA.100.032310_2019, Pirker_2019, Benjamin_2006, PhysRevA.75.042303_2007, PhysRevA.73.062328_2006, 9292429_2020, PhysRevA.105.0126082022, Coopmans2021, Nain2022, PhysRevA.86.042304, Epping_2016, YamasakiPirkerMuraoDuerKraus2018, PhysRevA.100.052333, Bugalho_2023, 
AvisRozpedekWehner2023, FischerAlexTowsleyDon_2021,MorelliUsuiAgudeloFriis2021}. 
A~powerful proposed approach is to use a central `factory' node~\cite{AvisRozpedekWehner2023, Pirker_2018, PhysRevA.73.062328_2006, 10.1145/3392141, PhysRevA.86.042304, PhysRevA.100.052333, Bugalho_2023, FischerAlexTowsleyDon_2021}: a register of at least $n$ qubits with photonic channels to each of $n$ end nodes. 
In the first step, the photonic channels are used to establish bipartite entanglement (Bell pairs) between a qubit in each end node and a different qubit in the factory node. Such pairs can be generated via a probabilistic process, which can be repeated until all pairs have been established, with a rate that falls only logarithmically with the number of end nodes~\footnote{Logarithmic rate reduction with end-node number can be achieved if attempts to make Bell-pairs with end nodes are run in parallel~\cite{AvisRozpedekWehner2023}, otherwise linear scaling can be achieved.} and linearly or better with the probability for establishing each Bell state~\cite{AvisRozpedekWehner2023}.  
Finally, the factory node implements either the remote state-preparation (RSP) protocol~\cite{Bennett_2001} (exemplified in Fig.~\ref{fig:scheme}) or the teleportation protocol~\cite{BennettBrassardCrepeauJozsaPeresWootters1993}, to deterministically establish a multipartite entangled state between the end nodes. 
A~factory node capable of universal quantum logic could be reprogrammed to distribute arbitrary quantum states. 

A key to the efficient scaling of the factory-node approach in end-node number is to first establish pairwise entanglement across the inevitably lossy photonic links{\textemdash}storing successfully generated pairs in qubit memories until all pairs are established. 
Directly distributing an $n$-photon multipartite entangled state to $n$ end nodes, requiring all $n$ photons to successfully traverse lossy channels simultaneously, will succeed with a rate that falls exponentially with the number of end nodes, even using recently developed near-deterministic  sources of multipartite photonic entanglement~\cite{Schwartz2016, Besse2020, Istrati2020,Thomas2022, Ferreira2024, Pont2024, Thomas2024, Huet2025}.

In this work, we demonstrate core functionalities of a factory node based on trapped ions. 
After establishing Bell states between each of three co-trapped ion qubits and a separate photon, deterministic  quantum logic gates are used to measure the ion qubits in a basis of  GHZ states~\cite{GreenbergerHorneZeilinger1989}. As a result, the three photonic qubits are prepared in GHZ-type states, which 
we verify with purpose-designed lower bounds on state fidelities and corresponding witnesses for genuine multipartite entanglement (GME). 
Our results thus demonstrate that the established tools for the generation of multipartite entanglement between trapped ions can be mapped across to traveling photons. Finally, we summarise how, in the future, 
the generated photons could be used to establish entanglement with remote matter-based end nodes, such that the distributed multipartite entanglement is stored and deterministically established between the end nodes.  
Indeed, ions in separate traps a few meters apart have previously been entangled~\cite{Moehring2007, Stephenson2020, Nichol2022, Main2025, main2025multipartitemixedspeciesentanglementquantum} and, using the
ion-cavity system of this work, ion-ion entanglement between traps at a distance of \SI{230}{m} has been achieved~\cite{Krut2022}. In future, our trapped-ion factory node prototype could be reprogrammed to distribute arbitrary states between end nodes a hundred kilometers part or further, given work on telecom-wavelength photon conversion~\cite{Bock2018, Walker2018, Krutyanskiy2019, KrutyanskiyCanteriMeranerKrcmarskyLanyon2024} and quantum repeater functionality~\cite{Krut2023, PhysRevA.110.032603}. 

\vspace*{-1mm}
\section{Setup and Approach}
\vspace*{-2mm}
    
\begin{figure}[t!]
\begin{center}
\includegraphics[angle = 0, width=0.45\textwidth,trim={0cm 0cm 0cm 0cm},clip]{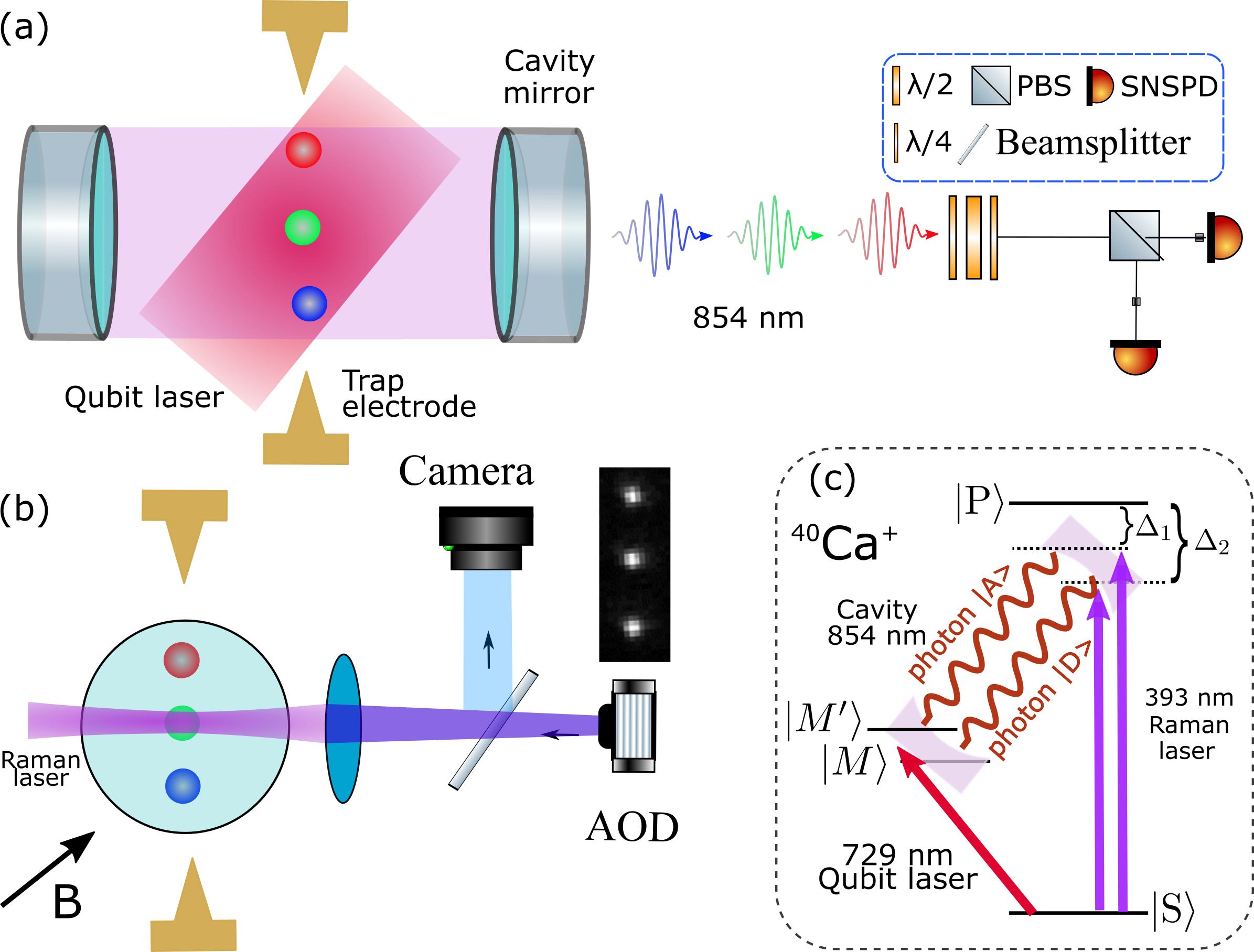}
\scriptsize{ \caption{\textbf{Experimental schematic.} 
(a) Three trapped $^{40}$Ca$^{+}$ ions at neighboring anti-nodes of an \SI{854}{nm} vacuum standing-wave mode in an optical cavity. Raman laser pulses trigger three photons to exit the cavity in a row, each polarization-entangled with the ion that emitted it (shown by colour). Polarization analysis involves half- ($\lambda/2$) and quarter-wave ($\lambda/4$) plates, 
a polarising beam splitter and single-mode fiber coupled photon detectors. 
(b) View along the cavity axis. An acousto-optic deflector (AOD) allows switching of the \SI{393}{\nano\meter} photon generation (Raman) laser along the string axis. Ion fluorescence at \SI{397}{\nano\meter} generated during ion-qubit readout is imaged onto a digital camera.  
(c) Atomic energy-level diagram. $\ket{S}{=}  \ket{\uparrow} =  \ket{4^{2} S_{1/2,m_j{=}{-}1/2}}$, $\ket{P}{=}\ket{4^{2} P_{3/2,m_j{=}-3/2}}$, $\ket{M} =\ket{\downarrow}{=}\ket{3^{2}D_{5/2,m_j{=}-5/2}}$, $\ket{M'}{=}\ket{3^{2}D_{5/2,m_j{=}-3/2}}$. 
\label{fig:experimental_schematic}}}
\end{center}
\end{figure}

{\noindent}The experimental setup, a conceptual schematic of which is presented in Fig.~\ref{fig:experimental_schematic}, comprises three $^{40}$Ca$^{+}$ ions confined in a three-dimensional linear Paul trap and placed at the position of the waist of a near-concentric optical Fabry{\textendash}Perot cavity of \SI{20}{mm} length for photon collection at \SI{854}{nm}~\cite{SchuppKrcmarskyKrutyanskiyMeranerNorthupLanyon2021, SchuppPhDthesis2021}.  The ions, \SI{5.26(10)}{\micro\meter} apart, are positioned at neighboring anti-nodes of a vacuum cavity mode. 
The polarisation of single \SI{854}{nm} photons exiting the cavity is analysed by a sequence of wave plates followed by a polarising beam splitter and finally two fiber-coupled single-photon detectors. 
More details on the experimental setup can be found in 
Appendix~\ref{appendix:experimental setup}.\\[-2mm]

Single \SI{854}{nm} photons are generated via a bichromatic cavity-mediated Raman transition (BCMRT)~\cite{StuteEtAl2012, SchuppKrcmarskyKrutyanskiyMeranerNorthupLanyon2021}, driven via a  \SI{393}{nm} Raman laser beam with a \SI{1.2}{\micro\meter} waist at the position of the ions~\cite{KrutyanskiyCanteriMeranerKrcmarskyLanyon2024}. 
A~Raman laser pulse on an ion in the state $\ket{S}{=}\ket{4^{2} S_{1/2,m_j{=}{-}1/2}}$ ideally generates the maximally entangled state $(\ket{\hspace*{1pt}M,D}+\ket{\hspace*{1pt}M',A})/\sqrt{2}$, where $\ket{\hspace*{1pt}M\hspace*{1pt}}{=}\ket{3^{2}D_{5/2,m_j{=}-5/2}}$ and $\ket{\hspace*{1pt}M'\hspace*{1pt}}{=}\ket{3^{2}D_{5/2,m_j{=}-3/2}}$ describe two metastable ionic energy levels, and $\ket{D}$ and $\ket{A}$ describe two orthogonal linear polarisations (diagonal and anti-diagonal, respectively) of a photon emitted into the cavity. 
After the photon exits through a preferred mirror~\cite{SchuppKrcmarskyKrutyanskiyMeranerNorthupLanyon2021}, the laser focus is moved to the neighbouring co-trapped ion, allowing sequential generation of three photons, each of which is ideally entangled with the ion that emitted it. 

%

\newpage
In cases in which either detector 
clicks during each of the three \SI{50}{\micro\second} time intervals during which a photon is expected (a triple coincidence is obtained, Appendix~\ref{appendix:rawdata}), the ion-readout sequence is triggered. 
The ion-readout sequence begins with manipulation of the ions' electronic states using a broadly focused \SI{729}{nm} laser beam, which couples equally to each ion.  A~first laser pulse maps $\ket{\hspace*{1pt}M'\hspace*{1pt}}$ to $\ket{S}$, thereby moving the ion-qubit encoding into superpositions of $\ket{S}$ and $\ket{\hspace*{1pt}M\hspace*{1pt}}$. From this point on we use spin notation for the logical states of the ion qubits, via the substitutions $\ket{S}=\ket{\hspace*{1pt}\uparrow\hspace*{1pt}}$ and $\ket{M}=\ket{\hspace*{1pt}\downarrow\hspace*{1pt}}$. A~second, trichromatic laser pulse implements a three-qubit M{\o}lmer{\textendash}S{\o}rensen (MS) quantum logic gate ~\cite{SoerensenMoelmer1999, SoerensenMoelmer2000, BenhelmKirchmairRoosBlatt2008}, described by the three-qubit unitary operator $U_{\mathrm{MS}}$, mediated by the axial center-of-mass mode of the ion string at \SI{0.871}{MHz}. A~third and final \SI{729}{\nano\meter} laser pulse implements the unitary rotation $U^{\otimes 3}$, corresponding to the same single-qubit rotation $U$ on each ion qubit. Additional information on the pulse sequence is provided in Appendix~\ref{appendix:pulse sequence}. 
Finally, the logical state of each ion qubit is measured individually via fluorescence state detection, in which scattered \SI{397}{\nano\meter} photons are collected by a lens and imaged onto a digital camera: $\ket{\hspace*{1pt}\uparrow\hspace*{1pt}}$ fluoresces, $\ket{\hspace*{1pt}\downarrow\hspace*{1pt}}$ does not. 
In combination with the preceding two laser pulses, observing one of the eight logical ion-qubit states $\ket{\hspace*{0.5pt}lmn}$ (with $l,m,n\in\{\downarrow,\uparrow\}$) is equivalent to projecting the ion qubits into a basis of GHZ states, that is 
$( U^{\otimes 3}\, {M\hspace*{-1pt}S})^{\dagger}\ket{\hspace*{0.5pt}lmn}\!\!\bra{lmn}({M\hspace*{-1pt}S}\,U^{\otimes 3})=\ket{\mathrm{GHZ}_{i\pm}}\!\!\bra{\mathrm{GHZ}_{i\pm}}$. 
The eight outcomes of the ion measurements, each occurring with ideally equal probability, 
lead to the following eight orthogonal ion-photon states,
\begin{subequations}
\begin{align}
\ket{
\hspace*{0.5pt}\downarrow
\hspace*{0.5pt}\downarrow
\hspace*{0.5pt}\downarrow
\hspace*{0.5pt}}
\ket{\mathrm{GHZ}_{1-}} &=\,
\ket{
\hspace*{0.5pt}\downarrow
\hspace*{0.5pt}\downarrow
\hspace*{0.5pt}\downarrow
\hspace*{0.5pt}}
\tfrac{1}{\sqrt{2}}\bigl(\ket{AAA}-\ket{DDD}\bigr),\label{eq:ddd}\\[1mm]
\ket{
\hspace*{0.5pt}\downarrow
\hspace*{0.5pt}\downarrow
\hspace*{0.5pt}\uparrow
\hspace*{0.5pt}}
\ket{\mathrm{GHZ}_{2+}} &=\,\ket{
\hspace*{0.5pt}\downarrow
\hspace*{0.5pt}\downarrow
\hspace*{0.5pt}\uparrow
\hspace*{0.5pt}}
\tfrac{1}{\sqrt{2}}\bigl(\ket{AAD}+\ket{DDA}\bigr),\label{eq:ddu}\\[1mm]
\ket{
\hspace*{0.5pt}\downarrow
\hspace*{0.5pt}\uparrow
\hspace*{0.5pt}\downarrow
\hspace*{0.5pt}}
\ket{\mathrm{GHZ}_{3+}} &=\,\ket{
\hspace*{0.5pt}\downarrow
\hspace*{0.5pt}\uparrow
\hspace*{0.5pt}\downarrow
\hspace*{0.5pt}}
\tfrac{1}{\sqrt{2}}\bigl(\ket{ADA}+\ket{DAD}\bigr),\\[1mm]
\ket{
\hspace*{0.5pt}\downarrow
\hspace*{0.5pt}\uparrow
\hspace*{0.5pt}\uparrow
\hspace*{0.5pt}}
\ket{\mathrm{GHZ}_{4-}} &=\,\ket{
\hspace*{0.5pt}\downarrow
\hspace*{0.5pt}\uparrow
\hspace*{0.5pt}\uparrow
\hspace*{0.5pt}}
\tfrac{1}{\sqrt{2}}\bigl(\ket{ADD}-\ket{DAA}\bigr),\\[1mm]
\ket{
\hspace*{0.5pt}\uparrow
\hspace*{0.5pt}\downarrow
\hspace*{0.5pt}\downarrow
\hspace*{0.5pt}}
\ket{\mathrm{GHZ}_{4+}} &=\,\ket{
\hspace*{0.5pt}\uparrow
\hspace*{0.5pt}\downarrow
\hspace*{0.5pt}\downarrow
\hspace*{0.5pt}}
\tfrac{1}{\sqrt{2}}\bigl(\ket{ADD}+\ket{DAA}\bigr),\\[1mm]
\ket{
\hspace*{0.5pt}\uparrow
\hspace*{0.5pt}\downarrow
\hspace*{0.5pt}\uparrow
\hspace*{0.5pt}}
\ket{\mathrm{GHZ}_{3-}} &=\,\ket{
\hspace*{0.5pt}\uparrow
\hspace*{0.5pt}\downarrow
\hspace*{0.5pt}\uparrow
\hspace*{0.5pt}}
\tfrac{1}{\sqrt{2}}\bigl(\ket{ADA}-\ket{DAD}\bigr),\\[1mm]
\ket{
\hspace*{0.5pt}\uparrow
\hspace*{0.5pt}\uparrow
\hspace*{0.5pt}\downarrow
\hspace*{0.5pt}}
\ket{\mathrm{GHZ}_{2-}} &=\,\ket{
\hspace*{0.5pt}\uparrow
\hspace*{0.5pt}\uparrow
\hspace*{0.5pt}\downarrow
\hspace*{0.5pt}}
\tfrac{1}{\sqrt{2}}\bigl(\ket{AAD}-\ket{DDA}\bigr),\label{eq:uud}\\[1mm]
\ket{
\hspace*{0.5pt}\uparrow
\hspace*{0.5pt}\uparrow
\hspace*{0.5pt}\uparrow
\hspace*{0.5pt}}
\ket{\mathrm{GHZ}_{1+}} &=\,\ket{
\hspace*{0.5pt}\uparrow
\hspace*{0.5pt}\uparrow
\hspace*{0.5pt}\uparrow
\hspace*{0.5pt}}
\tfrac{1}{\sqrt{2}}\bigl(\ket{AAA}+\ket{DDD}\bigr).\label{eq:uuu}
\end{align}
\end{subequations}

\begin{figure*}[ht]
\begin{center}
\includegraphics[angle = 0, width=\textwidth,trim={0cm 0cm 0cm 0cm},clip]{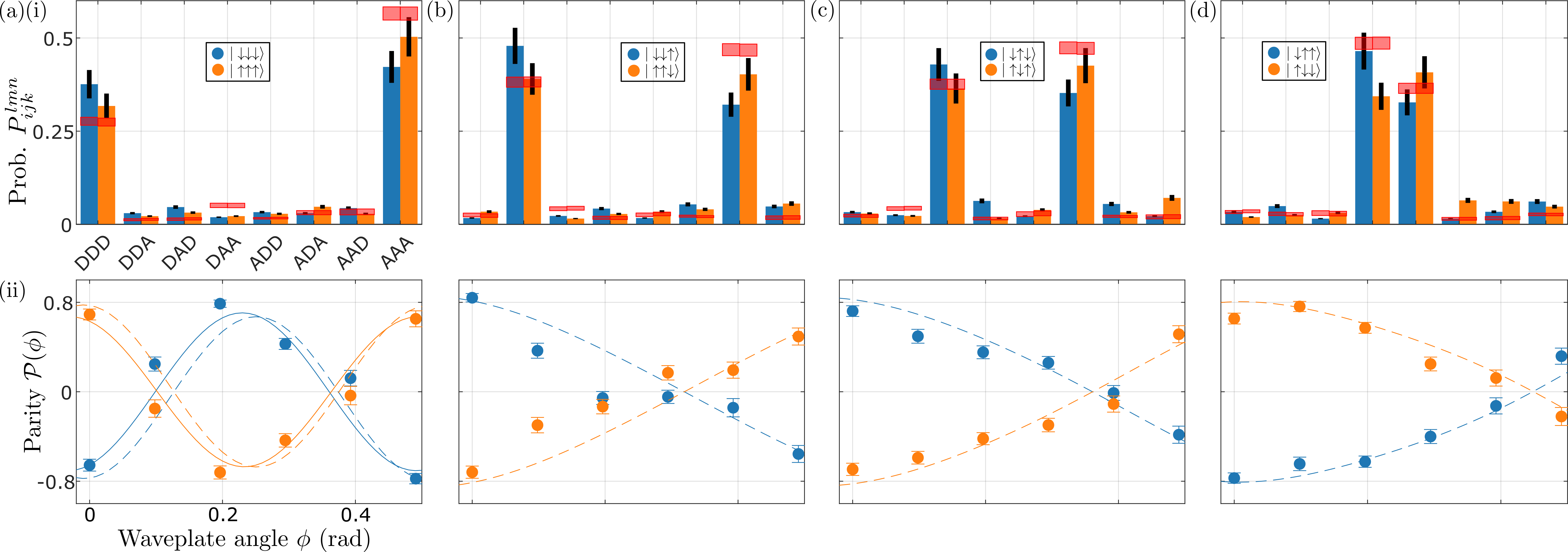}
\scriptsize{ \caption{\textbf{Logical polarisation probabilities and parities of three-photon GHZ states.} 
Top panels show logical polarisation probabilities $P^{lmn}_{ijk}$ for the eight ion-qubit outcomes $lmn$ as labeled in the insets, corresponding to the eight photon states in Eqs.~(\ref{eq:ddd}-\ref{eq:uuu}). Orange and blue bars show data. Red squares show simple model with height representing two standard deviations. D and A are diagonal and antidiagonal polarisations, respectively. 
Bottom panels show the corresponding parities $\mathcal{P}(\phi)$. Shapes show data with colours matching the bars in panels above. Solid lines:  fit to the function $\mathcal{P}=C\sin{(12\phi+\alpha)}$. Dashed lines: numerical model, see Appendix~\ref{appendix:simple model} for details. 
\label{fig:populationsandparities}}}
\end{center}
\end{figure*}

To verify the generation of the eight three-photon states $\ket{\mathrm{GHZ}_{i\pm}}$ for $i=1,2,3,4$, we perform polarization measurements described in more detail in Appendix~\ref{appendix:polarization measurements}. 
Although the wave plates have variable angles that allow us to switch between different measurement settings for the photonic qubits, the angles are not changed between the arrival of the three photons, such that all three are measured in the same polarisation basis. The first of these bases is the single-photon polarisation basis $\{\ket{A},\ket{D}\}$, which we designate as the logical basis for the polarization measurements. Measurements in the logical basis provide estimates of the logical polarisation probabilities $P_{ijk}^{\,lmn}$, where $i,j,k\in\{A,D\}$ denote the photonic state and $l,m,n\in\{\uparrow,\downarrow \}$ denote the ionic state. 
The logical probabilities are given by $P_{ijk}^{\,lmn}=\bra{ijk}\rho^{\,lmn}\ket{\hspace*{0.5pt}ijk}$ with $i,j,k\in\{A,D\}$, where $\rho^{\,lmn}$ is the density matrix of the generated three-photon state in the case of obtaining ion outcome $lmn$. 
Details on the calculation of the logical probabilities can be found in Appendix~\ref{app:popandparity}. 
 
For the other measurements settings, the wave plate angles are set to measure in the single-photon eigenbases $\{\ket{\pm_{\theta}}=(\ket{A}+e^{i\theta}\ket{D})/\sqrt{2}\}$ of $X_{\theta}=R_{\theta}XR_{\theta}^{\dagger}$ for different~$\theta$, where $X=\ket{A}\!\!\bra{D}+\ket{D}\!\!\bra{A}$ is the Pauli~$X$ operator for the logical basis, and $R_{\theta}=\ket{A}\!\!\bra{A}+e^{i\theta}\ket{D}\!\!\bra{D}$ is a rotation around the~$Z$ axis of the Bloch sphere. The angle~$\theta$ is determined by the angle~$\phi$ of the optical axis of the half-wave plate in the analysis path, see Fig.~\ref{fig:experimental_schematic}~(a), via the relationship~$\theta=4\phi$.
Measurements in the eigenbases of $X_{\theta}$ provide estimates of the parities $\mathcal{P}^{lmn}(\theta)=\tr(X_\theta^{\otimes 3}\rho^{\,lmn})$.

The described measurements allow the fidelity $F(\rho,\mathrm{GHZ}_{1\alpha})=\bra{\mathrm{GHZ}_{1\alpha}} \rho\ket{\mathrm{GHZ}_{1\alpha}}$ between any generated three-photon 
state $\rho$ and ideal GHZ states of the form $\ket{\mathrm{GHZ}_{1\alpha}}=(\ket{AAA}+e^{i\alpha}\ket{DDD})\sqrt{2}$ to be determined. If this fidelity exceeds the value~$\tfrac{1}{2}$, the corresponding three-qubit state must be GME (see Appendix~\ref{appendix:biseparability and GME} for a definition). For a proof of this well-known result, which has been widely employed in experiments~\cite{Sackett2000, Leibfried2005, MonzEtAl2011}, we refer to, e.g.,~\cite{GuehneToth2009,FriisVitaglianoMalikHuber2019} or Appendix~\ref{appendix:detecting GME with fid}.

The fidelity to the state $\ket{\mathrm{GHZ}_{1\alpha}}$ is given by
\begin{align}
F(\rho,\mathrm{GHZ}_{1\alpha}) & = \tfrac{1}{2}\bigl(\bra{AAA}\rho\ket{AAA}+
    \bra{DDD}\rho\ket{DDD}\bigr)\!+\!\tfrac{C}{2} \nonumber\\[1mm]
 & =\,  \tfrac{1}{2}(P_{DDD}+P_{AAA})+\tfrac{C}{2}.
 \label{eq:GHZ fidelity from parity}
\end{align}
Here, $C=2\operatorname{Re}\bigl(e^ {i\alpha}\bra{AAA}\rho\ket{DDD}\bigr)$, represents the off-diagonal element of $\rho$ corresponding to $\ket{\mathrm{GHZ}_{1\alpha}}$. This off-diagonal element is the only one that leads to a parity oscillation at angular frequency $3\theta$, whereas all others either result in parity oscillations at $\theta$ or do not contribute at all, as is explained in more detail in Appendix~\ref{appendix:parity measurements}. 
Fitting the measured parities to the function $\mathcal{P}(\theta)=C\cos{(3\theta+\alpha)}$ thus yields an estimate for the amplitude $C$. 
This method (or variants thereof) for determining the fidelity with GHZ states of the form $\ket{\mathrm{GHZ}_{1\alpha}}$ has been used extensively in the trapped-ion community (see, e.g.,~\cite{Sackett2000, Leibfried2005, MonzEtAl2011}), and can be applied to the photonic states in Eqs.~(\ref{eq:ddd}) and~(\ref{eq:uuu}) resulting from the ion measurement outcomes $\uparrow\hspace*{0.5pt}\uparrow\hspace*{0.5pt}\uparrow$ and $\downarrow\hspace*{0.5pt}\downarrow\hspace*{0.5pt}\downarrow$.

For the other ion measurement outcomes [Eqs.~(\ref{eq:ddu}) to~\ref{eq:uud})] we employ a newly developed method that generalizes the approach from \cite[Suppl. Inf., Sec. S.VI.]{BavarescoEtAl2018} and allows us to determine lower bounds on the fidelities $F(\rho,\mathrm{GHZ}_{i\pm})$ for all $i=1,2,3,4$ based only on computational-basis measurements $\{\ket{A}=\ket{0},\ket{D}=\ket{1}\}$ and the first parity measurement $\mathcal{P}^{lmn}(\theta=0)$. Using this notation and the index pair ($j$,$k$) to represent the previous indices $1=00, 2=01, 3=10, 4=11$ in binary (plus 1), along with the logical negation $\lnot$ ($\lnot 0=1$, $\lnot 1=0$) we can compactly write the state fidelities as
\begin{align}
      &F(\rho, \ket{\mathrm{GHZ}_{jk \pm}}) \geq \tfrac{1}{2} \left( \langle 0jk| \rho |0jk\rangle+\langle 1 \neg j \neg k| \rho | 1 \neg j \neg k\rangle \right)\nonumber\\[1mm]   
       &\ \pm  \tfrac{1}{2}\mathcal{P}^{lmn}(0)
          - \hspace*{-7mm}\sum_{\substack{j',k'=0,1 \\ (j',k')\neq(j,k)}}\hspace*{-7mm}
          \sqrt{
          \langle 0j'k'| \rho|0j'k' \rangle  
          \langle 1 \neg j' \neg k'| \rho|1 \neg j'  \neg k'\rangle
          }\,.\nonumber\\[-6mm]
           \label{eq:witness main}
\end{align}
A detailed derivation of these new witnesses is provided in Appendix~\ref{appendix:GME witness derivation}.

\vspace*{-1mm}
\section{Results}\label{sec:results}
\vspace*{-1mm}

{\noindent}First, we characterise the two-qubit density matrices $\rho_i$, with $i\in\{1,2,3\}$, of the generated ion-photon pairs in order of generation time. Ideally these are Bell states. State reconstruction is done in a separate calibration experiment via quantum state tomography (QST), as described in~\cite{KrutyanskiyCanteriMeranerKrcmarskyLanyon2024}. For QST, the MS gate and $U^{\otimes 3}$ operations are removed and replaced with ion-qubit analysis pulses, and ion readout is performed conditional on single-photon detection. 
We calculate the fidelity of the reconstructed ion-photon states $\rho_i^{i-p}$ with their nearest maximally entangled two-qubit states $\ket{A_i}$, via the expression $F_i=\tr(\ket{A_i}\!\!\bra{A_i}\rho_i^{i-p})$. The obtained values are 0.9281(84), 0.9467(70) and 0.9502(55), in order of their generation in time. Numbers in brackets represent uncertainties of one standard deviation in the given value. These fidelities are of comparable quality to those previously obtained~\cite{KrutyanskiyCanteriMeranerKrcmarskyLanyon2024}. The fidelities $F(\rho_j,\rho_i^{i-p})=\bigl(\tr\sqrt{\sqrt{\rho_j}\,\rho_i^{i-p} \sqrt{\rho_j}}\bigr)^{2}$ (see, e.g.,~\cite{NielsenChuang2011} or~\cite[Sec.~24.2.2]{BertlmannFriis2023}) between all pairwise combinations of the three reconstructed mixed states are 0.987(11), 0.975(9), 0.991(7),  
showing that the three states are  largely identical. The maximum fidelity of an arbitrary state $\rho$ with any pure state
is given by its largest eigenvalue, and is hence bounded from above by the square root of the purity of $\rho$. Consequently, for every $\ket{\psi}$, we have that
$F(\rho_i,\ket{\psi}) \le \sqrt{\tr(\rho_i^2)}$. The reconstructed states $\rho_i^{i-p}$ saturate the aforementioned bound for $\ket{\psi}=\ket{A_i}$, to within one standard deviation of uncertainty. Therefore, the infidelities in the generation ion-photon entangled states can be attributed to a lack of purity.
Possible causes are the cumulative effects of imperfections in the 729-nm laser pulses
used in the ion-qubit state analysis, as well as imperfections in the polarization analysis of the photons. A~detailed error budget is beyond the scope of this
work. The measured probabilities with which the single photons were detected are given by 24.1(3)\%, 22.9(3)\%, 20.7(3)\%, in the order of their generation time. The differences in these probabilities are attributed to a sub-optimal calibration of the Raman photon-generation process, compared to the one carrier out in~\cite{KrutyanskiyCanteriMeranerKrcmarskyLanyon2024}, as detailed in Appendix~\ref{appendix:probabilities}.\\[-3mm]

Next we run the full protocol, as described in the previous section. A~total of 1032565 attempts were made to generate three photons and 10037 cases were successful, yielding a success probability of 0.97\%. 
Successful cases correspond to asynchronous detection of a photon within each of three sequential \SI{50}{\micro\second}-wide time windows, which we henceforth refer to as a triple coincidence.   

The measured three-photon logical polarisation probabilities and parities, in the cases in which the ion outcomes $\ket{\hspace*{1pt}\downarrow\downarrow\downarrow\hspace*{1pt}}$ and $\ket{\hspace*{1pt}\uparrow\uparrow\uparrow\hspace*{1pt}}$ were obtained, are shown in Fig.~\ref{fig:populationsandparities}~(a) and are now discussed. 
We obtain $P^{\,\downarrow\downarrow\downarrow}_{DDD}=0.38(4)$, $P^{\,\downarrow\downarrow\downarrow}_{AAA}=0.42(4)$ and $P^{\,\uparrow\uparrow\uparrow}_{DDD}=0.32(3)$, $P^{\,\uparrow\uparrow\uparrow}_{AAA}=0.50(5)$.  
The amplitudes of the parity oscillations at the angular frequency $3\theta$, extracted from a fit in Fig.~\ref{fig:populationsandparities}~(a,lower panel), are $C^{\,\downarrow\downarrow\downarrow}=0.76(7)$ and $C^{\,\uparrow\uparrow\uparrow}=0.70(6)$.  
The difference in the fitted phase value between the two parity oscillations is \SI{3.0(2)}{\radian},
which is consistent with them being fully out of phase ($\pi$~\SI{}{\radian}), as expected from the orthogonality of the ideal states in Eqs.~(\ref{eq:ddd}) and~(\ref{eq:uuu}). 
Inserting the obtained values into Eq.~(\ref{eq:GHZ fidelity from parity}) yields three-photon GHZ-state fidelities of $F^{\,\downarrow\downarrow\downarrow}=0.78(4)$ and $F^{\,\uparrow\uparrow\uparrow}=0.76(4)$, thereby surpassing the 0.5 threshold by 7 and 6 standard deviations, respectively, and proving that genuine multipartite entanglement of GHZ type is present between the three photons.

A simple model for the photonic GHZ states is developed that considers imperfections only in the measured ion-photon states. Specifically, we take the measured ion-photon states $\rho_i$; numerically apply a perfect projection of the ions into the GHZ basis $\ket{\mathrm{GHZ}_{i\pm}}$ and; extract the associated eight photonic states. More details on the simple model are given in Appendix~\ref{appendix:simple model}. The simple  
model predicts photonic GHZ-state fidelities of $F^{\,\downarrow\downarrow\downarrow}=0.793(5)$ and $F^{\,\uparrow\uparrow\uparrow}=0.790(5)$, which are consistent with the 
measured values to within one standard deviation. 
Figure~\ref{fig:populationsandparities}~(b-d) presents logical polarisation probabilities and parities for the other six ion outcomes and compares them with the predictions of the simple model. Based on the agreement between data and simple model, we conclude that the imperfections in the generated photonic states are largely captured by those in the ion-photon states. 

\begin{figure}[t]
\begin{center}
\includegraphics[angle = 0, width=0.5\textwidth,trim={6.5cm 3.5cm 6.5cm 3.5cm},clip]{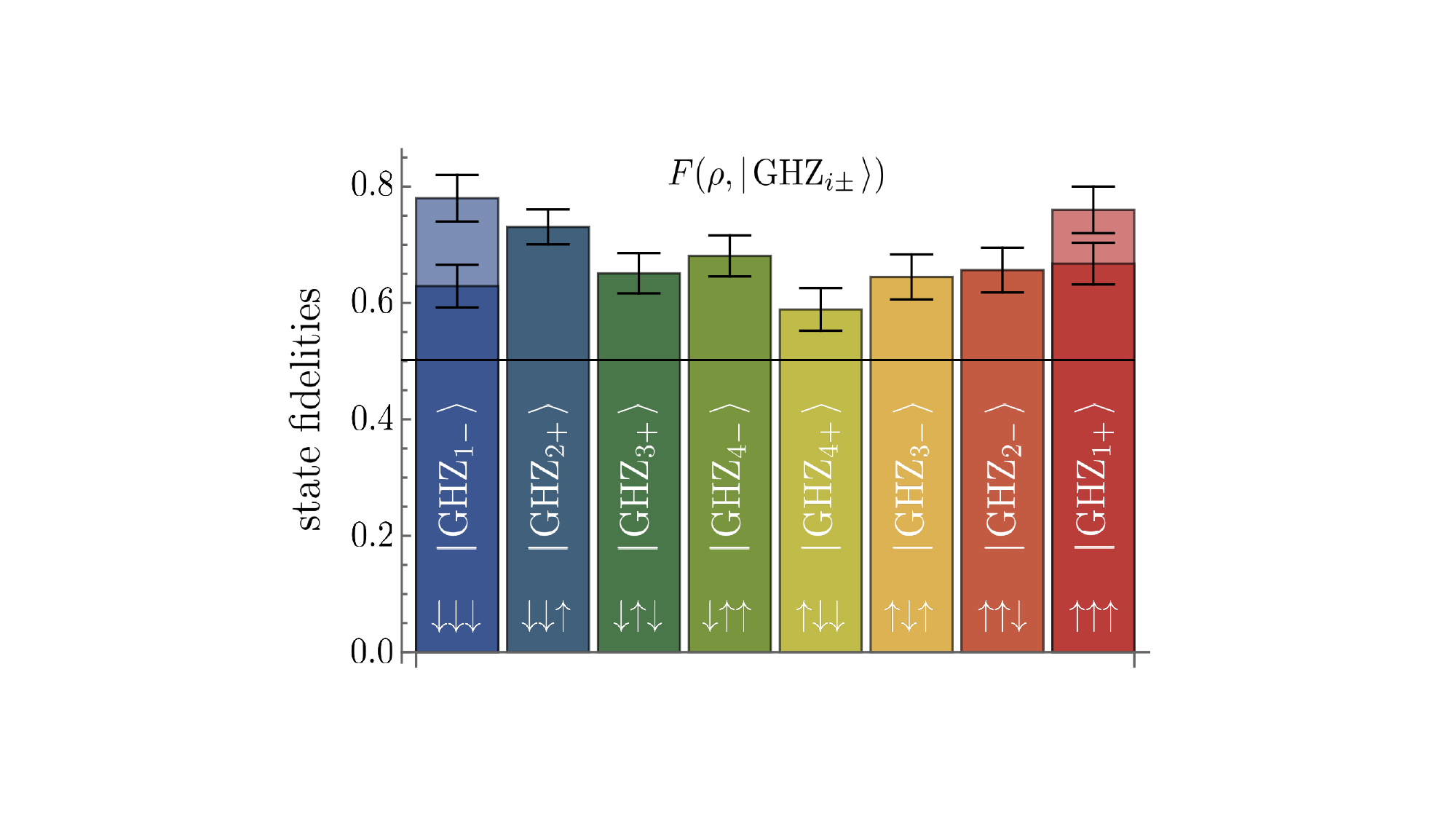}
\scriptsize{ \caption{\textbf{Photonic GHZ-state fidelities.} 
\label{fig:GHZfidelities} 
Lower bounds (darker-color bars) on the fidelities $F(\rho,\ket{\mathrm{GHZ}_{i\pm}})$ to all eight GHZ states $\ket{\mathrm{GHZ}_{i\pm}}$ and exact values for $F(\rho,\ket{\mathrm{GHZ}_{1\mp}})$ (light-blue and light-red bars, respectively) 
are shown in the order presented in Eqs.~(\ref{eq:ddd}){\textendash}(\ref{eq:uuu}). 
Error bars show one standard deviation of the mean.  
The labels $\downarrow\downarrow\downarrow$, $\downarrow\downarrow\uparrow$, etc. on the bars indicate the corresponding outcomes of the ion measurements. 
The horizontal black line indicates the threshold of $0.5$ above which the fidelity indicates that the state $\rho$ is GME.}}
\end{center}
\end{figure}

The results from the evaluation of the lower bounds on the fidelities (see Appendix~\ref{app:popandparity}) for all eight photonic GHZ states are presented in Fig.~\ref{fig:GHZfidelities}, and yield values for $F(\rho,\ket{\mathrm{GHZ}_{i\pm}})$ of at least [0.63(4),\ 0.73(3),\ 0.65(3),\ 0.68(4),\ 0.59(4),\ 0.64(4), 0.66(4),\ 0.67(4)] for the states $\ket{\mathrm{GHZ}_{i\pm}}$ 
in the order presented in Eqs.~(\ref{eq:ddd}){\textendash}(\ref{eq:uuu}). These values are above the 0.5 threshold required to detect GME by at least [3,\ 7,\ 4,\ 4,\ 2,\ 3,\ 4,\ 4] standard deviations, respectively, proving that all eight states are genuinely multipartite  entangled.


\vspace*{-1mm}
\section{Conclusion and outlook}
\vspace*{-1mm}
 
{\noindent}We have demonstrated key functionalities of a factory node based on a cavity-integrated three-qubit trapped-ion quantum processor, capable of generating and distributing multipartite-entangled photonic states.
In future, given more ion-qubits in the factory node, reprogramming such a factory node's logic-gate sequence could allow the distribution of broad classes of states, via the RSP protocol, or arbitrary states, via the teleportation protocol. 
The generated photons could be used to establish, heralded and stored multipartite entanglement between distributed matter-based end nodes, e.g., via the established two-photon click method~\cite{Moehring2007, Duan_2010, Stephenson2020, Krut2022}. Such end nodes need not be capable of universal quantum logic, could be hundreds of kilometers apart~\cite{KrutyanskiyCanteriMeranerKrcmarskyLanyon2024, Zhou2024, Krut2023, vanLeent2022} and could each contain either a single ion qubit or, via the photon wavelength-conversion technique~\cite{Walker2018, Krutyanskiy2019, Bock2018, Hannegan2022}, a qubit encoded into other optically compatible examples of quantum matter.
Once entanglement is stored in such end nodes, the final local correction rotations [Fig.~\ref{fig:scheme}~(d)] could be implemented to deterministically establish distributed stored multipartite entanglement. 

The ion system used in this work was recently employed to demonstrate how GHZ-type states of co-trapped ion qubits can enable noise-protected, optimal sensing of distributed fields at the micron scale \cite{BateHamannCanteriWinklerKoongKrutyanskiyDuerLanyon2025}. Building on this result, the present work establishes a direct route to distribute such probe states across a remote ion-trap network, paving the way for long-distance, optimal sensing of distributed fields over macroscopic distances~\cite{PhysRevResearch.2.023052}.

There are several immediate next steps in the development of the trapped-ion factory node. 
First, the number of photon-entangled qubits in the factory node should be increased, to allow for connection to more end nodes and the generation of more complex entangled states. This step could exploit recently-developed techniques involving ion-string shuttling, in our system~\cite{Canteri2025} and  others~\cite{you2024}, which have so far enabled a photon-interfaced register of up to ten trapped-ion qubits. Finally, for scaling efficiency in end-node number, the deterministic delivery of ion-entangled photons to their different target locations (and thus ultimately the deterministic establishment of remote stored Bell pairs) should be realised. Deterministic delivery could be achieved by storing the ion-qubit that successfully established a remote photon in protected memory states and continuing to establish the remaining Bell states until all are complete. There has been significant progress in the development of ion-qubit memories that are robust to the photon-generation process on co-trapped ions ~\cite{Drmota2023,Krut2023}. 

The data that support the findings of this article will be made publicly available on publication of this work in a peer-reviewed journal. The data are available upon reasonable request from the authors.

\begin{acknowledgments}

We are grateful to Marcus Huber and Wolfgang D\"ur for insightful discussions. 
This work was funded in part by; the Austrian Science Fund (FWF) [Grant DOIs: 10.55776/Y849, 10.55776/P34055, 10.55776/F71 and 10.55776/COE1]; the European Union under the DIGITAL-2021-QCI-01 Digital European Program under Project number No 101091642 and project name `QCI-CAT', and the European Union's Horizon Europe research and innovation programme under grant agreement No. 101102140 and project name `QIA-Phase 1' and NextGenerationEU; the \"Osterreichische Nationalstiftung f\"ur Forschung, Technologie und Entwicklung (AQUnet project).  
B.P.L. acknowledges funding by the CIFAR Quantum Information Science Program of Canada. The opinions expressed in this document reflect only the author's view and reflects in no way the European Commission's opinions. The European Commission is not responsible for any use that may be made of the information it contains. 
I.M. and N.F. acknowledge support from the Austrian Science Fund (FWF) through the project P 36478-N funded by the European Union{\textemdash}NextGenerationEU. N.F. further acknowledges support from the Austrian Federal Ministry of Education, Science and Research via the Austrian Research Promotion Agency (FFG) through the flagship project HPQC (FO999897481), the project FO999914030 (MUSIQ), and the project FO999921407 (HDcode) funded by the European Union{\textemdash}NextGenerationEU.
\end{acknowledgments}

\section*{Author contribution statement}

{\noindent}M.C., J.B. and V.K. contributed to the experimental setup. M.C. and J.B. took the experimental data. M.C. analysed the data. M.C., I.M., N.F. and B.P.L. performed theoretical modelling. B.P.L., M.C., N.F. and I.M. wrote the manuscript, with contributions from all authors. The project was conceived and supervised by B.P.L.

\hypertarget{sec:appendix}
\appendix
\section*{Appendix}

\renewcommand{\thesubsubsection}{A.\Roman{subsection}.\arabic{subsubsection}}
\renewcommand{\thesubsection}{A.\Roman{subsection}}
\renewcommand{\thesection}{}
\setcounter{equation}{0}
\numberwithin{equation}{section}
\setcounter{figure}{0}
\renewcommand{\theequation}{A.\arabic{equation}}
\renewcommand{\thefigure}{A.\arabic{figure}}

\noindent The structure of the appendix is now shortly summarised. Section~\ref{appendix:experimental setup} presents details on both the optical cavity used to collect photons from the ions and the employed single-photon detectors. Section~\ref{appendix:pulse sequence} details the experimental pulse sequence. Section~\ref{appendix:polarization measurements} describes the various settings of the wave plates used to analyse the photon states. Section~\ref{appendix:rawdata} explains how three-photon detection events (`triple coincidences') are extracted from raw data files. Section~\ref{app:popandparity} presents how logical probabilities and parities are extracted from triple coincidences. Section~\ref{appendix:simple model} details our simple model for the generated photon states, the predictions of which are presented in Fig.~\ref{fig:populationsandparities} of the main manuscript. Finally, Sec.~\ref{appendix:detection of GME} details the construction of the entanglement witnesses and their evaluation, the results of which are presented in Fig.~\ref{fig:GHZfidelities} of the main manuscript. 


\subsection{Experimental setup}\label{appendix:experimental setup}

{\noindent}Our cavity is a \SI{19.906(3)}{\milli\meter}-long near-concentric Fabry{\textendash}Perot optical with a \SI{12.31(8)}{\micro\meter} waist at the point of the ions and \SI{10}{\milli\meter} ion-mirror separation~\cite{SchuppKrcmarskyKrutyanskiyMeranerNorthupLanyon2021, SchuppPhDthesis2021}. The cavity axis is close to perpendicular to the ion-string axis, with a relative angle of \SI{85.3(1)}{\degree}~\cite{KrutyanskiyCanteriMeranerKrcmarskyLanyon2024}. The finesse of the mode at \SI{854}{\nano\meter} used for photon generation is $54(1)\times 10^3$. The cavity decay rate is given by $2\kappa = 2\pi \times 140(3)$~kHz. The maximum strength of the coherent coupling between a single photon in the cavity and a single ion is calculated to be $g_0 = 2\pi \times 1.53$~MHz in our system. For further details on our cavity see~\cite{SchuppPhDthesis2021} and for its use to achieve near-deterministic photon extraction see~\cite{SchuppKrcmarskyKrutyanskiyMeranerNorthupLanyon2021}.

The \SI{854}{\nano\meter} cavity photons are detected with two single-mode fiber-coupled superconducting nanowire single-photon detectors. One detector has an efficiency of 0.87(2) and free-running dark counts of 0.3(1) counts per second. The second detector has an efficiency of 0.88(2) with free-running dark counts of 0.5(1) counts per second. The detector efficiencies are calibrated during installation of the system by the manufacturer, using classical light, power meter, and calibrated attenuators. The calibration was cross checked by comparison with an independently calibrated single-photon avalanche photodiode.

\subsection{Pulse sequence}\label{appendix:pulse sequence}

{\noindent}The experimental sequence is now presented as a list, with the durations of each step given in parentheses. 

\begin{enumerate}
		\item Doppler cooling with \SI{397}{\nano\meter} laser and re-pump lasers at \SI{854}{\nano\meter} (\SI{6}{\milli\second}).
		\item\label{step2} Optical pumping to the state $\ket{S}{=}\ket{4^{2} S_{1/2,m_j{=}{-}1/2}}$, using a suitably polarised \SI{397}{\nano\meter} laser  (\SI{40}{\micro\second}).
		\item Sideband-resolved ground-state cooling of the axial center-of-mass (COM) mode on the $\ket{S}\leftrightarrow \ket{M}$ transition, where $\ket{\hspace*{1pt}M\hspace*{1pt}}{=}\ket{3^{2}D_{5/2,m_j{=}-5/2}}$, using a \SI{729}{\nano\meter} laser and \SI{854}{\nano\meter} re-pump laser  (\SI{2}{\milli\second}).
		\item The optical pumping of Step~\ref{step2} is repeated (\SI{40}{\micro\second}).
		\item The \SI{806}{\nano\meter} laser light that is otherwise sent into the optical cavity at all times for length stabilization is switched off (few microseconds).
		\item\label{step5} Start of a loop.		
		\item After every $9^{\mathrm{th}}$ instance of the loop: ground-state cooling for \SI{0.666}{\milli\second} followed by optical pumping for \SI{40}{\micro\second} (total \SI{0.706}{\milli\second}).
		\item Raman pulse (photon generation) on the first ion (\SI{50}{\micro\second}).
		\item Raman pulse on the second ion (\SI{50}{\micro\second}).
		\item Raman pulse on the third ion (\SI{50}{\micro\second}).
		\item End of $n^{\mathrm{th}}$ iteration of loop: go back to loop start (Step~\ref{step5}) if a triple photon coincidence was not detected, otherwise continue (maximum of 30 loops).
		\item Application of a \SI{729}{\nano\meter} $\pi$ pulse, ideally mapping $\ket{M'}\leftrightarrow   \ket{S}$, where $\ket{\hspace*{1pt}M'\hspace*{1pt}}{=}\ket{3^{2}D_{5/2,m_j{=}-3/2}}$ (\SI{9.72}{\micro\second}).
		\item MS gate on the axial COM sideband of the $\ket{S}\leftrightarrow   \ket{M}$ transition via a  \SI{729}{\nano\meter} laser pulse (\SI{167}{\micro\second}).
		\item Application of a \SI{729}{\nano\meter} $\pi/2$ pulse on the $\ket{S}\leftrightarrow \ket{M}$ transition, realising the operation $U^{\otimes 3}$   (\SI{6.41}{\micro\second}).
		\item The \SI{806}{\nano\meter} laser light is sent back into the cavity, for its length stablisation (few microseconds).
		\item Fluorescence state detection of the ion-qubit states using an EMCCD camera (\SI{7}{\milli\second}).
	\end{enumerate}



\subsection{Polarization measurements}\label{appendix:polarization measurements}

{\noindent}A sequence of wave plates followed by a polarising beam splitter (PBS) are used to analyse the photon polarisations. Different angular settings of the optical axes of these wave plates are used to measure in different polarisation bases. 
In this section we model the action of the wave plates on photon polarisation qubits and summarize the wave-plate settings for the different measurements made. 

The three wave plates are in order: a quarter-wave plate $Q_1$, a half-wave plate $H(\phi)$, and another quarter-wave plate $Q_2(\varphi)$, where $\phi$ and $\varphi$ are the angles of the wave plates relative to the respective optical axes. The angle of the first quarter-wave plate is never varied, as described later. We use Jones calculus for modeling the operation of the wave plates of the polarisation of a single photon. The matrix representing the half-wave plate is given by
 \begin{align}
     H(\phi)&= e^{ \frac{-\iota \pi}{2}}
      \begin{pmatrix}
             \cos^2{ \phi}- \sin^2{ \phi}&2 \cos{\phi} \sin{ \phi}\\
             2 \cos{ \phi} \sin{ \phi} & \sin^2{ \phi}-\cos^2{ \phi}
      \end{pmatrix}.
\end{align}
The action of the last quarter-wave plate is represented by
\begin{align}
          Q_2( \varphi)&=e^{ \frac{-\iota \pi}{4}}
          \begin{pmatrix}
          \cos^2{\varphi}+ \iota \sin^2{\varphi} & (1- \iota) \sin{  \varphi} \cos{\varphi}   \\
          (1- \iota) \sin{\varphi} \cos{\varphi}&\sin^2{ \varphi}+ \iota \cos^2{\varphi}
          \end{pmatrix}.
\end{align}
The first quarter-wave plate is always set to its optical axis, i.e., $Q_1=Q_2(0)$.

Experiments are carried out for eight different angular settings of the last two wave plates, as shown in  Table~\ref{settings}, corresponding to measurements of the photonic qubits in eight different polarisation bases. The first two angular settings are used to extract the logical polarisation probabilities. The last six angular settings are used to extract the parities and correspond to measuring in the $\{\ket{\pm_{\theta}}=(\ket{A}\pm e^{i\theta}\ket{D})/\sqrt{2}\}$ basis, by setting $H(\phi) ,Q_2(0)$ for six different angles $\phi = \theta/4$.

\begin{table}[]
\begin{tabular}{|l|l|l|}
\hline
Measurement basis & $\phi$ & $\varphi$  \\ \hline
D/A & 0 & +45 \\ \hline
A/D & 0 & -45 \\ \hline
$\pm_\theta$  & 0 & 0 \\ \hline
$\pm_\theta$  & 5.626 & 0 \\ \hline
$\pm_\theta$  & 11.25 & 0 \\ \hline
$\pm_\theta$  & 16.875 & 0 \\ \hline
$\pm_\theta$  & 22.5 & 0 \\ \hline
$\pm_\theta$  & 28.125 & 0 \\ \hline
\end{tabular}
\caption{Angular settings of the wave plates. All angles are in degrees. $\theta=\phi/4$. $D(A)$ is diagonal (anti-diagonal) linear polarisation. $\ket{\pm_{\theta}}=(\ket{A}\pm e^{i\theta}\ket{D})/\sqrt{2}$.} \label{settings}
\end{table}

\subsection{Extraction of triple coincidences from raw data files.}
\label{appendix:rawdata}
 
{\noindent}The results of measurements performed for any given wave-plate setting (Table~\ref{settings}) corresponds to one timetag file. Each entry in a timetag file contains the detection time of a photon and the particular channel (detector) in which it was detected. We extract triple coincidences as any event that led to the detection of at least one photon in each of the three consecutive time windows corresponding to three consecutive Raman pulses on the three ions (occurring during one iteration of the loop in the experimental pulse sequence, see Sec.~\ref{appendix:pulse sequence}). 

Each of the three photons can be detected either in the transmitted ($t$) or reflected ($r$) port of the polarising beam splitter (PBS). As such,  there are eight different temporal orders in which the detectors can fire and lead to a triple coincidence: $ttt,ttr, trt, trr, rtt, rtr,rrt,rrr$, which we label with the indices $o,p,q\in\{r,t\}$. 
The number of triple coincidences recorded, for a given ion-outcome $l,m,n\in\{\uparrow,\downarrow\}$, is denoted as $n_{opq,  \phi, \varphi}^{lmn}$, where $ \mathrm{\phi}$ and 
$ \mathrm{\varphi}$ refer to the orientation of the half-wave plate and second quarter-wave plate respectively as explained in the previous section.
For example $n_{rtr,  \phi, \varphi}^{\downarrow\uparrow\uparrow}$ is the number of triple coincidences recorded in which: the ion outcome $\ket{\downarrow\uparrow\uparrow}$ was obtained; the photons were detected in the order reflected-transmitted-reflected port; and the wave plate angles are $ \mathrm{\phi}$ and $\mathrm{\varphi}.$

\subsection{Calculation of the logical probabilities, parities, and witnesses}
\label{app:popandparity}

\noindent \textbf{Logical probabilities}. The bar charts in Fig.~\ref{fig:populationsandparities} of the main text present the measured logical polarisation probabilities for the three photon states. The notation introduced in the main text for those probabilities is $P_{ijk}^{\,lmn}$ with $i,j,k\in\{A,D\}$, where $l,m,n\in \{\uparrow,\downarrow\}$ labels the ion outcome.

As now described, the probabilities $P_{ijk}^{\,lmn}$ are calculated from the data by adding up the photon counts obtained from the first two wave plate settings shown in Table~\ref{settings}. This adding up eliminates  bias in the data caused by any differences in the detector efficiencies. Although the quantum efficiency of the detectors are the same up to percent level, the optical paths between the PBS and the detector front facets can have different efficiencies.

The polarisation analysis setting for the first measurement basis in Table~\ref{settings} implements the following mapping of the incoming single-photon polarisation state: $\ket{D}\rightarrow t$ and $\ket{A}\rightarrow r$, where $t$ and $r$ are the transmitted and reflected ports of the polarising beam splitter. The settings for the second measurement basis in Table~\ref{settings} implements the following `flipped' mapping: $\ket{A}\rightarrow t$ and $\ket{D}\rightarrow r$.

Recall from the previous section that the  number of triple coincidences recorded, for a given ion-outcome $lmn\in\{\uparrow,\downarrow\}$, is denoted as $n_{opq,  \phi, \varphi}^{lmn}$, where $opq\in\{r,t\}$. Under the mapping $\ket{D}\rightarrow t$ and $\ket{A}\rightarrow r$, the logical polarisation probabilities are calculated from the recorded coincidences via
\begin{align}
\label{eq:pop}
    P^{lmn}_{ijk} = \frac{n_{opq, 0, +45}^{lmn} + n_{\lnot o \lnot p \lnot q, 0, -45}^{lmn}}{\sum_{o,p,q} n_{opq, 0, +45}^{lmn}+\sum_{o,p,q} n_{opq, 0, -45}^{lmn}}\,,
\end{align}
where the negation $\lnot$ is defined as ($\lnot t=r$, $\lnot r=t$). For example,  
\begin{align}
    P^{\uparrow\uparrow\uparrow}_{DAD} = \frac{n_{trt, 0, +45}^{\uparrow\uparrow\uparrow} + n_{rtr, 0, -45}^{\uparrow\uparrow\uparrow}}{\sum_{o,p,q} n_{opq, 0, +45}^{\uparrow\uparrow\uparrow}+\sum_{o,p,q} n_{opq, 0, -45}^{\uparrow\uparrow\uparrow}}\,.
\end{align}
The errors in the logical probabilities are propagated from the error on the triple coincidence numbers assuming a Poissonian distribution with standard deviation $\sigma \left(n_{opq, \phi, \varphi}^{lmn}\right) = \sqrt{n_{opq, \phi, \varphi}^{lmn}}$\,.\\

\noindent \textbf{Parities}. The oscillating graphs at the bottom of Fig.~\ref{fig:populationsandparities} of the main text present the measured parities ($\mathcal{P})$ for the three-photon states. The notation introduced in the main text for those parities is $\mathcal{P}^{lmn}(\theta)$, where the angle $\theta$ is related to the half-wave-plate angle $\phi$ via $\theta=4\phi$. Unlike the logical probabilities, the parities are not corrected for detector efficiency imbalance by adding up measurements made in flipped bases. Instead, a correction factor is included, as described below. The parities are calculated from the triple coincidences via 

\begin{align}
\label{eq:parity1}
&{\mathcal{P}}^{lmn}(\theta) =  \tfrac{1}{N}\Big(\beta^3 n_{ttt, \phi,0}^{lmn}+\beta n_{rtr, \phi,0}^{lmn}+ 
\beta n_{rrt, \phi,0}^{lmn} + \beta n_{trr, \phi,0}^{lmn}\nonumber\\
& \quad-\beta^2 n_{ttr, \phi,0}^{lmn}-\beta^2 n_{trt, \phi,0}^{lmn} - \beta^2 n_{rtt, \phi,0}^{lmn}-n_{rrr, \phi,0}^{lmn}\Big)\,,
\end{align}
where
\begin{align}\label{eq:parity2}
&N =  \beta^3 n_{ttt, \phi,0}^{lmn}+\beta^2 n_{ttr, \phi,0}^{lmn}+\beta^2 n_{trt, \phi,0}^{lmn}+ \beta n_{trr, \phi,0}^{lmn}+ \nonumber\\
& \quad  \ \beta^2 n_{rtt, \phi,0}^{lmn}+\beta n_{rtr, \phi,0}^{lmn}+ 
\beta n_{rrt, \phi,0}^{lmn}+n_{rrr, \phi,0}^{lmn}\,,
\end{align}
and where $\beta=1.25(1)$ is a factor that corrects for the detector-efficiency imbalance. Specifically, a single photon in the reflected port of the PBS has a higher probability of being detected, by a factor of 1.25, than a photon in the transmitted port. The lower probability in the transmitted port is likely caused by a poorer coupling into the optical fiber  than in the reflected path. 

The $\beta$ parameter is extracted from the single-photon counts recorded for the first two measurements settings (Table~\ref{settings}). We denote the total number of single-photon counts recorded by a given detector (summed over all ion outcomes) as $S_{\mathrm{\phi},\mathrm{\varphi}}^{\,o}$, where $o\in \{r,t\}$ indicates the reflected or transmitted port of the PBS. The value $\beta$ is calculated as
\begin{align}\beta = \frac{S_{0,+45}^{\,r} + S_{0,-45}^{\,r}}{S_{0,+45}^{\,t} + S_{0,-45}^{\,t}}.\end{align}

We now present the main results of the paper for uncorrected counts showing that our conclusions are not affected by this correction. As a reminder, the  values presented in the main text are: the amplitudes of the parity oscillations $C^{\,\downarrow\downarrow\downarrow}=0.76(7)$ and $C^{\,\uparrow\uparrow\uparrow}=0.70(6)$; the difference in the fitted phase value~$\alpha$ between the two parity oscillations, which is \SI{3.0(2)}{\radian}; the three-photon GHZ-state fidelities are $F^{\,\downarrow\downarrow\downarrow}=0.78(4)$ and $F^{\,\uparrow\uparrow\uparrow}=0.76(4)$, surpassing the threshold of 0.5 for genuine multipartite entanglement (GME) by 7 and 6 standard deviations, respectively. Since $\alpha$ only affects parity data, the populations remain the same $P^{\,\downarrow\downarrow\downarrow}_{DDD}=0.38(4)$, $P^{\,\downarrow\downarrow\downarrow}_{AAA}=0.42(4)$ and $P^{\,\uparrow\uparrow\uparrow}_{DDD}=0.32(3)$, $P^{\,\uparrow\uparrow\uparrow}_{AAA}=0.50(5)$.

For the uncorrected values we obtain: $C^{\,\downarrow\downarrow\downarrow}=0.73(7)$ and $C^{\,\uparrow\uparrow\uparrow}=0.72(2)$; a phase difference between the two states $\ket{\mathrm{GHZ}_{1\pm}}$ of $ \SI{3.2(2)}{\radian}$; and the three-photon GHZ-state fidelities are $F^{\,\downarrow\downarrow\downarrow}=0.76(4)$ and $F^{\,\uparrow\uparrow\uparrow}=0.77(4)$, surpassing the 0.5 thresholds by 6 standard deviations in both cases. All the values are the same within a standard deviation.\\

\noindent \textbf{Witnesses:} We will further use these logical probabilities to calculate fidelity witnesses with genuinely multipartite entangled states $\ket{\mathrm{GHZ}_{i\pm}}$ according to Eq.~(\ref{eq:witness}). 
For the witnesses, the computational basis $ \{ \ket{A}, \ket{D}\}$ is denoted as $ \{ \ket{0}, \ket{1}\}$, and the index pair ($j$=0, $k$=1) is used to represent the index $i$ in the state $\ket{\mathrm{GHZ}_{i\pm}}$ such that $1 = 00$, $2=01$, $3=10$, $4=11$. 
For ion outcome $lmn$ the logical probabilities given by $\bra{0ij}\rho\ket{0ij}$ 
can be calculated directly from Eq.~(\ref{eq:pop}) as 
\begin{align}
    \bra{0ij}\rho\ket{0ij} = P_{0ij}^{lmn}\,.
\end{align}

For the fidelity witness we require the parity only for the measurement in the Pauli-X basis, i.e., $\phi=0,  \varphi=0$ denoted by
$\mathcal{P}^{lmn}(0)$, which can be calculated directly from Eq.~(\ref{eq:parity1}).

The lower bounds on the fidelity values presented in the main text and in Appendix~\ref{appendix:GME witness derivation} are determined by including the correction term $ \beta=1.25$. Here, we report the fidelity values without the correction term, i.e., for $\beta=1$: We obtain the values [$0.64(3)$, $0.73(3)$, $0.66(3)$, $0.680(35)$, $0.58(4)$, $0.66(4)$, $0.66(4)$, $0.65(4)$], where the values are ordered according to the ordering of the ion measurement outcomes $\ket{
\hspace*{0.5pt}\downarrow
\hspace*{0.5pt}\downarrow
\hspace*{0.5pt}\downarrow
\hspace*{0.5pt}}$,  
$\ket{
\hspace*{0.5pt}\downarrow
\hspace*{0.5pt}\downarrow
\hspace*{0.5pt}\uparrow
\hspace*{0.5pt}}$, 
$\ket{
\hspace*{0.5pt}\downarrow
\hspace*{0.5pt}\uparrow
\hspace*{0.5pt}\downarrow
\hspace*{0.5pt}}$, 
$\ket{
\hspace*{0.5pt}\downarrow
\hspace*{0.5pt}\uparrow
\hspace*{0.5pt}\uparrow
\hspace*{0.5pt}}$, 
$\ket{
\hspace*{0.5pt}\uparrow
\hspace*{0.5pt}\downarrow
\hspace*{0.5pt}\downarrow
\hspace*{0.5pt}}$, 
$\ket{
\hspace*{0.5pt}\uparrow
\hspace*{0.5pt}\downarrow
\hspace*{0.5pt}\uparrow
\hspace*{0.5pt}}$, 
$\ket{
\hspace*{0.5pt}\uparrow
\hspace*{0.5pt}\uparrow
\hspace*{0.5pt}\downarrow
\hspace*{0.5pt}}$, and 
$\ket{
\hspace*{0.5pt}\uparrow
\hspace*{0.5pt}\uparrow
\hspace*{0.5pt}\uparrow
\hspace*{0.5pt}}$.  
These values are above
the 0.5 threshold required to detect GME by at least
[3,\ 7,\ 4,\ 4,\ 2,\ 3,\ 4,\ 4] standard deviations, proving that all eight states are genuinely multipartite entangled also in the case of $\beta=1$.


\subsection{Simple model}\label{appendix:simple model}
{\noindent}In this section we describe a simple theoretical model for predicting the generated three-photon states. This model was used to predict the logical populations and parities presented in Fig.~\ref{fig:populationsandparities}, as well as the photonic GHZ-state fidelities presented in the main paper. 
The only imperfections considered in the simple model are those in the measured two-qubit density matrices $\rho_i^{i-p}$, with $i\in\{1,2,3\}$, of the generated ion-photon pairs in order of the generation time. Those density matrices were tomographically reconstructed in the calibration experiment reported at the beginning of Sec.~\ref{sec:results}. 

The initial state for the model is a tensor product of the three ion-photon states, i.e., $\rho =  \bigotimes_i \rho_i^{i-p}$. Next a sequence of operations is applied to the initial state, which together corresponds to perfectly projecting the ion qubits into a basis of GHZ states.  
First, we apply an ideal MS gate to the ion qubits, given by the unitary operator
\begin{align}
U_{\mathrm{MS}} = e^{-\frac{\iota \pi}{4} H_{\mathrm{MS}}},
\end{align}
where $H_{\mathrm{MS}} = \sum_{i\neq j}X^i X^j$, and $X^k$ is the Pauli-$X$ operator on ion qubit $k$ with sum running over all pairs of ion qubits. Second, we apply the unitary rotation $U^{\otimes 3}$, corresponding to the same single-qubit rotation $U$ on each ion qubit given by $U = e^{-\iota \pi/4 X}$.
At this point, the resulting state is $\rho^{\mathrm{tot}} = U^{\otimes 3} U_{\mathrm{MS}}\rho U_{\mathrm{MS}}^\dagger(U^{\otimes 3})^\dagger$. Third, the ion qubits are projected into the logical basis, described by the projective operators $M_p = \ket{ijk}\bra{ijk}$
, where $i,j,k \in \{\uparrow, \downarrow\}$. 
The state after the projective measurement is 
\begin{align}
\rho^{\,ijk} = \frac{M_p\rho^{\mathrm{tot}}M_p^\dagger}{\tr\left(M_p^\dagger M_p \rho^{\mathrm{tot}}\right)}.
\end{align}
Fourth, the state of the three photonic qubits is obtained by tracing out the ion states via
\begin{align}
\rho^{\mathrm{GHZ}} = \tr_{\mathrm{ion}}(\rho^{\,ijk}). 
\end{align}
Finally, the photonic state is rotated by the action of the wave plates via $\rho^{\mathrm{theory}} = Q_2(\varphi)H(\phi)\rho^{\mathrm{GHZ}}H(\phi)^\dagger Q_2(\varphi)^\dagger$ with angles taken from Table~\ref{settings}. The final photonic state $\rho^{\mathrm{theory}}$ is used to extract the logical probabilities and the parities, as explained in the previous section. 
Note that the first quarter-wave plate is not modeled as we extract the density matrices $\rho^{i-p}_i$ already in rotated basis after $Q_1$, i.e., the tomographical reconstruction is done for the state after~$Q_1$.

\subsection{Detection of Genuine Multipartite Entanglement}\label{appendix:detection of GME}

{\noindent}In this section of the appendix we first provide definitions for biseparability and genuine multipartite entanglement (GME) in Sec.~\ref{appendix:biseparability and GME} and briefly explain how GME can be detected via the fidelity to pure GME states in Sec.~\ref{appendix:detecting GME with fid}. We then explain how the fidelity to standard (up to the choice of relative phase) GHZ states can be estimated from parity measurements in Sec.~\ref{appendix:parity measurements}, 
before deriving lower bounds for the fidelity to all eight GHZ-type states (with relative phases $n\pi$) based on measurements in two bases in Sec.~\ref{appendix:GME witness derivation}.
\subsubsection{Biseparability and Genuine Multipartite Entanglement}\label{appendix:biseparability and GME}

{\noindent}Multipartite states can be characterized on the basis of their separability with respect to different groupings of subsystems (partitions).
For instance, three-qubit pure states $ \ket{\phi_{ABC}}$ in a Hilbert space $\mathcal{H}_{A} \otimes \mathcal{H}_{B} \otimes \mathcal{H}_{C}$, can be classified as fully separable, biseparable, or genuinely multipartite entangled (GME). Pure states are called \emph{fully separable} if they are product states with respect to all subsystems, $\ket{ \phi_{A|B|C}}= \ket{ \psi_{A}} \otimes\ket{ \eta_{B}} \otimes\ket{ \zeta_{C}}$. 
Pure states are \emph{biseparable} 
when the state can be written as a tensor product of two factors, i.e., two of the three subsystems may form an entangled state but the third subsystem is separable from the other two, e.g., $\ket{ \phi_{A|BC}}=\ket{\psi_A}\otimes\ket{\chi_{BC}}$, and similarly for $\ket{ \phi_{AB|C}}$ and $\ket{ \phi_{AC|B}}$. States that are not biseparable are called \emph{genuinely multipartite entangled}. 
For mixed states, biseparable states include incoherent mixtures of states that are separable with respect to different partitions. As such, mixed biseparable states can be entangled across all partitions, and can thus be multipartite entangled, but are not GME, since they can be formed by mixing states with only bipartite entanglement. For a more detailed introduction to multipartite entanglement, we refer to~\cite[Chapter~18]{BertlmannFriis2023}.


\subsubsection{Detection of Genuine Multipartite Entanglement from Fidelity}\label{appendix:detecting GME with fid}

{\noindent}For three-qubit states, one example of pure GME states are \emph{Greenberger{\textendash}Horne{\textendash}Zeilinger} (GHZ) states, the most prominent representative being
\begin{align}
      \ket{\mathrm{GHZ}_{1+}}= \frac{1}{ \sqrt{2}}( \ket{000}+ \ket{111})\,.
       \label{eq:GHZ}
\end{align}
Here, $\ket{0}$ and $\ket{1}$ are the computational-basis states, which correspond to the linear-polarization states $\ket{A}=\ket{0}$ and $\ket{D}=\ket{1}$ in our experiment. However, for the sake of presenting the techniques for the estimation of fidelity and detection of GME in a system-agnostic and compact fashion, we state general results in terms of the computational basis $\{\ket{0},\ket{1}\}$ in the following, unless stated otherwise. 

For every biseparable pure state in the Hilbert space $\mathcal{H}_{ABC}$, e.g., for $\ket{\phi_{A|BC}}=\ket{\eta_A}\otimes\ket{\chi_{BC}}$, the fidelity $F(\ket{ \phi_{A|BC}},\ket{\psi_{ABC}})$ with an arbitrary pure state $\ket{\psi_{ABC}}$ is bounded as
\begin{align}
 F(\ket{ \phi_{A|BC}},\ket{\psi_{ABC}})\,=\, 
 | \bra{ \eta_{A}} \langle \chi_{BC}|\psi_{ABC} \rangle|^2  \leq \lambda^2,
\end{align}
where $\lambda$ is the largest Schmidt coefficient of $\ket{\psi_{ABC}}$
with respect to the bipartition that separates subsystem $A$ from the subsystem formed by $B$ and $C$ together. That is, $\ket{\psi_{ABC}}=\sum_i \lambda_i \ket{\eta_A^i}\otimes\ket{\chi_{BC}^i}$ for some bases (the Schmidt bases) $\{\ket{\eta_A^i}\}_i$ and $\{\ket{\chi_{BC}^j}\}_j$ of $\mathcal{H}_A$ and $\mathcal{H}_B\otimes\mathcal{H}_C$, respectively. Furthermore, the Schmidt coefficients can be chosen so that $\lambda_i\geq0$ for all~$i$ and $\lambda=\max_i \lambda_i$. If the fidelity of an unknown pure state $\ket{\tilde{\phi}_{ABC}}$ to any given state $\ket{\psi_{ABC}}$ exceeds the value of the largest squared Schmidt coefficients for all bipartitions, then it is not separable with respect to any bipartition, and hence GME. Since any mixed biseparable state is a convex sum of pure biseparable states, this bound extends to mixed states by convexity. This well-known result constitutes a standard technique for the detection of GME, see, e.g.,~\cite{GuehneToth2009,FriisVitaglianoMalikHuber2019} or \cite[Sec.~18.3.3]{BertlmannFriis2023}.

The GHZ state $\ket{\mathrm{GHZ}_{1+}}$ is symmetric with respect to the exchange of the three subsystems, so the largest Schmidt coefficient is the same for all bipartitions and has the value $\lambda= \frac{1}{ \sqrt{2}}$. Therefore, every 
mixed biseparable state $\rho$ satisfies the inequality
\begin{align}
    F(\rho,\ket{\mathrm{GHZ}_{1+}}) &=\,
  \bra{\mathrm{GHZ}_{1+}}\rho\ket{\mathrm{GHZ}_{1+}}\,\leq\, \tfrac{1}{2}\,.
\end{align}
Conversely, every state $\rho$ with fidelity greater than $\tfrac{1}{2}$ must be GME. 

This result can be seen to hold not just for the GHZ state above, but for all pure tripartite states $U_A\otimes V_B\otimes W_C\ket{\mathrm{GHZ}_{1+}}$ that are equivalent to $\ket{\mathrm{GHZ}_{1+}}$ up to local unitaries $U_A$, $V_B$, and $W_C$. In particular, it is true for all eight states in the basis formed by
\begin{subequations}
\begin{align}
    \ket{\mathrm{GHZ}_{1\pm}}= \frac{1}{ \sqrt{2}}( \ket{000}\pm \ket{111})\,,
       \label{eq:GHZ1pm}\\[1mm]
       \ket{\mathrm{GHZ}_{2\pm}}= \frac{1}{ \sqrt{2}}( \ket{001}\pm \ket{110})\,,
       \label{eq:GHZ2pm}\\[1mm]
       \ket{\mathrm{GHZ}_{3\pm}}= \frac{1}{ \sqrt{2}}( \ket{010}\pm \ket{101})\,,
       \label{eq:GHZ3pm}\\[1mm]
       \ket{\mathrm{GHZ}_{4\pm}}= \frac{1}{ \sqrt{2}}( \ket{011}\pm \ket{100})\,.
       \label{eq:GHZ4pm}
\end{align}
\end{subequations}
That is, every state $\rho$ for which 
\begin{align}
    F(\rho,\ket{\mathrm{GHZ}_{i\pm}}) &=\,
  \bra{\mathrm{GHZ}_{i\pm}}\rho\ket{\mathrm{GHZ}_{i\pm}}\,>\, \tfrac{1}{2}\,
\end{align}
for any combination of the index $i=1,2,3,4$ and the sign~$\pm$ following the subscript index is GME.

\subsubsection{Estimating the GHZ Fidelity Using Parity Measurements}\label{appendix:parity measurements}

{\noindent}Employing the witnesses for GME discussed in the previous section requires estimates of, or at least lower bounds on the fidelity to GHZ-type states. 
For states of the form $\ket{\mathrm{GHZ}_{1\alpha}}=(\ket{000}+e^{i\alpha}\ket{111})/\sqrt{2}$ the fidelity can be estimated by measuring three-qubit observables of the form $X_{\theta}^{\otimes 3}$ for different~$\theta$, where $X_{\theta}=R_{\theta}XR_{\theta}^{\dagger}$, where $X=\ket{0}\!\!\bra{1}+\ket{1}\!\!\bra{0}$ is the Pauli~$X$ operator, and $R_{\theta}=\ket{0}\!\!\bra{0}+e^{i\theta}\ket{1}\!\!\bra{1}$ is a rotation around the~$Z$ axis of the Bloch sphere. 
Although there are other ways of estimating the fidelity $F(\rho,\ket{\mathrm{GHZ}_{1\alpha}})$, this approach, which we will explain in the following, thus only requires measuring all qubits in the same basis. In our setup this has the advantage that the wave plates do not need to be adjusted between the arrival of different photons, but this is also advantageous when measuring trapped ions, since no individual addressing of ions is required.

The key to the estimation of the fidelity in this way lies in the observation that the operators $X_{\theta}^{\otimes 3}$ for all $\theta$ have non-zero entries only on the main anti-diagonal with respect to the computational basis. As such, the only density-matrix elements that contribute to expectation values $\tr(X_{\theta}^{\otimes 3}\rho)$ are 
\begin{subequations}
\begin{align}
    \bra{000}\rho\ket{111}  &=\,C\exp(-i\alpha)/2,\\[1mm]
    \bra{001}\rho\ket{110}  &=\,\tilde{C}_1\exp(-i\beta_1)/2,\\[1mm]
    \bra{010}\rho\ket{101}  &=\,\tilde{C}_2\exp(-i\beta_2)/2,\\[1mm]
    \bra{011}\rho\ket{100}  &=\,\tilde{C}_3\exp(-i\beta_3)/2,
\end{align} 
\end{subequations}
and their complex conjugates, which we parameterize with amplitudes $C,\tilde{C}_i\geq 0$ and phases $\alpha$ and $\beta_i$ for $i=1,2,3$. Specifically, the expectation value takes the form
\begin{align}
    \mathcal{P}(\theta)\coloneqq\tr(X_{\theta}^{\otimes 3}\rho) &=\,
    C\cos(3\theta+\alpha)+\sum\limits_i \tilde{C}_i \cos(\theta+\beta_i).
\end{align}
Consequently, $\mathcal{P}(\theta)$ has contributions from oscillations with two different angular frequencies, $3\theta$ and $\theta$. 

The amplitude $C$ of the component of the $3\theta$ oscillation, which can be obtained, from a suitable fitting function or by averaging the amplitude of $\mathcal{P}(\theta)$ over many periods, thus gives an estimate of the off-diagonal element $\bra{000}\rho\ket{111}$. 
In combination with measurements of $Z^{\otimes3}$, which provides estimates of the diagonal elements $\bra{000}\rho\ket{000}$ and $\bra{111}\rho\ket{111}$, we can obtain the fidelity
\begin{align}
F(\rho,\mathrm{GHZ}_{1\alpha}) & = \, \tfrac{1}{2}\bigl(\bra{000}\rho\ket{000}+
    \bra{111}\rho\ket{111}\bigr)+C/2\nonumber\\[1mm]
 & =\,  \tfrac{1}{2}(P_{000}+P_{111})+C/2 \,.
 \label{eq:GHZ fidelity from parity appendix}
\end{align}
However, for GHZ-type states in subspaces that are not spanned by $\ket{000}$ and $\ket{111}$, this method does not work, since the contributions to $\mathcal{P}(\theta)$ arising from such states all oscillate with the same frequency and therefore cannot be isolated. For such states, we present a method for bounding the fidelity from below from similar, but fewer measurements in the next section.


\subsubsection{Lower Bounds on the Fidelities to GHZ-Type States Using Measurements in Two Bases}\label{appendix:GME witness derivation}

{\noindent}To construct fidelity witnesses for GHZ-type states from fewer measurements we again note that the fidelity of an unknown (measured in the lab) state $\rho$ with, e.g., the state $\ket{\mathrm{GHZ}_{1+}}= \frac{1}{\sqrt{2}}
\left (\ket{000}+\ket{111}\right)$ is:
\begin{align}
     F(\rho, \ket{\mathrm{GHZ}_{1+}}) = \tfrac{1}{2}  ( \langle 000| \rho |000\rangle+\langle 111| \rho |111 \rangle+
     \nonumber\\  \quad
     \langle 000| \rho |111 \rangle+
     \langle 111| \rho |000 \rangle) 
      \label{eq:fid}
\end{align}
The first two terms correspond to diagonal elements of the density matrix and can be obtained from measurements in the computational basis, i.e., measuring each qubit in the basis $\{\ket{0},\ket{1}\}$. The remaining off-diagonal terms cannot be obtained from measurements in the computational basis, but can be bounded from parity measurements in the Pauli-$X$ basis. That is, denoting the eigenstates of $X=\ket{0}\!\!\bra{1}+\ket{1}\!\!\bra{0}$ as $\ket{\pm}=\bigl(\ket{0}\pm\ket{1}\bigr)/\sqrt{2}$, we can write the expectation value of $X^{\otimes 3}$ as the difference of the probabilities $\mathcal{P}_{+}$ and $\mathcal{P}_{-}$ of obtaining even and odd numbers of outcomes, i.e.,
\begin{align}
&\mathcal{P}_{+}-\mathcal{P}_{-} =\tr( \rho X\otimes X\otimes X)= \sum_{i,j,k= \pm}i\cdot j\cdot k\, \ket{\hspace*{0.5pt}i,j,k} \bra{i,j,k}
      \nonumber\\ 
      &\ \ \ =\,\langle +++| \rho |+++ \rangle+\langle +--| \rho |+-- \rangle + \langle --+| \rho |--+ \rangle \nonumber\\[1mm] 
      &\ \ \ \ +\langle -+-| \rho |-+- \rangle 
          - \langle +-+| \rho |+-+ \rangle-\langle -++| \rho |-++ \rangle \nonumber\\[1mm] 
    & \ \ \ \ -\langle ++-| \rho |++- \rangle-\langle ---| \rho |---\rangle\\[2mm] 
      &\ \ \ =\,\langle 000| \rho |111 \rangle+\langle 111| \rho |000 \rangle +\langle 001| \rho |110 \rangle 
      +
          \langle 110| \rho |001  \rangle\nonumber\\[1mm]  
      &\ \ \ \  
          +\langle 101| \rho |010 \rangle+\langle 010| \rho |101 \rangle + \langle 100| \rho |011 \rangle+\langle 011| \rho |100\rangle,
          \nonumber
\end{align}
such that we have 
 \begin{align}
     &F(\rho, \ket{\mathrm{GHZ}_{1+}}) = \tfrac{1}{2} \bigl( \langle 000| \rho |000\rangle\!+\!\langle 111| \rho |111 \rangle \bigr)\!+\!\tfrac{1}{2}\bigl(\mathcal{P}_{+}\!-\!\mathcal{P}_{-}\bigr)
      \nonumber \\[1mm]
     & -\bigl[\operatorname{Re} \bigl(\langle 001| \rho |110 \rangle \bigr)\!+\!
      \operatorname{Re} \bigl(\langle 010| \rho |101 \rangle \bigr)\!+\!\operatorname{Re} \bigl(\langle 100| \rho |011 \rangle\bigr)\bigr].
 \end{align}
We can bound the real parts by including a sign change $\pm \operatorname{Re}(\langle abc| \rho|xyz \rangle) \leq|\langle abc| \rho|xyz \rangle|$, for expressions appearing in other GHZ-type states like $ \ket{\mathrm{GHZ}_{1-}}$ and using the Cauchy-Schwarz inequality $| \langle abc| \rho|xyz \rangle| \leq \sqrt{ \langle abc| \rho|abc \rangle \langle xyz| \rho|xyz \rangle}$. 
Putting everything together, and recalling that $\mathcal{P}( \phi, \varphi=0)=\mathcal{P}_{+}-\mathcal{P}_{-}$ we obtain the fidelity bound
 \begin{align}
   &   F(\rho, \ket{\mathrm{GHZ}_{1+}})  \geq \tfrac{1}{2} \bigl( \langle 000| \rho |000\rangle+\langle 111| \rho |111 \rangle \bigr)+\tfrac{1}{2}\mathcal{P}(0)
      \nonumber\\[1mm]
      &\quad -\,\bigl
            ( \sqrt{ \langle 001| \rho|001 \rangle \langle 110| \rho|110 \rangle} +\,\sqrt{ \langle 010| \rho|010 \rangle \langle 101| \rho|101 \rangle}\nonumber\\ 
   & \quad +\,\sqrt{ \langle 011| \rho|011\rangle \langle 100| \rho|100 \rangle}\,\bigr).
    \label{eq:fidwithghz+}
\end{align}
For the other GHZ states we just have to exchange the computational basis matrix elements and adjust the sign in front of the term $ \frac{1}{2}(\mathcal{P}_{+}-\mathcal{P}_{-})$, so that we can compactly write

\begin{align}
      &F(\rho, \ket{\mathrm{GHZ}_{jk \pm}}) \geq \tfrac{1}{2} \left( \langle 0jk| \rho |0jk\rangle+\langle 1 \neg j \neg k| \rho | 1 \neg j \neg k\rangle \right)\nonumber\\[1mm]   
       &\ \pm  \tfrac{1}{2}\mathcal{P}^{lmn}(0)
          - \hspace*{-7mm}\sum_{\substack{j',k'=0,1 \\ (j',k')\neq(j,k)}}\hspace*{-7mm}
          \sqrt{
          \langle 0j'k'| \rho|0j'k' \rangle  
          \langle 1 \neg j' \neg k'| \rho|1 \neg j'  \neg k'\rangle
          }\,.\nonumber\\[-6mm]
           \label{eq:witness}
\end{align}
where we have used the index pair $(j,k)$ to represent the previous indices $1=00, 2=01, 3=10, 4=11$ in binary (plus 1), and $ \neg$ is the logical negation, $\neg 0=1, \neg 1=0$.
Via the results of Sec.~\ref{appendix:detecting GME with fid}, the fidelity bounds in Eq.~(\ref{eq:witness}) provide  witnesses for the detection of a complete set of genuinely multipartite entangled states in the experiment.

The photonic GHZ-state fidelities obtained from theoretical modeling and parity fitting, as well as the estimated lower bounds are compared in Table~\ref{table:labelforthetable}.\\

\onecolumngrid
\setlength\extrarowheight{2.0pt}
\begin{medboxtable}{Fidelities of three-photon states}{table:labelforthetable}
\begin{center}
\begin{tabular}{|c|c|c|c|c|c|c|c|c|}
\hline
     \makecell{Ion result} & 
     \makecell{$\downarrow\downarrow\downarrow$}  &   
     \makecell{$\downarrow\downarrow\uparrow$} &   
     \makecell{$\downarrow\uparrow\downarrow$} &   
     \makecell{$\downarrow\uparrow\uparrow$} & 
     \makecell{$\uparrow\downarrow\downarrow$}  &   
     \makecell{$\uparrow\downarrow\uparrow$} &   
     \makecell{$\uparrow\uparrow\downarrow$} &   
     \makecell{$\uparrow\uparrow\uparrow$}  \\ \hline\hline
    \makecell{Three-photon state} &
    \makecell{$\ket{\mathrm{GHZ}_{1-}}$} &
    \makecell{$\ket{\mathrm{GHZ}_{2+}}$} &
    \makecell{$\ket{\mathrm{GHZ}_{3+}}$} &
    \makecell{$\ket{\mathrm{GHZ}_{4-}}$} &
    \makecell{$\ket{\mathrm{GHZ}_{4+}}$} &
    \makecell{$\ket{\mathrm{GHZ}_{3-}}$} &
    \makecell{$\ket{\mathrm{GHZ}_{2-}}$} &
    \makecell{$\ket{\mathrm{GHZ}_{1+}}$}
     \\ \hline\hline
     \makecell{Fidelity model $F_{\mathrm{th.}}$} & 
     $0.79$  & $0.81$ & $0.82$  & $0.81$ & $0.81$  & $0.82$ & $0.81$  & $0.79$ \\ \hline\hline
     \makecell{Fidelity parity  $F_{\mathcal{P}}$} & 
     $0.78(4)$  & $-$ & $-$  & $-$ & $-$  & $-$ & $-$  & $0.76(4)$ \\ \hline\hline
     \makecell{Lower bound $F_{\beta=1.25}$} & 
     0.63(4) & 0.73(3)& 0.65(3)& 0.68(4)& 0.59(4)&
0.64(4)& 0.66(4)& 0.67(4)
     \\ \hline
\end{tabular}
\end{center}
\small{Summary of fidelities and witnesses obtained. $F_{\mathrm{th}}$ is the expected theoretical fidelity from the simple model that considers imperfections in the measured ion-photon Bell states. $F_{\mathcal{P}}$ is the fidelity measured in the experiment via parity fitting. 
$F_{\beta=1.25}$ are the lower bounds from the entanglement witnesses, shown in Figure 4 of the main text.
}
\end{medboxtable}
\twocolumngrid


\subsection{Single-photon wave packets and detection probabilities}\label{appendix:probabilities}
\vspace*{-2mm}

\begin{figure}[t!]
\begin{center}
\includegraphics[width=0.5\textwidth]{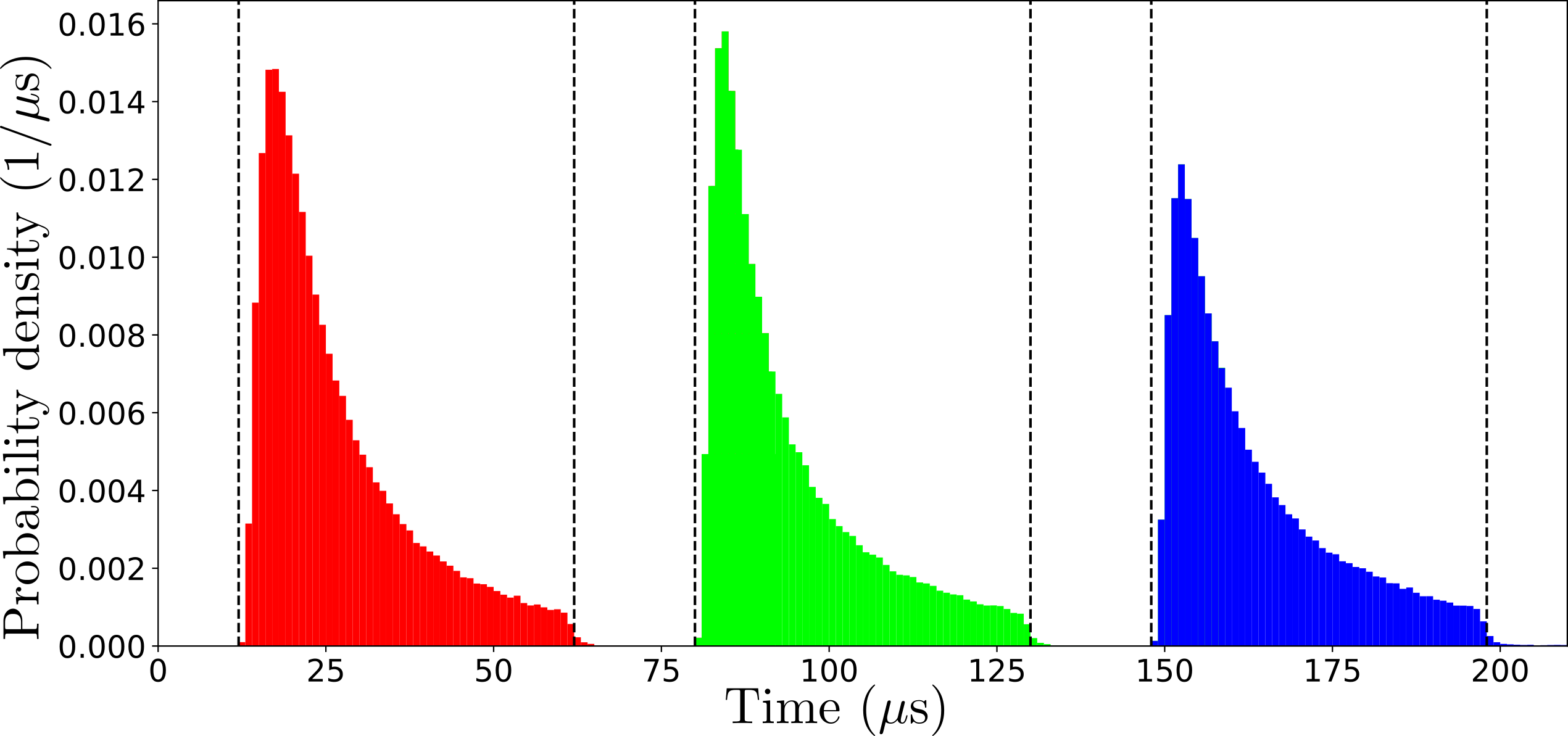}
\scriptsize{ \caption{\textbf{Histograms of photon-arrival times.} Data are taken from the results of measurements in the eight bases presented in Table~\ref{settings}. Probability densities are shown on the vertical axis: The number of counts has been normalized by the number of attempts and by the \SI{1}{\micro\second} time-bin width. Three single-photon wave packets are visible: The color denotes the ion that is expected to have produced the photon (c.f. Fig.~\ref{fig:experimental_schematic}). Black dashed vertical lines indicate the three \SI{50}{\micro\second}-long time windows in which data is subsequently analysed. 
\label{fig:wavepackets}}}
\end{center}
\end{figure}

{\noindent}Histograms of all single-photon detection events recorded over the eight measurement settings, summarized in Table~\ref{settings}, are presented in Fig.~\ref{fig:wavepackets}. The measured probabilities for detecting a photon in the first, second, and third window are 23.18(5)\%, 20.95(5)\%, and 18.89(4)\%, respectively, obtained by integrating over the corresponding time windows in Fig.~\ref{fig:wavepackets}. These probabilities are associated with the generation of a photon from ions 1, 2, and 3, respectively. The probability of detecting a triple coincidence over the eight data sets is 0.97(1)\%, corresponding to the probability to get at least one photon in each of the three time windows, normalized by the number of attempts. This probability differs from the product of the three single-photon probabilities reported above (0.917(3)\%).  
We attribute this difference to drifting single-photon probabilities (photon efficiencies from the three ions) over the measurement set, as discussed below. In the case in which the single-photon probabilities are constant over the measurement time, one expects these two numbers to be equal.  

In our previous work~\cite{KrutyanskiyCanteriMeranerKrcmarskyLanyon2024}, higher
single-photon detection probabilities from three co-trapped ions of 0.315(3)\%, 0.347(3)\%, and 0.320(3)\% were achieved and matched well with a detailed model that quantifies the sources of various inefficiencies. In the present work, the focus was not on achieving optimal single-photon  efficiency. Causes of the lower efficiencies in the present work are now shortly described. First, the optical coupling efficiencies into the fibers connected to the detectors were not optimised (Fig.~\ref{fig:experimental_schematic}~a). Second, the length of the Raman pulse was not chosen to optimise the photon efficiency: a longer pulse length would lead to higher efficiencies, as evidenced by the abrupt decay of the single-photon wave packets at the end of the \SI{50}{\micro\second}-long windows in Fig.~\ref{fig:wavepackets}. Finally, we expect that the focus of the Raman laser beam was not optimally aligned with the position of each ion, as described below. 

When taking the data reported in the present work, the spatial overlap between the location of each ion and the location of the focus of the Raman laser beam (when aimed at a specific ion) had not been optimised for several days. It is therefore likely that the ions were not positioned at the intensity maxima of the Gaussian beam, but on the slopes: displaced in a direction transverse to the beam-propagation direction. In that case, the Rabi frequency (relevant laser-ion coupling strength) is extremely sensitive to any further relative displacements over the time during which data is taken. Such further displacements, causing a change in Rabi frequency, lead to a change in the photon-generation probability as the BCMRT process is driven off-resonantly. For example, when ions are placed off-centre, simulations show that further displacements of the ions on the order of \SI{50}{\nano\meter} can reduce the probability of photon production fractionally by tens of percent (see Appendix II.C of~\cite{Canteri2025}). Further displacements between the ion's positions and laser-beam focus on this length scale can occur due to, e.g., the ion string moving along the string-axis direction as the ion trap rethermalises following loading from a hot atomic oven, or by the alignment drifts of the optical system projecting the focused Raman laser beam. 

In addition, a calibration of the Raman photon-generation process performed before taking the data in the presented work likely lead to a higher photon efficiency from ion 1 as compared with the other ions, and more instability in the efficiencies from the others. This process entails finding the Raman laser frequency that maximises the photon-detection efficiency from a given ion, and updating the value if needed. The process was carried out on ion 1 and an adjustment was made. The same adjustment was made to the Raman beam frequency sent to all three ions, potentially leading to the Raman processes on the second and third ions being driven off-resonantly. The photon-generation efficiency of an off-resonantly driven Raman process is more sensitive to fluctuations in system parameters during data taking than otherwise. 

The measurements made in the eight different bases (Table~\ref{settings}) were taken over a period of one hour. The single-photon detection probabilities recorded over the eight bases changed significantly. For example, during measurements in the first two bases the average single-photon probabilities were [23.793(97)\%, 21.799(93)\%, 19.735(80)\%]. 
During measurements in the fourth and fifth basis those values were 
[23.720(97)\%, 23.267(97)\%, 21.180(92)\%].
During measurements in the last two bases the average single-photon probabilities had dropped to [21.966(90)\%, 18.225(83)\%, 16.526(78)\%]. In each case, probabilities are estimated by dividing the total number of recorded counts and dividing by the number of photon-generation attempts. The uncertainty is the standard deviation calculated as the square root of the number of recorded counts.  
A description of the calibration and optimisation processes for the Raman photon-generation process that leads to stable and higher single-photon detection efficiencies for the three-ion configuration employed in the present work is described in~\cite{KrutyanskiyCanteriMeranerKrcmarskyLanyon2024}.


\bibliographystyle{apsrev4-1fixed_with_article_titles_full_names_new}
\bibliography{Master_Bib_File}

\begin{thebibliography}{88}%
\makeatletter
\providecommand \@ifxundefined [1]{%
 \@ifx{#1\undefined}
}%
\providecommand \@ifnum [1]{%
 \ifnum #1\expandafter \@firstoftwo
 \else \expandafter \@secondoftwo
 \fi
}%
\providecommand \@ifx [1]{%
 \ifx #1\expandafter \@firstoftwo
 \else \expandafter \@secondoftwo
 \fi
}%
\providecommand \natexlab [1]{#1}%
\providecommand \enquote  [1]{#1}%
\providecommand \bibnamefont  [1]{#1}%
\providecommand \bibfnamefont [1]{#1}%
\providecommand \citenamefont [1]{#1}%
\providecommand \href@noop [0]{\@secondoftwo}%
\providecommand \href [0]{\begingroup \@sanitize@url \@href}%
\providecommand \@href[1]{\@@startlink{#1}\@@href}%
\providecommand \@@href[1]{\endgroup#1\@@endlink}%
\providecommand \@sanitize@url [0]{\catcode `\\12\catcode `\$12\catcode
  `\&12\catcode `\#12\catcode `\^12\catcode `\_12\catcode `\%12\relax}%
\providecommand \@@startlink[1]{}%
\providecommand \@@endlink[0]{}%
\providecommand \url  [0]{\begingroup\@sanitize@url \@url }%
\providecommand \@url [1]{\endgroup\@href {#1}{\urlprefix }}%
\providecommand \urlprefix  [0]{URL }%
\providecommand \Eprint [0]{\href }%
\providecommand \doibase [0]{https://doi.org/}%
\providecommand \selectlanguage [0]{\@gobble}%
\providecommand \bibinfo  [0]{\@secondoftwo}%
\providecommand \bibfield  [0]{\@secondoftwo}%
\providecommand \translation [1]{[#1]}%
\providecommand \BibitemOpen [0]{}%
\providecommand \bibitemStop [0]{}%
\providecommand \bibitemNoStop [0]{.\EOS\space}%
\providecommand \EOS [0]{\spacefactor3000\relax}%
\providecommand \BibitemShut  [1]{\csname bibitem#1\endcsname}%
\let\auto@bib@innerbib\@empty
\bibitem [{\citenamefont {Kimble}(2008)}]{Kimble2008}%
  \BibitemOpen
  \bibfield  {author} {\bibinfo {author} {\bibfnamefont {H.~Jeff}\ \bibnamefont
  {Kimble}},\ }\emph {\enquote {\bibinfo {title} {The quantum internet},}\
  }\href {https://doi.org/10.1038/nature07127} {\bibfield  {journal} {\bibinfo
  {journal} {Nature}\ }\textbf {\bibinfo {volume} {453}},\ \bibinfo {pages}
  {1023} (\bibinfo {year} {2008})},\ \Eprint {http://arxiv.org/abs/0806.4195}
  {arXiv:0806.4195}\BibitemShut {NoStop}%
\bibitem [{\citenamefont {Duan}\ and\ \citenamefont
  {Monroe}(2010)}]{Duan_2010}%
  \BibitemOpen
  \bibfield  {author} {\bibinfo {author} {\bibfnamefont {Lu-Ming}\ \bibnamefont
  {Duan}}\ and\ \bibinfo {author} {\bibfnamefont {Christopher}\ \bibnamefont
  {Monroe}},\ }\emph {\enquote {\bibinfo {title} {{Colloquium: Quantum networks
  with trapped ions}},}\ }\href {https://doi.org/10.1103/RevModPhys.82.1209}
  {\bibfield  {journal} {\bibinfo  {journal} {Rev. Mod. Phys.}\ }\textbf
  {\bibinfo {volume} {82}},\ \bibinfo {pages} {1209} (\bibinfo {year}
  {2010})}\BibitemShut {NoStop}%
\bibitem [{\citenamefont {Reiserer}\ and\ \citenamefont
  {Rempe}(2015)}]{Reiserer_2015}%
  \BibitemOpen
  \bibfield  {author} {\bibinfo {author} {\bibfnamefont {Andreas}\ \bibnamefont
  {Reiserer}}\ and\ \bibinfo {author} {\bibfnamefont {Gerhard}\ \bibnamefont
  {Rempe}},\ }\emph {\enquote {\bibinfo {title} {Cavity-based quantum networks
  with single atoms and optical photons},}\ }\href
  {https://doi.org/10.1103/RevModPhys.87.1379} {\bibfield  {journal} {\bibinfo
  {journal} {Rev. Mod. Phys.}\ }\textbf {\bibinfo {volume} {87}},\ \bibinfo
  {pages} {1379} (\bibinfo {year} {2015})},\ \Eprint
  {http://arxiv.org/abs/1412.2889} {arXiv:1412.2889}\BibitemShut {NoStop}%
\bibitem [{\citenamefont {Wehner}\ \emph {et~al.}(2018)\citenamefont {Wehner},
  \citenamefont {Elkouss},\ and\ \citenamefont
  {Hanson}}]{WehnerElkoussHanson2018}%
  \BibitemOpen
  \bibfield  {author} {\bibinfo {author} {\bibfnamefont {Stephanie}\
  \bibnamefont {Wehner}}, \bibinfo {author} {\bibfnamefont {David}\
  \bibnamefont {Elkouss}}, \ and\ \bibinfo {author} {\bibfnamefont {Ronald}\
  \bibnamefont {Hanson}},\ }\emph {\enquote {\bibinfo {title} {Quantum
  internet: A vision for the road ahead},}\ }\href
  {https://doi.org/10.1126/science.aam9288} {\bibfield  {journal} {\bibinfo
  {journal} {Science}\ }\textbf {\bibinfo {volume} {362}},\ \bibinfo {pages}
  {eaam9288} (\bibinfo {year} {2018})}\BibitemShut {NoStop}%
\bibitem [{\citenamefont {Wei}\ \emph {et~al.}(2022)\citenamefont {Wei},
  \citenamefont {Jing}, \citenamefont {Zhang}, \citenamefont {Liao},
  \citenamefont {Yuan}, \citenamefont {Fan}, \citenamefont {Lyu}, \citenamefont
  {Zhou}, \citenamefont {Wang}, \citenamefont {Deng}, \citenamefont {Song},
  \citenamefont {Oblak}, \citenamefont {Guo},\ and\ \citenamefont
  {Zhou}}]{Wei_2022}%
  \BibitemOpen
  \bibfield  {author} {\bibinfo {author} {\bibfnamefont {Shi-Hai}\ \bibnamefont
  {Wei}}, \bibinfo {author} {\bibfnamefont {Bo}~\bibnamefont {Jing}}, \bibinfo
  {author} {\bibfnamefont {Xue-Ying}\ \bibnamefont {Zhang}}, \bibinfo {author}
  {\bibfnamefont {Jin-Yu}\ \bibnamefont {Liao}}, \bibinfo {author}
  {\bibfnamefont {Chen-Zhi}\ \bibnamefont {Yuan}}, \bibinfo {author}
  {\bibfnamefont {Bo-Yu}\ \bibnamefont {Fan}}, \bibinfo {author} {\bibfnamefont
  {Chen}\ \bibnamefont {Lyu}}, \bibinfo {author} {\bibfnamefont {Dian-Li}\
  \bibnamefont {Zhou}}, \bibinfo {author} {\bibfnamefont {You}\ \bibnamefont
  {Wang}}, \bibinfo {author} {\bibfnamefont {Guang-Wei}\ \bibnamefont {Deng}},
  \bibinfo {author} {\bibfnamefont {Hai-Zhi}\ \bibnamefont {Song}}, \bibinfo
  {author} {\bibfnamefont {Daniel}\ \bibnamefont {Oblak}}, \bibinfo {author}
  {\bibfnamefont {Guang-Can}\ \bibnamefont {Guo}}, \ and\ \bibinfo {author}
  {\bibfnamefont {Qiang}\ \bibnamefont {Zhou}},\ }\emph {\enquote {\bibinfo
  {title} {{Towards Real-World Quantum Networks: A Review}},}\ }\href
  {https://doi.org/10.1002/lpor.202100219} {\bibfield  {journal} {\bibinfo
  {journal} {Laser Photonics Rev.}\ }\textbf {\bibinfo {volume} {16}} (\bibinfo
  {year} {2022})},\ \Eprint {http://arxiv.org/abs/2201.04802}
  {arXiv:2201.04802}\BibitemShut {NoStop}%
\bibitem [{\citenamefont {Covey}\ \emph {et~al.}(2023)\citenamefont {Covey},
  \citenamefont {Weinfurter},\ and\ \citenamefont {Bernien}}]{Covey2023}%
  \BibitemOpen
  \bibfield  {author} {\bibinfo {author} {\bibfnamefont {Jacob~P.}\
  \bibnamefont {Covey}}, \bibinfo {author} {\bibfnamefont {Harald}\
  \bibnamefont {Weinfurter}}, \ and\ \bibinfo {author} {\bibfnamefont {Hannes}\
  \bibnamefont {Bernien}},\ }\emph {\enquote {\bibinfo {title} {Quantum
  networks with neutral atom processing nodes},}\ }\href
  {https://doi.org/10.1038/s41534-023-00759-9} {\bibfield  {journal} {\bibinfo
  {journal} {npj Quantum Inf.}\ }\textbf {\bibinfo {volume} {9}},\ \bibinfo
  {pages} {90} (\bibinfo {year} {2023})},\ \Eprint
  {http://arxiv.org/abs/2304.02088} {arXiv:2304.02088}\BibitemShut {NoStop}%
\bibitem [{\citenamefont {Azuma}\ \emph {et~al.}(2023)\citenamefont {Azuma},
  \citenamefont {Economou}, \citenamefont {Elkouss}, \citenamefont {Hilaire},
  \citenamefont {Jiang}, \citenamefont {Lo},\ and\ \citenamefont
  {Tzitrin}}]{Azuma2023}%
  \BibitemOpen
  \bibfield  {author} {\bibinfo {author} {\bibfnamefont {Koji}\ \bibnamefont
  {Azuma}}, \bibinfo {author} {\bibfnamefont {Sophia~E.}\ \bibnamefont
  {Economou}}, \bibinfo {author} {\bibfnamefont {David}\ \bibnamefont
  {Elkouss}}, \bibinfo {author} {\bibfnamefont {Paul}\ \bibnamefont {Hilaire}},
  \bibinfo {author} {\bibfnamefont {Liang}\ \bibnamefont {Jiang}}, \bibinfo
  {author} {\bibfnamefont {Hoi-Kwong}\ \bibnamefont {Lo}}, \ and\ \bibinfo
  {author} {\bibfnamefont {Ilan}\ \bibnamefont {Tzitrin}},\ }\emph {\enquote
  {\bibinfo {title} {Quantum repeaters: From quantum networks to the quantum
  internet},}\ }\href {https://doi.org/10.1103/RevModPhys.95.045006} {\bibfield
   {journal} {\bibinfo  {journal} {Rev. Mod. Phys.}\ }\textbf {\bibinfo
  {volume} {95}},\ \bibinfo {pages} {045006} (\bibinfo {year} {2023})},\
  \Eprint {http://arxiv.org/abs/2212.10820} {arXiv:2212.10820}\BibitemShut
  {NoStop}%
\bibitem [{\citenamefont {Tittel}\ \emph {et~al.}(2025)\citenamefont {Tittel},
  \citenamefont {Afzelius}, \citenamefont {Kinos}, \citenamefont {Rippe},\ and\
  \citenamefont {Walther}}]{Tittel_2025}%
  \BibitemOpen
  \bibfield  {author} {\bibinfo {author} {\bibfnamefont {Wolfgang}\
  \bibnamefont {Tittel}}, \bibinfo {author} {\bibfnamefont {Mikael}\
  \bibnamefont {Afzelius}}, \bibinfo {author} {\bibfnamefont {Adam}\
  \bibnamefont {Kinos}}, \bibinfo {author} {\bibfnamefont {Lars}\ \bibnamefont
  {Rippe}}, \ and\ \bibinfo {author} {\bibfnamefont {Andreas}\ \bibnamefont
  {Walther}},\ }\emph {\enquote {\bibinfo {title} {Quantum networks using
  rare-earth ions},}\ }\href {https://doi.org/10.1088/2058-9565/addd93}
  {\bibfield  {journal} {\bibinfo  {journal} {Quantum Sci. Technol.}\ }\textbf
  {\bibinfo {volume} {10}},\ \bibinfo {pages} {033002} (\bibinfo {year}
  {2025})},\ \Eprint {http://arxiv.org/abs/2501.06110}
  {arXiv:2501.06110}\BibitemShut {NoStop}%
\bibitem [{\citenamefont {Monroe}\ \emph {et~al.}(2014)\citenamefont {Monroe},
  \citenamefont {Raussendorf}, \citenamefont {Ruthven}, \citenamefont {Brown},
  \citenamefont {Maunz}, \citenamefont {Duan},\ and\ \citenamefont
  {Kim}}]{Monroe_2014}%
  \BibitemOpen
  \bibfield  {author} {\bibinfo {author} {\bibfnamefont {Christopher}\
  \bibnamefont {Monroe}}, \bibinfo {author} {\bibfnamefont {Robert}\
  \bibnamefont {Raussendorf}}, \bibinfo {author} {\bibfnamefont {Angela}\
  \bibnamefont {Ruthven}}, \bibinfo {author} {\bibfnamefont {Kenneth~R.}\
  \bibnamefont {Brown}}, \bibinfo {author} {\bibfnamefont {Peter}\ \bibnamefont
  {Maunz}}, \bibinfo {author} {\bibfnamefont {Lu-Ming}\ \bibnamefont {Duan}}, \
  and\ \bibinfo {author} {\bibfnamefont {Jungsang}\ \bibnamefont {Kim}},\
  }\emph {\enquote {\bibinfo {title} {Large-scale modular quantum-computer
  architecture with atomic memory and photonic interconnects},}\ }\href
  {https://doi.org/10.1103/PhysRevA.89.022317} {\bibfield  {journal} {\bibinfo
  {journal} {Phys. Rev. A}\ }\textbf {\bibinfo {volume} {89}},\ \bibinfo
  {pages} {022317} (\bibinfo {year} {2014})},\ \Eprint
  {http://arxiv.org/abs/1208.0391} {arXiv:1208.0391}\BibitemShut {NoStop}%
\bibitem [{\citenamefont {Brown}\ \emph {et~al.}(2016)\citenamefont {Brown},
  \citenamefont {Kim},\ and\ \citenamefont {Monroe}}]{Brown2016}%
  \BibitemOpen
  \bibfield  {author} {\bibinfo {author} {\bibfnamefont {Kenneth~R.}\
  \bibnamefont {Brown}}, \bibinfo {author} {\bibfnamefont {Jungsang}\
  \bibnamefont {Kim}}, \ and\ \bibinfo {author} {\bibfnamefont {Christopher}\
  \bibnamefont {Monroe}},\ }\emph {\enquote {\bibinfo {title} {Co-designing a
  scalable quantum computer with trapped atomic ions},}\ }\href
  {https://doi.org/10.1038/npjqi.2016.34} {\bibfield  {journal} {\bibinfo
  {journal} {npj Quantum Inf.}\ }\textbf {\bibinfo {volume} {2}},\ \bibinfo
  {pages} {16034} (\bibinfo {year} {2016})},\ \Eprint
  {http://arxiv.org/abs/1602.02840} {arXiv:1602.02840}\BibitemShut {NoStop}%
\bibitem [{\citenamefont {K{\'o}m{\'a}r}\ \emph {et~al.}(2014)\citenamefont
  {K{\'o}m{\'a}r}, \citenamefont {Kessler}, \citenamefont {Bishof},
  \citenamefont {Jiang}, \citenamefont {S{\o}rensen}, \citenamefont {Ye},\ and\
  \citenamefont {Lukin}}]{Komar2014}%
  \BibitemOpen
  \bibfield  {author} {\bibinfo {author} {\bibfnamefont {Peter}\ \bibnamefont
  {K{\'o}m{\'a}r}}, \bibinfo {author} {\bibfnamefont {Eric~M.}\ \bibnamefont
  {Kessler}}, \bibinfo {author} {\bibfnamefont {Michael}\ \bibnamefont
  {Bishof}}, \bibinfo {author} {\bibfnamefont {Liang}\ \bibnamefont {Jiang}},
  \bibinfo {author} {\bibfnamefont {Anders~S.}\ \bibnamefont {S{\o}rensen}},
  \bibinfo {author} {\bibfnamefont {Jun}\ \bibnamefont {Ye}}, \ and\ \bibinfo
  {author} {\bibfnamefont {Mikhail~D.}\ \bibnamefont {Lukin}},\ }\emph
  {\enquote {\bibinfo {title} {A quantum network of clocks},}\ }\href
  {https://doi.org/10.1038/nphys3000} {\bibfield  {journal} {\bibinfo
  {journal} {Nat. Phys.}\ }\textbf {\bibinfo {volume} {10}},\ \bibinfo {pages}
  {582} (\bibinfo {year} {2014})},\ \Eprint {http://arxiv.org/abs/1310.6045}
  {arXiv:1310.6045}\BibitemShut {NoStop}%
\bibitem [{\citenamefont {Muralidharan}\ \emph {et~al.}(2016)\citenamefont
  {Muralidharan}, \citenamefont {Li}, \citenamefont {Kim}, \citenamefont
  {L{\"u}tkenhaus}, \citenamefont {Lukin},\ and\ \citenamefont
  {Jiang}}]{Muralidharan2016}%
  \BibitemOpen
  \bibfield  {author} {\bibinfo {author} {\bibfnamefont {Sreraman}\
  \bibnamefont {Muralidharan}}, \bibinfo {author} {\bibfnamefont {Linshu}\
  \bibnamefont {Li}}, \bibinfo {author} {\bibfnamefont {Jungsang}\ \bibnamefont
  {Kim}}, \bibinfo {author} {\bibfnamefont {Norbert}\ \bibnamefont
  {L{\"u}tkenhaus}}, \bibinfo {author} {\bibfnamefont {Mikhail~D.}\
  \bibnamefont {Lukin}}, \ and\ \bibinfo {author} {\bibfnamefont {Liang}\
  \bibnamefont {Jiang}},\ }\emph {\enquote {\bibinfo {title} {Optimal
  architectures for long distance quantum communication},}\ }\href
  {https://doi.org/10.1038/srep20463} {\bibfield  {journal} {\bibinfo
  {journal} {Sci. Rep.}\ }\textbf {\bibinfo {volume} {6}},\ \bibinfo {pages}
  {20463} (\bibinfo {year} {2016})},\ \Eprint {http://arxiv.org/abs/1509.08435}
  {arXiv:1509.08435}\BibitemShut {NoStop}%
\bibitem [{\citenamefont {Zapatero}\ \emph {et~al.}(2023)\citenamefont
  {Zapatero}, \citenamefont {van Leent}, \citenamefont {Arnon-Friedman},
  \citenamefont {Liu}, \citenamefont {Zhang}, \citenamefont {Weinfurter},\ and\
  \citenamefont {Curty}}]{Zapatero2023}%
  \BibitemOpen
  \bibfield  {author} {\bibinfo {author} {\bibfnamefont {V{\'i}ctor}\
  \bibnamefont {Zapatero}}, \bibinfo {author} {\bibfnamefont {Tim}\
  \bibnamefont {van Leent}}, \bibinfo {author} {\bibfnamefont {Rotem}\
  \bibnamefont {Arnon-Friedman}}, \bibinfo {author} {\bibfnamefont {Wen-Zhao}\
  \bibnamefont {Liu}}, \bibinfo {author} {\bibfnamefont {Qiang}\ \bibnamefont
  {Zhang}}, \bibinfo {author} {\bibfnamefont {Harald}\ \bibnamefont
  {Weinfurter}}, \ and\ \bibinfo {author} {\bibfnamefont {Marcos}\ \bibnamefont
  {Curty}},\ }\emph {\enquote {\bibinfo {title} {Advances in device-independent
  quantum key distribution},}\ }\href
  {https://doi.org/10.1038/s41534-023-00684-x} {\bibfield  {journal} {\bibinfo
  {journal} {npj Quantum Inf.}\ }\textbf {\bibinfo {volume} {9}},\ \bibinfo
  {pages} {10} (\bibinfo {year} {2023})},\ \Eprint
  {http://arxiv.org/abs/2208.12842} {arXiv:2208.12842}\BibitemShut {NoStop}%
\bibitem [{\citenamefont {Zhang}\ and\ \citenamefont
  {Zhuang}(2021)}]{Zhang_2021}%
  \BibitemOpen
  \bibfield  {author} {\bibinfo {author} {\bibfnamefont {Zheshen}\ \bibnamefont
  {Zhang}}\ and\ \bibinfo {author} {\bibfnamefont {Quntao}\ \bibnamefont
  {Zhuang}},\ }\emph {\enquote {\bibinfo {title} {Distributed quantum
  sensing},}\ }\href {https://doi.org/10.1088/2058-9565/abd4c3} {\bibfield
  {journal} {\bibinfo  {journal} {Quantum Sci. Technol.}\ }\textbf {\bibinfo
  {volume} {6}},\ \bibinfo {pages} {043001} (\bibinfo {year} {2021})},\ \Eprint
  {http://arxiv.org/abs/2010.14744} {arXiv:2010.14744}\BibitemShut {NoStop}%
\bibitem [{\citenamefont {Proctor}\ \emph {et~al.}(2018)\citenamefont
  {Proctor}, \citenamefont {Knott},\ and\ \citenamefont
  {Dunningham}}]{Proctor_2018}%
  \BibitemOpen
  \bibfield  {author} {\bibinfo {author} {\bibfnamefont {Timothy~J.}\
  \bibnamefont {Proctor}}, \bibinfo {author} {\bibfnamefont {Paul~A.}\
  \bibnamefont {Knott}}, \ and\ \bibinfo {author} {\bibfnamefont {Jacob~A.}\
  \bibnamefont {Dunningham}},\ }\emph {\enquote {\bibinfo {title}
  {{Multiparameter Estimation in Networked Quantum Sensors}},}\ }\href
  {https://doi.org/10.1103/PhysRevLett.120.080501} {\bibfield  {journal}
  {\bibinfo  {journal} {Phys. Rev. Lett.}\ }\textbf {\bibinfo {volume} {120}},\
  \bibinfo {pages} {080501} (\bibinfo {year} {2018})},\ \Eprint
  {http://arxiv.org/abs/1707.06252} {arXiv:1707.06252}\BibitemShut {NoStop}%
\bibitem [{\citenamefont {Ekert}(1991)}]{Ekert1991}%
  \BibitemOpen
  \bibfield  {author} {\bibinfo {author} {\bibfnamefont {Artur~K.}\
  \bibnamefont {Ekert}},\ }\emph {\enquote {\bibinfo {title} {{Quantum
  cryptography based on Bell's theorem}},}\ }\href {\doibase
  10.1103/PhysRevLett.67.661} {\bibfield  {journal} {\bibinfo  {journal} {Phys.
  Rev. Lett.}\ }\textbf {\bibinfo {volume} {67}},\ \bibinfo {pages}
  {661{\textendash}663} (\bibinfo {year} {1991})}\BibitemShut {NoStop}%
\bibitem [{\citenamefont {Broadbent}\ \emph {et~al.}(2009)\citenamefont
  {Broadbent}, \citenamefont {Fitzsimons},\ and\ \citenamefont
  {Kashefi}}]{BroadbentFitzsimonsKashefi2009}%
  \BibitemOpen
  \bibfield  {author} {\bibinfo {author} {\bibfnamefont {Anne}\ \bibnamefont
  {Broadbent}}, \bibinfo {author} {\bibfnamefont {Joseph}\ \bibnamefont
  {Fitzsimons}}, \ and\ \bibinfo {author} {\bibfnamefont {Elham}\ \bibnamefont
  {Kashefi}},\ }\emph {\enquote {\bibinfo {title} {{Universal Blind Quantum
  Computation}},}\ }in\ \href {https://doi.org/10.1109/FOCS.2009.36} {\emph
  {\bibinfo {booktitle} {2009 50th Annual IEEE Symposium on Foundations of
  Computer Science, Atlanta, GA, USA}}}\ (\bibinfo {year} {2009})\ pp.\
  \bibinfo {pages} {517--526},\ \Eprint {http://arxiv.org/abs/0807.4154}
  {arXiv:0807.4154}\BibitemShut {NoStop}%
\bibitem [{\citenamefont {Ruskuc}\ \emph {et~al.}(2025)\citenamefont {Ruskuc},
  \citenamefont {Wu}, \citenamefont {Green}, \citenamefont {Hermans},
  \citenamefont {Pajak}, \citenamefont {Choi},\ and\ \citenamefont
  {Faraon}}]{Ruskuc2025}%
  \BibitemOpen
  \bibfield  {author} {\bibinfo {author} {\bibfnamefont {Andrei}\ \bibnamefont
  {Ruskuc}}, \bibinfo {author} {\bibfnamefont {Chun-Ju}\ \bibnamefont {Wu}},
  \bibinfo {author} {\bibfnamefont {Emanuel}\ \bibnamefont {Green}}, \bibinfo
  {author} {\bibfnamefont {Sophie L.~N.}\ \bibnamefont {Hermans}}, \bibinfo
  {author} {\bibfnamefont {William}\ \bibnamefont {Pajak}}, \bibinfo {author}
  {\bibfnamefont {Joonhee}\ \bibnamefont {Choi}}, \ and\ \bibinfo {author}
  {\bibfnamefont {Andrei}\ \bibnamefont {Faraon}},\ }\emph {\enquote {\bibinfo
  {title} {Multiplexed entanglement of multi-emitter quantum network nodes},}\
  }\href {https://doi.org/10.1038/s41586-024-08537-z} {\bibfield  {journal}
  {\bibinfo  {journal} {Nature}\ }\textbf {\bibinfo {volume} {639}},\ \bibinfo
  {pages} {54} (\bibinfo {year} {2025})},\ \Eprint
  {http://arxiv.org/abs/2402.16224} {arXiv:2402.16224}\BibitemShut {NoStop}%
\bibitem [{\citenamefont {Shi}\ \emph {et~al.}(2025)\citenamefont {Shi},
  \citenamefont {Zhang}, \citenamefont {Wu}, \citenamefont {Sun}, \citenamefont
  {Liang}, \citenamefont {Wang}, \citenamefont {Pu},\ and\ \citenamefont
  {Duan}}]{shi2025}%
  \BibitemOpen
  \bibfield  {author} {\bibinfo {author} {\bibfnamefont {Jixuan}\ \bibnamefont
  {Shi}}, \bibinfo {author} {\bibfnamefont {Sheng}\ \bibnamefont {Zhang}},
  \bibinfo {author} {\bibfnamefont {Yukai}\ \bibnamefont {Wu}}, \bibinfo
  {author} {\bibfnamefont {Yuedong}\ \bibnamefont {Sun}}, \bibinfo {author}
  {\bibfnamefont {Yibo}\ \bibnamefont {Liang}}, \bibinfo {author}
  {\bibfnamefont {Hai}\ \bibnamefont {Wang}}, \bibinfo {author} {\bibfnamefont
  {Yunfei}\ \bibnamefont {Pu}}, \ and\ \bibinfo {author} {\bibfnamefont
  {Luming}\ \bibnamefont {Duan}},\ }\emph {\enquote {\bibinfo {title}
  {{Scalable and Modular Generation of $W$-State Entanglements via
  Memory-Enhanced Fusion}},}\ }\href {https://doi.org/10.1103/w4w4-lf6r}
  {\bibfield  {journal} {\bibinfo  {journal} {Phys. Rev. Lett.}\ }\textbf
  {\bibinfo {volume} {135}},\ \bibinfo {pages} {150802} (\bibinfo {year}
  {2025})},\ \Eprint {http://arxiv.org/abs/2504.16399}
  {arXiv:2504.16399}\BibitemShut {NoStop}%
\bibitem [{\citenamefont {Jing}\ \emph {et~al.}(2019)\citenamefont {Jing},
  \citenamefont {Wang}, \citenamefont {Yu}, \citenamefont {Sun}, \citenamefont
  {Jiang}, \citenamefont {Yang}, \citenamefont {Jiang}, \citenamefont {Luo},
  \citenamefont {Zhang}, \citenamefont {Jiang}, \citenamefont {Bao},\ and\
  \citenamefont {Pan}}]{Jing2019}%
  \BibitemOpen
  \bibfield  {author} {\bibinfo {author} {\bibfnamefont {Bo}~\bibnamefont
  {Jing}}, \bibinfo {author} {\bibfnamefont {Xu-Jie}\ \bibnamefont {Wang}},
  \bibinfo {author} {\bibfnamefont {Yong}\ \bibnamefont {Yu}}, \bibinfo
  {author} {\bibfnamefont {Peng-Fei}\ \bibnamefont {Sun}}, \bibinfo {author}
  {\bibfnamefont {Yan}\ \bibnamefont {Jiang}}, \bibinfo {author} {\bibfnamefont
  {Sheng-Jun}\ \bibnamefont {Yang}}, \bibinfo {author} {\bibfnamefont
  {Wen-Hao}\ \bibnamefont {Jiang}}, \bibinfo {author} {\bibfnamefont {Xi-Yu}\
  \bibnamefont {Luo}}, \bibinfo {author} {\bibfnamefont {Jun}\ \bibnamefont
  {Zhang}}, \bibinfo {author} {\bibfnamefont {Xiao}\ \bibnamefont {Jiang}},
  \bibinfo {author} {\bibfnamefont {Xiao-Hui}\ \bibnamefont {Bao}}, \ and\
  \bibinfo {author} {\bibfnamefont {Jian-Wei}\ \bibnamefont {Pan}},\ }\emph
  {\enquote {\bibinfo {title} {Entanglement of three quantum memories via
  interference of three single photons},}\ }\href
  {https://doi.org/10.1038/s41566-018-0342-x} {\bibfield  {journal} {\bibinfo
  {journal} {Nat. Photonics}\ }\textbf {\bibinfo {volume} {13}},\ \bibinfo
  {pages} {210} (\bibinfo {year} {2019})},\ \Eprint
  {http://arxiv.org/abs/1808.05393} {arXiv:1808.05393}\BibitemShut {NoStop}%
\bibitem [{\citenamefont {Pompili}\ \emph {et~al.}(2021)\citenamefont
  {Pompili}, \citenamefont {Hermans}, \citenamefont {Baier}, \citenamefont
  {Beukers}, \citenamefont {Humphreys}, \citenamefont {Schouten}, \citenamefont
  {Vermeulen}, \citenamefont {Tiggelman}, \citenamefont {dos Santos~Martins},
  \citenamefont {Dirkse}, \citenamefont {Wehner},\ and\ \citenamefont
  {Hanson}}]{PompiliEtAl2021}%
  \BibitemOpen
  \bibfield  {author} {\bibinfo {author} {\bibfnamefont {Matteo}\ \bibnamefont
  {Pompili}}, \bibinfo {author} {\bibfnamefont {Sophie L.~N.}\ \bibnamefont
  {Hermans}}, \bibinfo {author} {\bibfnamefont {Simon}\ \bibnamefont {Baier}},
  \bibinfo {author} {\bibfnamefont {Hans K.~C.}\ \bibnamefont {Beukers}},
  \bibinfo {author} {\bibfnamefont {Peter~C.}\ \bibnamefont {Humphreys}},
  \bibinfo {author} {\bibfnamefont {Raymond~N.}\ \bibnamefont {Schouten}},
  \bibinfo {author} {\bibfnamefont {Raymond F.~L.}\ \bibnamefont {Vermeulen}},
  \bibinfo {author} {\bibfnamefont {Marijn~J.}\ \bibnamefont {Tiggelman}},
  \bibinfo {author} {\bibfnamefont {Laura}\ \bibnamefont {dos Santos~Martins}},
  \bibinfo {author} {\bibfnamefont {Bas}\ \bibnamefont {Dirkse}}, \bibinfo
  {author} {\bibfnamefont {Stephanie}\ \bibnamefont {Wehner}}, \ and\ \bibinfo
  {author} {\bibfnamefont {Ronald}\ \bibnamefont {Hanson}},\ }\emph {\enquote
  {\bibinfo {title} {{Realization of a multi-node quantum network of remote
  solid-state qubits}},}\ }\href {https://doi.org/10.1126/science.abg1919}
  {\bibfield  {journal} {\bibinfo  {journal} {Science}\ }\textbf {\bibinfo
  {volume} {372}},\ \bibinfo {pages} {259} (\bibinfo {year} {2021})},\ \Eprint
  {http://arxiv.org/abs/2102.04471} {arXiv:2102.04471}\BibitemShut {NoStop}%
\bibitem [{\citenamefont {Beukers}\ \emph {et~al.}(2024)\citenamefont
  {Beukers}, \citenamefont {Pasini}, \citenamefont {Choi}, \citenamefont
  {Englund}, \citenamefont {Hanson},\ and\ \citenamefont
  {Borregaard}}]{Beukers_2024}%
  \BibitemOpen
  \bibfield  {author} {\bibinfo {author} {\bibfnamefont {Hans~K.C.}\
  \bibnamefont {Beukers}}, \bibinfo {author} {\bibfnamefont {Matteo}\
  \bibnamefont {Pasini}}, \bibinfo {author} {\bibfnamefont {Hyeongrak}\
  \bibnamefont {Choi}}, \bibinfo {author} {\bibfnamefont {Dirk}\ \bibnamefont
  {Englund}}, \bibinfo {author} {\bibfnamefont {Ronald}\ \bibnamefont
  {Hanson}}, \ and\ \bibinfo {author} {\bibfnamefont {Johannes}\ \bibnamefont
  {Borregaard}},\ }\emph {\enquote {\bibinfo {title} {{Remote-Entanglement
  Protocols for Stationary Qubits with Photonic Interfaces}},}\ }\href
  {https://doi.org/10.1103/PRXQuantum.5.010202} {\bibfield  {journal} {\bibinfo
   {journal} {PRX Quantum}\ }\textbf {\bibinfo {volume} {5}},\ \bibinfo {pages}
  {010202} (\bibinfo {year} {2024})},\ \Eprint
  {http://arxiv.org/abs/2310.19878} {arXiv:2310.19878}\BibitemShut {NoStop}%
\bibitem [{\citenamefont {Walln\"ofer}\ \emph {et~al.}(2016)\citenamefont
  {Walln\"ofer}, \citenamefont {Zwerger}, \citenamefont {Muschik},
  \citenamefont {Sangouard},\ and\ \citenamefont {D\"ur}}]{PhysRevA.94.052307}%
  \BibitemOpen
  \bibfield  {author} {\bibinfo {author} {\bibfnamefont {Julius}\ \bibnamefont
  {Walln\"ofer}}, \bibinfo {author} {\bibfnamefont {Michael}\ \bibnamefont
  {Zwerger}}, \bibinfo {author} {\bibfnamefont {Christine}\ \bibnamefont
  {Muschik}}, \bibinfo {author} {\bibfnamefont {Nicolas}\ \bibnamefont
  {Sangouard}}, \ and\ \bibinfo {author} {\bibfnamefont {Wolfgang}\
  \bibnamefont {D\"ur}},\ }\emph {\enquote {\bibinfo {title} {Two-dimensional
  quantum repeaters},}\ }\href {https://doi.org/10.1103/PhysRevA.94.052307}
  {\bibfield  {journal} {\bibinfo  {journal} {Phys. Rev. A}\ }\textbf {\bibinfo
  {volume} {94}},\ \bibinfo {pages} {052307} (\bibinfo {year} {2016})},\
  \Eprint {http://arxiv.org/abs/1604.05352} {arXiv:1604.05352}\BibitemShut
  {NoStop}%
\bibitem [{\citenamefont {Van~Meter}\ \emph {et~al.}(2011)\citenamefont
  {Van~Meter}, \citenamefont {Touch},\ and\ \citenamefont
  {Horsman}}]{METER_2011}%
  \BibitemOpen
  \bibfield  {author} {\bibinfo {author} {\bibfnamefont {Rodney}\ \bibnamefont
  {Van~Meter}}, \bibinfo {author} {\bibfnamefont {Joe}\ \bibnamefont {Touch}},
  \ and\ \bibinfo {author} {\bibfnamefont {Clare}\ \bibnamefont {Horsman}},\
  }\emph {\enquote {\bibinfo {title} {Recursive quantum repeater networks},}\
  }\href {https://doi.org/10.2201/NiiPi.2011.8.8} {\bibfield  {journal}
  {\bibinfo  {journal} {Progress in Informatics}\ ,\ \bibinfo {pages} {65}}
  (\bibinfo {year} {2011})},\ \Eprint {http://arxiv.org/abs/1105.1238}
  {arXiv:1105.1238}\BibitemShut {NoStop}%
\bibitem [{\citenamefont {Pirker}\ \emph {et~al.}(2018)\citenamefont {Pirker},
  \citenamefont {Walln{\"o}fer},\ and\ \citenamefont {D{\"u}r}}]{Pirker_2018}%
  \BibitemOpen
  \bibfield  {author} {\bibinfo {author} {\bibfnamefont {Alexander}\
  \bibnamefont {Pirker}}, \bibinfo {author} {\bibfnamefont {Julius}\
  \bibnamefont {Walln{\"o}fer}}, \ and\ \bibinfo {author} {\bibfnamefont
  {Wolfgang}\ \bibnamefont {D{\"u}r}},\ }\emph {\enquote {\bibinfo {title}
  {Modular architectures for quantum networks},}\ }\href
  {https://doi.org/10.1088/1367-2630/aac2aa} {\bibfield  {journal} {\bibinfo
  {journal} {New J. Phys.}\ }\textbf {\bibinfo {volume} {20}},\ \bibinfo
  {pages} {053054} (\bibinfo {year} {2018})},\ \Eprint
  {http://arxiv.org/abs/1711.02606} {arXiv:1711.02606}\BibitemShut {NoStop}%
\bibitem [{\citenamefont {Caprara~Vivoli}\ \emph {et~al.}(2019)\citenamefont
  {Caprara~Vivoli}, \citenamefont {Ribeiro},\ and\ \citenamefont
  {Wehner}}]{PhysRevA.100.032310_2019}%
  \BibitemOpen
  \bibfield  {author} {\bibinfo {author} {\bibfnamefont {Valentina}\
  \bibnamefont {Caprara~Vivoli}}, \bibinfo {author} {\bibfnamefont
  {J\'er\'emy}\ \bibnamefont {Ribeiro}}, \ and\ \bibinfo {author}
  {\bibfnamefont {Stephanie}\ \bibnamefont {Wehner}},\ }\emph {\enquote
  {\bibinfo {title} {{High-fidelity Greenberger-Horne-Zeilinger state
  generation within nearby nodes}},}\ }\href
  {https://doi.org/10.1103/PhysRevA.100.032310} {\bibfield  {journal} {\bibinfo
   {journal} {Phys. Rev. A}\ }\textbf {\bibinfo {volume} {100}},\ \bibinfo
  {pages} {032310} (\bibinfo {year} {2019})},\ \Eprint
  {http://arxiv.org/abs/1805.10663} {arXiv:1805.10663}\BibitemShut {NoStop}%
\bibitem [{\citenamefont {Pirker}\ and\ \citenamefont
  {D{\"u}r}(2019)}]{Pirker_2019}%
  \BibitemOpen
  \bibfield  {author} {\bibinfo {author} {\bibfnamefont {Alexander}\
  \bibnamefont {Pirker}}\ and\ \bibinfo {author} {\bibfnamefont {Wolfgang}\
  \bibnamefont {D{\"u}r}},\ }\emph {\enquote {\bibinfo {title} {A quantum
  network stack and protocols for reliable entanglement-based networks},}\
  }\href {https://doi.org/10.1088/1367-2630/ab05f7} {\bibfield  {journal}
  {\bibinfo  {journal} {New J. Phys.}\ }\textbf {\bibinfo {volume} {21}},\
  \bibinfo {pages} {033003} (\bibinfo {year} {2019})},\ \Eprint
  {http://arxiv.org/abs/1810.03556} {arXiv:1810.03556}\BibitemShut {NoStop}%
\bibitem [{\citenamefont {Benjamin}\ \emph {et~al.}(2006)\citenamefont
  {Benjamin}, \citenamefont {Browne}, \citenamefont {Fitzsimons},\ and\
  \citenamefont {Morton}}]{Benjamin_2006}%
  \BibitemOpen
  \bibfield  {author} {\bibinfo {author} {\bibfnamefont {Simon~C.}\
  \bibnamefont {Benjamin}}, \bibinfo {author} {\bibfnamefont {Dan~E.}\
  \bibnamefont {Browne}}, \bibinfo {author} {\bibfnamefont {Joe}\ \bibnamefont
  {Fitzsimons}}, \ and\ \bibinfo {author} {\bibfnamefont {John J.~L.}\
  \bibnamefont {Morton}},\ }\emph {\enquote {\bibinfo {title} {Brokered
  graph-state quantum computation},}\ }\href
  {https://doi.org/10.1088/1367-2630/8/8/141} {\bibfield  {journal} {\bibinfo
  {journal} {New J. Phys.}\ }\textbf {\bibinfo {volume} {8}},\ \bibinfo {pages}
  {141} (\bibinfo {year} {2006})},\ \Eprint
  {http://arxiv.org/abs/quant-ph/0509209} {arXiv:quant-ph/0509209}\BibitemShut
  {NoStop}%
\bibitem [{\citenamefont {Campbell}\ \emph {et~al.}(2007)\citenamefont
  {Campbell}, \citenamefont {Fitzsimons}, \citenamefont {Benjamin},\ and\
  \citenamefont {Kok}}]{PhysRevA.75.042303_2007}%
  \BibitemOpen
  \bibfield  {author} {\bibinfo {author} {\bibfnamefont {Earl~T.}\ \bibnamefont
  {Campbell}}, \bibinfo {author} {\bibfnamefont {Joseph}\ \bibnamefont
  {Fitzsimons}}, \bibinfo {author} {\bibfnamefont {Simon~C.}\ \bibnamefont
  {Benjamin}}, \ and\ \bibinfo {author} {\bibfnamefont {Pieter}\ \bibnamefont
  {Kok}},\ }\emph {\enquote {\bibinfo {title} {Adaptive strategies for
  graph-state growth in the presence of monitored errors},}\ }\href
  {https://doi.org/10.1103/PhysRevA.75.042303} {\bibfield  {journal} {\bibinfo
  {journal} {Phys. Rev. A}\ }\textbf {\bibinfo {volume} {75}},\ \bibinfo
  {pages} {042303} (\bibinfo {year} {2007})},\ \Eprint
  {http://arxiv.org/abs/quant-ph/0606199} {arXiv:quant-ph/0606199}\BibitemShut
  {NoStop}%
\bibitem [{\citenamefont {Kruszynska}\ \emph {et~al.}(2006)\citenamefont
  {Kruszynska}, \citenamefont {Anders}, \citenamefont {D\"ur},\ and\
  \citenamefont {Briegel}}]{PhysRevA.73.062328_2006}%
  \BibitemOpen
  \bibfield  {author} {\bibinfo {author} {\bibfnamefont {Caroline}\
  \bibnamefont {Kruszynska}}, \bibinfo {author} {\bibfnamefont {Simon}\
  \bibnamefont {Anders}}, \bibinfo {author} {\bibfnamefont {Wolfgang}\
  \bibnamefont {D\"ur}}, \ and\ \bibinfo {author} {\bibfnamefont {Hans~J.}\
  \bibnamefont {Briegel}},\ }\emph {\enquote {\bibinfo {title} {Quantum
  communication cost of preparing multipartite entanglement},}\ }\href
  {https://doi.org/10.1103/PhysRevA.73.062328} {\bibfield  {journal} {\bibinfo
  {journal} {Phys. Rev. A}\ }\textbf {\bibinfo {volume} {73}},\ \bibinfo
  {pages} {062328} (\bibinfo {year} {2006})},\ \Eprint
  {http://arxiv.org/abs/quant-ph/0512218} {arXiv:quant-ph/0512218}\BibitemShut
  {NoStop}%
\bibitem [{\citenamefont {de~Bone}\ \emph {et~al.}(2020)\citenamefont
  {de~Bone}, \citenamefont {Ouyang}, \citenamefont {Goodenough},\ and\
  \citenamefont {Elkouss}}]{9292429_2020}%
  \BibitemOpen
  \bibfield  {author} {\bibinfo {author} {\bibfnamefont {Sebastian}\
  \bibnamefont {de~Bone}}, \bibinfo {author} {\bibfnamefont {Runsheng}\
  \bibnamefont {Ouyang}}, \bibinfo {author} {\bibfnamefont {Kenneth}\
  \bibnamefont {Goodenough}}, \ and\ \bibinfo {author} {\bibfnamefont {David}\
  \bibnamefont {Elkouss}},\ }\emph {\enquote {\bibinfo {title} {{Protocols for
  Creating and Distilling Multipartite GHZ States With Bell Pairs}},}\ }\href
  {https://doi.org/10.1109/TQE.2020.3044179} {\bibfield  {journal} {\bibinfo
  {journal} {IEEE Trans. Quantum Eng.}\ }\textbf {\bibinfo {volume} {1}},\
  \bibinfo {pages} {1} (\bibinfo {year} {2020})},\ \Eprint
  {http://arxiv.org/abs/2010.12259} {arXiv:2010.12259}\BibitemShut {NoStop}%
\bibitem [{\citenamefont {Coopmans}\ \emph {et~al.}(2022)\citenamefont
  {Coopmans}, \citenamefont {Brand},\ and\ \citenamefont
  {Elkouss}}]{PhysRevA.105.0126082022}%
  \BibitemOpen
  \bibfield  {author} {\bibinfo {author} {\bibfnamefont {Tim}\ \bibnamefont
  {Coopmans}}, \bibinfo {author} {\bibfnamefont {Sebastiaan}\ \bibnamefont
  {Brand}}, \ and\ \bibinfo {author} {\bibfnamefont {David}\ \bibnamefont
  {Elkouss}},\ }\emph {\enquote {\bibinfo {title} {Improved analytical bounds
  on delivery times of long-distance entanglement},}\ }\href
  {https://doi.org/10.1103/PhysRevA.105.012608} {\bibfield  {journal} {\bibinfo
   {journal} {Phys. Rev. A}\ }\textbf {\bibinfo {volume} {105}},\ \bibinfo
  {pages} {012608} (\bibinfo {year} {2022})},\ \Eprint
  {http://arxiv.org/abs/2103.11454} {arXiv:2103.11454}\BibitemShut {NoStop}%
\bibitem [{\citenamefont {Coopmans}\ \emph {et~al.}(2021)\citenamefont
  {Coopmans}, \citenamefont {Knegjens}, \citenamefont {Dahlberg}, \citenamefont
  {Maier}, \citenamefont {Nijsten}, \citenamefont {de~Oliveira~Filho},
  \citenamefont {Papendrecht}, \citenamefont {Rabbie}, \citenamefont
  {Rozp{\c{e}}dek}, \citenamefont {Skrzypczyk}, \citenamefont {Wubben},
  \citenamefont {de~Jong}, \citenamefont {Podareanu}, \citenamefont
  {Torres-Knoop}, \citenamefont {Elkouss},\ and\ \citenamefont
  {Wehner}}]{Coopmans2021}%
  \BibitemOpen
  \bibfield  {author} {\bibinfo {author} {\bibfnamefont {Tim}\ \bibnamefont
  {Coopmans}}, \bibinfo {author} {\bibfnamefont {Robert}\ \bibnamefont
  {Knegjens}}, \bibinfo {author} {\bibfnamefont {Axel}\ \bibnamefont
  {Dahlberg}}, \bibinfo {author} {\bibfnamefont {David}\ \bibnamefont {Maier}},
  \bibinfo {author} {\bibfnamefont {Loek}\ \bibnamefont {Nijsten}}, \bibinfo
  {author} {\bibfnamefont {Julio}\ \bibnamefont {de~Oliveira~Filho}}, \bibinfo
  {author} {\bibfnamefont {Martijn}\ \bibnamefont {Papendrecht}}, \bibinfo
  {author} {\bibfnamefont {Julian}\ \bibnamefont {Rabbie}}, \bibinfo {author}
  {\bibfnamefont {Filip}\ \bibnamefont {Rozp{\c{e}}dek}}, \bibinfo {author}
  {\bibfnamefont {Matthew}\ \bibnamefont {Skrzypczyk}}, \bibinfo {author}
  {\bibfnamefont {Leon}\ \bibnamefont {Wubben}}, \bibinfo {author}
  {\bibfnamefont {Walter}\ \bibnamefont {de~Jong}}, \bibinfo {author}
  {\bibfnamefont {Damian}\ \bibnamefont {Podareanu}}, \bibinfo {author}
  {\bibfnamefont {Ariana}\ \bibnamefont {Torres-Knoop}}, \bibinfo {author}
  {\bibfnamefont {David}\ \bibnamefont {Elkouss}}, \ and\ \bibinfo {author}
  {\bibfnamefont {Stephanie}\ \bibnamefont {Wehner}},\ }\emph {\enquote
  {\bibinfo {title} {{NetSquid, a NETwork Simulator for QUantum Information
  using Discrete events}},}\ }\href
  {https://doi.org/10.1038/s42005-021-00647-8} {\bibfield  {journal} {\bibinfo
  {journal} {Commun. Phys.}\ }\textbf {\bibinfo {volume} {4}},\ \bibinfo
  {pages} {164} (\bibinfo {year} {2021})},\ \Eprint
  {http://arxiv.org/abs/2010.12535} {arXiv:2010.12535}\BibitemShut {NoStop}%
\bibitem [{\citenamefont {Nain}\ \emph {et~al.}(2022)\citenamefont {Nain},
  \citenamefont {Vardoyan}, \citenamefont {Guha},\ and\ \citenamefont
  {Towsley}}]{Nain2022}%
  \BibitemOpen
  \bibfield  {author} {\bibinfo {author} {\bibfnamefont {Philippe}\
  \bibnamefont {Nain}}, \bibinfo {author} {\bibfnamefont {Gayane}\ \bibnamefont
  {Vardoyan}}, \bibinfo {author} {\bibfnamefont {Saikat}\ \bibnamefont {Guha}},
  \ and\ \bibinfo {author} {\bibfnamefont {Don}\ \bibnamefont {Towsley}},\
  }\emph {\enquote {\bibinfo {title} {Analysis of a tripartite entanglement
  distribution switch},}\ }\href {https://doi.org/10.1007/s11134-021-09731-w}
  {\bibfield  {journal} {\bibinfo  {journal} {Queueing Syst.}\ }\textbf
  {\bibinfo {volume} {101}},\ \bibinfo {pages} {291} (\bibinfo {year}
  {2022})}\BibitemShut {NoStop}%
\bibitem [{\citenamefont {Cuquet}\ and\ \citenamefont
  {Calsamiglia}(2012)}]{PhysRevA.86.042304}%
  \BibitemOpen
  \bibfield  {author} {\bibinfo {author} {\bibfnamefont {Mart\'{\i}}\
  \bibnamefont {Cuquet}}\ and\ \bibinfo {author} {\bibfnamefont {John}\
  \bibnamefont {Calsamiglia}},\ }\emph {\enquote {\bibinfo {title} {Growth of
  graph states in quantum networks},}\ }\href
  {https://doi.org/10.1103/PhysRevA.86.042304} {\bibfield  {journal} {\bibinfo
  {journal} {Phys. Rev. A}\ }\textbf {\bibinfo {volume} {86}},\ \bibinfo
  {pages} {042304} (\bibinfo {year} {2012})},\ \Eprint
  {http://arxiv.org/abs/1208.0710} {arXiv:1208.0710}\BibitemShut {NoStop}%
\bibitem [{\citenamefont {Epping}\ \emph {et~al.}(2016)\citenamefont {Epping},
  \citenamefont {Kampermann},\ and\ \citenamefont {Bru{\ss}}}]{Epping_2016}%
  \BibitemOpen
  \bibfield  {author} {\bibinfo {author} {\bibfnamefont {Michael}\ \bibnamefont
  {Epping}}, \bibinfo {author} {\bibfnamefont {Hermann}\ \bibnamefont
  {Kampermann}}, \ and\ \bibinfo {author} {\bibfnamefont {Dagmar}\ \bibnamefont
  {Bru{\ss}}},\ }\emph {\enquote {\bibinfo {title} {Large-scale quantum
  networks based on graphs},}\ }\href
  {https://doi.org/10.1088/1367-2630/18/5/053036} {\bibfield  {journal}
  {\bibinfo  {journal} {New J. Phys.}\ }\textbf {\bibinfo {volume} {18}},\
  \bibinfo {pages} {053036} (\bibinfo {year} {2016})},\ \Eprint
  {http://arxiv.org/abs/1504.06599} {arXiv:1504.06599}\BibitemShut {NoStop}%
\bibitem [{\citenamefont {Yamasaki}\ \emph {et~al.}(2018)\citenamefont
  {Yamasaki}, \citenamefont {Pirker}, \citenamefont {Murao}, \citenamefont
  {D{\"u}r},\ and\ \citenamefont {Kraus}}]{YamasakiPirkerMuraoDuerKraus2018}%
  \BibitemOpen
  \bibfield  {author} {\bibinfo {author} {\bibfnamefont {Hayata}\ \bibnamefont
  {Yamasaki}}, \bibinfo {author} {\bibfnamefont {Alexander}\ \bibnamefont
  {Pirker}}, \bibinfo {author} {\bibfnamefont {Mio}\ \bibnamefont {Murao}},
  \bibinfo {author} {\bibfnamefont {Wolfgang}\ \bibnamefont {D{\"u}r}}, \ and\
  \bibinfo {author} {\bibfnamefont {Barbara}\ \bibnamefont {Kraus}},\ }\emph
  {\enquote {\bibinfo {title} {Multipartite entanglement outperforming
  bipartite entanglement under limited quantum system sizes},}\ }\href
  {https://doi.org/10.1103/PhysRevA.98.052313} {\bibfield  {journal} {\bibinfo
  {journal} {Phys. Rev. A}\ }\textbf {\bibinfo {volume} {98}},\ \bibinfo
  {pages} {052313} (\bibinfo {year} {2018})},\ \Eprint
  {http://arxiv.org/abs/1808.00005} {arXiv:1808.00005}\BibitemShut {NoStop}%
\bibitem [{\citenamefont {Meignant}\ \emph {et~al.}(2019)\citenamefont
  {Meignant}, \citenamefont {Markham},\ and\ \citenamefont
  {Grosshans}}]{PhysRevA.100.052333}%
  \BibitemOpen
  \bibfield  {author} {\bibinfo {author} {\bibfnamefont {Cl\'ement}\
  \bibnamefont {Meignant}}, \bibinfo {author} {\bibfnamefont {Damian}\
  \bibnamefont {Markham}}, \ and\ \bibinfo {author} {\bibfnamefont
  {Fr\'ed\'eric}\ \bibnamefont {Grosshans}},\ }\emph {\enquote {\bibinfo
  {title} {Distributing graph states over arbitrary quantum networks},}\ }\href
  {https://doi.org/10.1103/PhysRevA.100.052333} {\bibfield  {journal} {\bibinfo
   {journal} {Phys. Rev. A}\ }\textbf {\bibinfo {volume} {100}},\ \bibinfo
  {pages} {052333} (\bibinfo {year} {2019})},\ \Eprint
  {http://arxiv.org/abs/1811.05445} {arXiv:1811.05445}\BibitemShut {NoStop}%
\bibitem [{\citenamefont {Bugalho}\ \emph {et~al.}(2023)\citenamefont
  {Bugalho}, \citenamefont {Coutinho}, \citenamefont {Monteiro},\ and\
  \citenamefont {Omar}}]{Bugalho_2023}%
  \BibitemOpen
  \bibfield  {author} {\bibinfo {author} {\bibfnamefont {Lu{\'i}s}\
  \bibnamefont {Bugalho}}, \bibinfo {author} {\bibfnamefont {Bruno~C.}\
  \bibnamefont {Coutinho}}, \bibinfo {author} {\bibfnamefont {Francisco~A.}\
  \bibnamefont {Monteiro}}, \ and\ \bibinfo {author} {\bibfnamefont {Yasser}\
  \bibnamefont {Omar}},\ }\emph {\enquote {\bibinfo {title} {{Distributing
  Multipartite Entanglement over Noisy Quantum Networks}},}\ }\href
  {https://doi.org/10.22331/q-2023-02-09-920} {\bibfield  {journal} {\bibinfo
  {journal} {Quantum}\ }\textbf {\bibinfo {volume} {7}},\ \bibinfo {pages}
  {920} (\bibinfo {year} {2023})},\ \Eprint {http://arxiv.org/abs/2103.14759}
  {arXiv:2103.14759}\BibitemShut {NoStop}%
\bibitem [{\citenamefont {Avis}\ \emph {et~al.}(2023)\citenamefont {Avis},
  \citenamefont {Rozp\c{e}dek},\ and\ \citenamefont
  {Wehner}}]{AvisRozpedekWehner2023}%
  \BibitemOpen
  \bibfield  {author} {\bibinfo {author} {\bibfnamefont {Guus}\ \bibnamefont
  {Avis}}, \bibinfo {author} {\bibfnamefont {Filip}\ \bibnamefont
  {Rozp\c{e}dek}}, \ and\ \bibinfo {author} {\bibfnamefont {Stephanie}\
  \bibnamefont {Wehner}},\ }\emph {\enquote {\bibinfo {title} {Analysis of
  multipartite entanglement distribution using a central quantum-network
  node},}\ }\href {https://doi.org/10.1103/PhysRevA.107.012609} {\bibfield
  {journal} {\bibinfo  {journal} {Phys. Rev. A}\ }\textbf {\bibinfo {volume}
  {107}},\ \bibinfo {pages} {012609} (\bibinfo {year} {2023})},\ \Eprint
  {http://arxiv.org/abs/2203.05517} {arXiv:2203.05517}\BibitemShut {NoStop}%
\bibitem [{\citenamefont {Fischer}\ and\ \citenamefont
  {Towsley}(2021)}]{FischerAlexTowsleyDon_2021}%
  \BibitemOpen
  \bibfield  {author} {\bibinfo {author} {\bibfnamefont {Alex}\ \bibnamefont
  {Fischer}}\ and\ \bibinfo {author} {\bibfnamefont {Don}\ \bibnamefont
  {Towsley}},\ }\emph {\enquote {\bibinfo {title} {{Distributing Graph States
  Across Quantum Networks}},}\ }in\ \href
  {https://doi.org/10.1109/QCE52317.2021.00049} {\emph {\bibinfo {booktitle}
  {2021 IEEE International Conference on Quantum Computing and Engineering
  (QCE)}}}\ (\bibinfo {year} {2021})\ pp.\ \bibinfo {pages} {324--333},\
  \Eprint {http://arxiv.org/abs/2009.10888} {arXiv:2009.10888}\BibitemShut
  {NoStop}%
\bibitem [{\citenamefont {Morelli}\ \emph {et~al.}(2021)\citenamefont
  {Morelli}, \citenamefont {Usui}, \citenamefont {Agudelo},\ and\ \citenamefont
  {Friis}}]{MorelliUsuiAgudeloFriis2021}%
  \BibitemOpen
  \bibfield  {author} {\bibinfo {author} {\bibfnamefont {Simon}\ \bibnamefont
  {Morelli}}, \bibinfo {author} {\bibfnamefont {Ayaka}\ \bibnamefont {Usui}},
  \bibinfo {author} {\bibfnamefont {Elizabeth}\ \bibnamefont {Agudelo}}, \ and\
  \bibinfo {author} {\bibfnamefont {Nicolai}\ \bibnamefont {Friis}},\ }\emph
  {\enquote {\bibinfo {title} {{Bayesian parameter estimation using Gaussian
  states and measurements}},}\ }\href
  {https://doi.org/10.1088/2058-9565/abd83d} {\bibfield  {journal} {\bibinfo
  {journal} {Quantum Sci. Technol.}\ }\textbf {\bibinfo {volume} {6}},\
  \bibinfo {pages} {025018} (\bibinfo {year} {2021})},\ \Eprint
  {http://arxiv.org/abs/2009.03709} {arXiv:2009.03709}\BibitemShut {NoStop}%
\bibitem [{\citenamefont {Nain}\ \emph {et~al.}(2020)\citenamefont {Nain},
  \citenamefont {Vardoyan}, \citenamefont {Guha},\ and\ \citenamefont
  {Towsley}}]{10.1145/3392141}%
  \BibitemOpen
  \bibfield  {author} {\bibinfo {author} {\bibfnamefont {Philippe}\
  \bibnamefont {Nain}}, \bibinfo {author} {\bibfnamefont {Gayane}\ \bibnamefont
  {Vardoyan}}, \bibinfo {author} {\bibfnamefont {Saikat}\ \bibnamefont {Guha}},
  \ and\ \bibinfo {author} {\bibfnamefont {Don}\ \bibnamefont {Towsley}},\
  }\emph {\enquote {\bibinfo {title} {{On the Analysis of a Multipartite
  Entanglement Distribution Switch}},}\ }\href
  {https://doi.org/10.1145/3392141} {\bibfield  {journal} {\bibinfo  {journal}
  {Proc. ACM Meas. Anal. Comput. Syst.}\ }\textbf {\bibinfo {volume} {4}}
  (\bibinfo {year} {2020})},\ \Eprint {http://arxiv.org/abs/2212.01784}
  {arXiv:2212.01784}\BibitemShut {NoStop}%
\bibitem [{\citenamefont {Bennett}\ \emph {et~al.}(2001)\citenamefont
  {Bennett}, \citenamefont {DiVincenzo}, \citenamefont {Shor}, \citenamefont
  {Smolin}, \citenamefont {Terhal},\ and\ \citenamefont
  {Wootters}}]{Bennett_2001}%
  \BibitemOpen
  \bibfield  {author} {\bibinfo {author} {\bibfnamefont {Charles~H.}\
  \bibnamefont {Bennett}}, \bibinfo {author} {\bibfnamefont {David~P.}\
  \bibnamefont {DiVincenzo}}, \bibinfo {author} {\bibfnamefont {Peter~W.}\
  \bibnamefont {Shor}}, \bibinfo {author} {\bibfnamefont {John~A.}\
  \bibnamefont {Smolin}}, \bibinfo {author} {\bibfnamefont {Barbara~M.}\
  \bibnamefont {Terhal}}, \ and\ \bibinfo {author} {\bibfnamefont {William~K.}\
  \bibnamefont {Wootters}},\ }\emph {\enquote {\bibinfo {title} {{Remote State
  Preparation}},}\ }\href {https://doi.org/10.1103/PhysRevLett.87.077902}
  {\bibfield  {journal} {\bibinfo  {journal} {Phys. Rev. Lett.}\ }\textbf
  {\bibinfo {volume} {87}},\ \bibinfo {pages} {077902} (\bibinfo {year}
  {2001})},\ \Eprint {http://arxiv.org/abs/quant-ph/0006044}
  {arXiv:quant-ph/0006044}\BibitemShut {NoStop}%
\bibitem [{\citenamefont {Bennett}\ \emph {et~al.}(1993)\citenamefont
  {Bennett}, \citenamefont {Brassard}, \citenamefont {Cr{\'e}peau},
  \citenamefont {Jozsa}, \citenamefont {Peres},\ and\ \citenamefont
  {Wootters}}]{BennettBrassardCrepeauJozsaPeresWootters1993}%
  \BibitemOpen
  \bibfield  {author} {\bibinfo {author} {\bibfnamefont {Charles~H.}\
  \bibnamefont {Bennett}}, \bibinfo {author} {\bibfnamefont {Gilles}\
  \bibnamefont {Brassard}}, \bibinfo {author} {\bibfnamefont {Claude}\
  \bibnamefont {Cr{\'e}peau}}, \bibinfo {author} {\bibfnamefont {Richard}\
  \bibnamefont {Jozsa}}, \bibinfo {author} {\bibfnamefont {Asher}\ \bibnamefont
  {Peres}}, \ and\ \bibinfo {author} {\bibfnamefont {William~K.}\ \bibnamefont
  {Wootters}},\ }\emph {\enquote {\bibinfo {title} {{Teleporting an Unknown
  Quantum State via Dual Classical and {E}instein-{P}odolsky-{R}osen
  Channels}},}\ }\href {https://doi.org/10.1103/PhysRevLett.70.1895} {\bibfield
   {journal} {\bibinfo  {journal} {Phys. Rev. Lett.}\ }\textbf {\bibinfo
  {volume} {70}},\ \bibinfo {pages} {1895\textendash1899} (\bibinfo {year}
  {1993})}\BibitemShut {NoStop}%
\bibitem [{\citenamefont {Schwartz}\ \emph {et~al.}(2016)\citenamefont
  {Schwartz}, \citenamefont {Cogan}, \citenamefont {Schmidgall}, \citenamefont
  {Don}, \citenamefont {Gantz}, \citenamefont {Kenneth}, \citenamefont
  {Lindner},\ and\ \citenamefont {Gershoni}}]{Schwartz2016}%
  \BibitemOpen
  \bibfield  {author} {\bibinfo {author} {\bibfnamefont {Ido}\ \bibnamefont
  {Schwartz}}, \bibinfo {author} {\bibfnamefont {Dan}\ \bibnamefont {Cogan}},
  \bibinfo {author} {\bibfnamefont {Emma~R.}\ \bibnamefont {Schmidgall}},
  \bibinfo {author} {\bibfnamefont {Yaroslav}\ \bibnamefont {Don}}, \bibinfo
  {author} {\bibfnamefont {Liron}\ \bibnamefont {Gantz}}, \bibinfo {author}
  {\bibfnamefont {Oded}\ \bibnamefont {Kenneth}}, \bibinfo {author}
  {\bibfnamefont {Netanel~H.}\ \bibnamefont {Lindner}}, \ and\ \bibinfo
  {author} {\bibfnamefont {David}\ \bibnamefont {Gershoni}},\ }\emph {\enquote
  {\bibinfo {title} {Deterministic generation of a cluster state of entangled
  photons},}\ }\href {https://doi.org/10.1126/science.aah4758} {\bibfield
  {journal} {\bibinfo  {journal} {Science}\ }\textbf {\bibinfo {volume}
  {354}},\ \bibinfo {pages} {434} (\bibinfo {year} {2016})},\ \Eprint
  {http://arxiv.org/abs/1606.07492} {arXiv:1606.07492}\BibitemShut {NoStop}%
\bibitem [{\citenamefont {Besse}\ \emph {et~al.}(2020)\citenamefont {Besse},
  \citenamefont {Reuer}, \citenamefont {Collodo}, \citenamefont {Wulff},
  \citenamefont {Wernli}, \citenamefont {Copetudo}, \citenamefont {Malz},
  \citenamefont {Magnard}, \citenamefont {Akin}, \citenamefont {Gabureac},
  \citenamefont {Norris}, \citenamefont {Cirac}, \citenamefont {Wallraff},\
  and\ \citenamefont {Eichler}}]{Besse2020}%
  \BibitemOpen
  \bibfield  {author} {\bibinfo {author} {\bibfnamefont {Jean-Claude}\
  \bibnamefont {Besse}}, \bibinfo {author} {\bibfnamefont {Kevin}\ \bibnamefont
  {Reuer}}, \bibinfo {author} {\bibfnamefont {Michele~C.}\ \bibnamefont
  {Collodo}}, \bibinfo {author} {\bibfnamefont {Arne}\ \bibnamefont {Wulff}},
  \bibinfo {author} {\bibfnamefont {Lucien}\ \bibnamefont {Wernli}}, \bibinfo
  {author} {\bibfnamefont {Adrian}\ \bibnamefont {Copetudo}}, \bibinfo {author}
  {\bibfnamefont {Daniel}\ \bibnamefont {Malz}}, \bibinfo {author}
  {\bibfnamefont {Paul}\ \bibnamefont {Magnard}}, \bibinfo {author}
  {\bibfnamefont {Abdulkadir}\ \bibnamefont {Akin}}, \bibinfo {author}
  {\bibfnamefont {Mihai}\ \bibnamefont {Gabureac}}, \bibinfo {author}
  {\bibfnamefont {Graham~J.}\ \bibnamefont {Norris}}, \bibinfo {author}
  {\bibfnamefont {J.~Ignacio}\ \bibnamefont {Cirac}}, \bibinfo {author}
  {\bibfnamefont {Andreas}\ \bibnamefont {Wallraff}}, \ and\ \bibinfo {author}
  {\bibfnamefont {Christopher}\ \bibnamefont {Eichler}},\ }\emph {\enquote
  {\bibinfo {title} {Realizing a deterministic source of multipartite-entangled
  photonic qubits},}\ }\href {https://doi.org/10.1038/s41467-020-18635-x}
  {\bibfield  {journal} {\bibinfo  {journal} {Nat. Commun.}\ }\textbf {\bibinfo
  {volume} {11}},\ \bibinfo {pages} {4877} (\bibinfo {year} {2020})},\ \Eprint
  {http://arxiv.org/abs/2005.07060} {arXiv:2005.07060}\BibitemShut {NoStop}%
\bibitem [{\citenamefont {Istrati}\ \emph {et~al.}(2020)\citenamefont
  {Istrati}, \citenamefont {Pilnyak}, \citenamefont {Loredo}, \citenamefont
  {Ant{\'o}n}, \citenamefont {Somaschi}, \citenamefont {Hilaire}, \citenamefont
  {Ollivier}, \citenamefont {Esmann}, \citenamefont {Cohen}, \citenamefont
  {Vidro}, \citenamefont {Millet}, \citenamefont {Lema{\^i}tre}, \citenamefont
  {Sagnes}, \citenamefont {Harouri}, \citenamefont {Lanco}, \citenamefont
  {Senellart},\ and\ \citenamefont {Eisenberg}}]{Istrati2020}%
  \BibitemOpen
  \bibfield  {author} {\bibinfo {author} {\bibfnamefont {Daniel}\ \bibnamefont
  {Istrati}}, \bibinfo {author} {\bibfnamefont {Yehuda}\ \bibnamefont
  {Pilnyak}}, \bibinfo {author} {\bibfnamefont {Juan~C.}\ \bibnamefont
  {Loredo}}, \bibinfo {author} {\bibfnamefont {Carlos}\ \bibnamefont
  {Ant{\'o}n}}, \bibinfo {author} {\bibfnamefont {Niccolo}\ \bibnamefont
  {Somaschi}}, \bibinfo {author} {\bibfnamefont {Paul}\ \bibnamefont
  {Hilaire}}, \bibinfo {author} {\bibfnamefont {Harold}\ \bibnamefont
  {Ollivier}}, \bibinfo {author} {\bibfnamefont {Martin}\ \bibnamefont
  {Esmann}}, \bibinfo {author} {\bibfnamefont {Lior}\ \bibnamefont {Cohen}},
  \bibinfo {author} {\bibfnamefont {Leonid}\ \bibnamefont {Vidro}}, \bibinfo
  {author} {\bibfnamefont {Cl{\'e}ment}\ \bibnamefont {Millet}}, \bibinfo
  {author} {\bibfnamefont {Aristide}\ \bibnamefont {Lema{\^i}tre}}, \bibinfo
  {author} {\bibfnamefont {Isabelle}\ \bibnamefont {Sagnes}}, \bibinfo {author}
  {\bibfnamefont {Abdelmounaim}\ \bibnamefont {Harouri}}, \bibinfo {author}
  {\bibfnamefont {Lo{\"i}c}\ \bibnamefont {Lanco}}, \bibinfo {author}
  {\bibfnamefont {Pascale}\ \bibnamefont {Senellart}}, \ and\ \bibinfo {author}
  {\bibfnamefont {Hagai~S.}\ \bibnamefont {Eisenberg}},\ }\emph {\enquote
  {\bibinfo {title} {Sequential generation of linear cluster states from a
  single photon emitter},}\ }\href {https://doi.org/10.1038/s41467-020-19341-4}
  {\bibfield  {journal} {\bibinfo  {journal} {Nat. Commun.}\ }\textbf {\bibinfo
  {volume} {11}},\ \bibinfo {pages} {5501} (\bibinfo {year} {2020})},\ \Eprint
  {http://arxiv.org/abs/1912.04375} {arXiv:1912.04375}\BibitemShut {NoStop}%
\bibitem [{\citenamefont {Thomas}\ \emph {et~al.}(2022)\citenamefont {Thomas},
  \citenamefont {Ruscio}, \citenamefont {Morin},\ and\ \citenamefont
  {Rempe}}]{Thomas2022}%
  \BibitemOpen
  \bibfield  {author} {\bibinfo {author} {\bibfnamefont {Philip}\ \bibnamefont
  {Thomas}}, \bibinfo {author} {\bibfnamefont {Leonardo}\ \bibnamefont
  {Ruscio}}, \bibinfo {author} {\bibfnamefont {Olivier}\ \bibnamefont {Morin}},
  \ and\ \bibinfo {author} {\bibfnamefont {Gerhard}\ \bibnamefont {Rempe}},\
  }\emph {\enquote {\bibinfo {title} {Efficient generation of entangled
  multiphoton graph states from a single atom},}\ }\href
  {https://doi.org/10.1038/s41586-022-04987-5} {\bibfield  {journal} {\bibinfo
  {journal} {Nature}\ }\textbf {\bibinfo {volume} {608}},\ \bibinfo {pages}
  {677} (\bibinfo {year} {2022})},\ \Eprint {http://arxiv.org/abs/2205.12736}
  {arXiv:2205.12736}\BibitemShut {NoStop}%
\bibitem [{\citenamefont {Ferreira}\ \emph {et~al.}(2024)\citenamefont
  {Ferreira}, \citenamefont {Kim}, \citenamefont {Butler}, \citenamefont
  {Pichler},\ and\ \citenamefont {Painter}}]{Ferreira2024}%
  \BibitemOpen
  \bibfield  {author} {\bibinfo {author} {\bibfnamefont {Vinicius~S.}\
  \bibnamefont {Ferreira}}, \bibinfo {author} {\bibfnamefont {Gihwan}\
  \bibnamefont {Kim}}, \bibinfo {author} {\bibfnamefont {Andreas}\ \bibnamefont
  {Butler}}, \bibinfo {author} {\bibfnamefont {Hannes}\ \bibnamefont
  {Pichler}}, \ and\ \bibinfo {author} {\bibfnamefont {Oskar}\ \bibnamefont
  {Painter}},\ }\emph {\enquote {\bibinfo {title} {Deterministic generation of
  multidimensional photonic cluster states with a single quantum emitter},}\
  }\href {https://doi.org/10.1038/s41567-024-02408-0} {\bibfield  {journal}
  {\bibinfo  {journal} {Nat. Phys.}\ }\textbf {\bibinfo {volume} {20}},\
  \bibinfo {pages} {865} (\bibinfo {year} {2024})},\ \Eprint
  {http://arxiv.org/abs/2206.10076} {arXiv:2206.10076}\BibitemShut {NoStop}%
\bibitem [{\citenamefont {Pont}\ \emph {et~al.}(2024)\citenamefont {Pont},
  \citenamefont {Corrielli}, \citenamefont {Fyrillas}, \citenamefont {Agresti},
  \citenamefont {Carvacho}, \citenamefont {Maring}, \citenamefont {Emeriau},
  \citenamefont {Ceccarelli}, \citenamefont {Albiero}, \citenamefont
  {Dias~Ferreira}, \citenamefont {Somaschi}, \citenamefont {Senellart},
  \citenamefont {Sagnes}, \citenamefont {Morassi}, \citenamefont
  {Lema{\^i}tre}, \citenamefont {Senellart}, \citenamefont {Sciarrino},
  \citenamefont {Liscidini}, \citenamefont {Belabas},\ and\ \citenamefont
  {Osellame}}]{Pont2024}%
  \BibitemOpen
  \bibfield  {author} {\bibinfo {author} {\bibfnamefont {Mathias}\ \bibnamefont
  {Pont}}, \bibinfo {author} {\bibfnamefont {Giacomo}\ \bibnamefont
  {Corrielli}}, \bibinfo {author} {\bibfnamefont {Andreas}\ \bibnamefont
  {Fyrillas}}, \bibinfo {author} {\bibfnamefont {Iris}\ \bibnamefont
  {Agresti}}, \bibinfo {author} {\bibfnamefont {Gonzalo}\ \bibnamefont
  {Carvacho}}, \bibinfo {author} {\bibfnamefont {Nicolas}\ \bibnamefont
  {Maring}}, \bibinfo {author} {\bibfnamefont {Pierre-Emmanuel}\ \bibnamefont
  {Emeriau}}, \bibinfo {author} {\bibfnamefont {Francesco}\ \bibnamefont
  {Ceccarelli}}, \bibinfo {author} {\bibfnamefont {Ricardo}\ \bibnamefont
  {Albiero}}, \bibinfo {author} {\bibfnamefont {Paulo~Henrique}\ \bibnamefont
  {Dias~Ferreira}}, \bibinfo {author} {\bibfnamefont {Niccolo}\ \bibnamefont
  {Somaschi}}, \bibinfo {author} {\bibfnamefont {Jean}\ \bibnamefont
  {Senellart}}, \bibinfo {author} {\bibfnamefont {Isabelle}\ \bibnamefont
  {Sagnes}}, \bibinfo {author} {\bibfnamefont {Martina}\ \bibnamefont
  {Morassi}}, \bibinfo {author} {\bibfnamefont {Aristide}\ \bibnamefont
  {Lema{\^i}tre}}, \bibinfo {author} {\bibfnamefont {Pascale}\ \bibnamefont
  {Senellart}}, \bibinfo {author} {\bibfnamefont {Fabio}\ \bibnamefont
  {Sciarrino}}, \bibinfo {author} {\bibfnamefont {Marco}\ \bibnamefont
  {Liscidini}}, \bibinfo {author} {\bibfnamefont {Nadia}\ \bibnamefont
  {Belabas}}, \ and\ \bibinfo {author} {\bibfnamefont {Roberto}\ \bibnamefont
  {Osellame}},\ }\emph {\enquote {\bibinfo {title} {High-fidelity four-photon
  ghz states on chip},}\ }\href {https://doi.org/10.1038/s41534-024-00830-z}
  {\bibfield  {journal} {\bibinfo  {journal} {npj Quantum Inf.}\ }\textbf
  {\bibinfo {volume} {10}},\ \bibinfo {pages} {50} (\bibinfo {year} {2024})},\
  \Eprint {http://arxiv.org/abs/2211.15626} {arXiv:2211.15626}\BibitemShut
  {NoStop}%
\bibitem [{\citenamefont {Thomas}\ \emph {et~al.}(2024)\citenamefont {Thomas},
  \citenamefont {Ruscio}, \citenamefont {Morin},\ and\ \citenamefont
  {Rempe}}]{Thomas2024}%
  \BibitemOpen
  \bibfield  {author} {\bibinfo {author} {\bibfnamefont {Philip}\ \bibnamefont
  {Thomas}}, \bibinfo {author} {\bibfnamefont {Leonardo}\ \bibnamefont
  {Ruscio}}, \bibinfo {author} {\bibfnamefont {Olivier}\ \bibnamefont {Morin}},
  \ and\ \bibinfo {author} {\bibfnamefont {Gerhard}\ \bibnamefont {Rempe}},\
  }\emph {\enquote {\bibinfo {title} {Fusion of deterministically generated
  photonic graph states},}\ }\href {https://doi.org/10.1038/s41586-024-07357-5}
  {\bibfield  {journal} {\bibinfo  {journal} {Nature}\ }\textbf {\bibinfo
  {volume} {629}},\ \bibinfo {pages} {567} (\bibinfo {year} {2024})},\ \Eprint
  {http://arxiv.org/abs/2403.11950} {arXiv:2403.11950}\BibitemShut {NoStop}%
\bibitem [{\citenamefont {Huet}\ \emph {et~al.}(2025)\citenamefont {Huet},
  \citenamefont {Ramesh}, \citenamefont {Wein}, \citenamefont {Coste},
  \citenamefont {Hilaire}, \citenamefont {Somaschi}, \citenamefont {Morassi},
  \citenamefont {Lema{\^i}tre}, \citenamefont {Sagnes}, \citenamefont {Doty},
  \citenamefont {Krebs}, \citenamefont {Lanco}, \citenamefont {Fioretto},\ and\
  \citenamefont {Senellart}}]{Huet2025}%
  \BibitemOpen
  \bibfield  {author} {\bibinfo {author} {\bibfnamefont {H{\^e}lio}\
  \bibnamefont {Huet}}, \bibinfo {author} {\bibfnamefont {Prashant~R.}\
  \bibnamefont {Ramesh}}, \bibinfo {author} {\bibfnamefont {Stephen~C.}\
  \bibnamefont {Wein}}, \bibinfo {author} {\bibfnamefont {Nathan}\ \bibnamefont
  {Coste}}, \bibinfo {author} {\bibfnamefont {Paul}\ \bibnamefont {Hilaire}},
  \bibinfo {author} {\bibfnamefont {Niccolo}\ \bibnamefont {Somaschi}},
  \bibinfo {author} {\bibfnamefont {Martina}\ \bibnamefont {Morassi}}, \bibinfo
  {author} {\bibfnamefont {Aristide}\ \bibnamefont {Lema{\^i}tre}}, \bibinfo
  {author} {\bibfnamefont {Isabelle}\ \bibnamefont {Sagnes}}, \bibinfo {author}
  {\bibfnamefont {Matthew~F.}\ \bibnamefont {Doty}}, \bibinfo {author}
  {\bibfnamefont {Olivier}\ \bibnamefont {Krebs}}, \bibinfo {author}
  {\bibfnamefont {Lo{\"i}c}\ \bibnamefont {Lanco}}, \bibinfo {author}
  {\bibfnamefont {Dario~Alessandro}\ \bibnamefont {Fioretto}}, \ and\ \bibinfo
  {author} {\bibfnamefont {Pascale}\ \bibnamefont {Senellart}},\ }\emph
  {\enquote {\bibinfo {title} {Deterministic and reconfigurable graph state
  generation with a single solid-state quantum emitter},}\ }\href
  {https://doi.org/10.1038/s41467-025-59693-3} {\bibfield  {journal} {\bibinfo
  {journal} {Nat. Commun.}\ }\textbf {\bibinfo {volume} {16}},\ \bibinfo
  {pages} {4337} (\bibinfo {year} {2025})},\ \Eprint
  {http://arxiv.org/abs/2410.23518} {arXiv:2410.23518}\BibitemShut {NoStop}%
\bibitem [{\citenamefont {Greenberger}\ \emph {et~al.}(1989)\citenamefont
  {Greenberger}, \citenamefont {Horne},\ and\ \citenamefont
  {Zeilinger}}]{GreenbergerHorneZeilinger1989}%
  \BibitemOpen
  \bibfield  {author} {\bibinfo {author} {\bibfnamefont {Daniel~M.}\
  \bibnamefont {Greenberger}}, \bibinfo {author} {\bibfnamefont {Michael~A.}\
  \bibnamefont {Horne}}, \ and\ \bibinfo {author} {\bibfnamefont {Anton}\
  \bibnamefont {Zeilinger}},\ }\emph {\enquote {\bibinfo {title} {Going beyond
  {B}ell's theorem},}\ }in\ \href@noop {} {\emph {\bibinfo {booktitle} {Bell's
  {T}heorem, {Q}uantum {T}heory, and {C}onceptions of the {U}niverse}}},\
  \bibinfo {editor} {edited by\ \bibinfo {editor} {\bibfnamefont
  {M.}~\bibnamefont {Kafalatos}}}\ (\bibinfo  {publisher} {Kluwer Academics},\
  \bibinfo {address} {Dordrecht, The {N}etherlands},\ \bibinfo {year} {1989})\
  p.\ \bibinfo {pages} {73{\textendash}76},\ \Eprint
  {http://arxiv.org/abs/0712.0921} {arXiv:0712.0921}\BibitemShut {NoStop}%
\bibitem [{\citenamefont {Moehring}\ \emph {et~al.}(2007)\citenamefont
  {Moehring}, \citenamefont {Maunz}, \citenamefont {Olmschenk}, \citenamefont
  {Younge}, \citenamefont {Matsukevich}, \citenamefont {Duan},\ and\
  \citenamefont {Monroe}}]{Moehring2007}%
  \BibitemOpen
  \bibfield  {author} {\bibinfo {author} {\bibfnamefont {David~L.}\
  \bibnamefont {Moehring}}, \bibinfo {author} {\bibfnamefont {Peter}\
  \bibnamefont {Maunz}}, \bibinfo {author} {\bibfnamefont {Steven}\
  \bibnamefont {Olmschenk}}, \bibinfo {author} {\bibfnamefont {Kelly~Cooper}\
  \bibnamefont {Younge}}, \bibinfo {author} {\bibfnamefont {Dzmitry~N.}\
  \bibnamefont {Matsukevich}}, \bibinfo {author} {\bibfnamefont {Lu-Ming}\
  \bibnamefont {Duan}}, \ and\ \bibinfo {author} {\bibfnamefont {Christopher}\
  \bibnamefont {Monroe}},\ }\emph {\enquote {\bibinfo {title} {Entanglement of
  single-atom quantum bits at a distance},}\ }\href
  {https://doi.org/10.1038/nature06118} {\bibfield  {journal} {\bibinfo
  {journal} {Nature}\ }\textbf {\bibinfo {volume} {449}},\ \bibinfo {pages}
  {68} (\bibinfo {year} {2007})}\BibitemShut {NoStop}%
\bibitem [{\citenamefont {Stephenson}\ \emph {et~al.}(2020)\citenamefont
  {Stephenson}, \citenamefont {Nadlinger}, \citenamefont {Nichol},
  \citenamefont {An}, \citenamefont {Drmota}, \citenamefont {Ballance},
  \citenamefont {Thirumalai}, \citenamefont {Goodwin}, \citenamefont {Lucas},\
  and\ \citenamefont {Ballance}}]{Stephenson2020}%
  \BibitemOpen
  \bibfield  {author} {\bibinfo {author} {\bibfnamefont {Laurent~J.}\
  \bibnamefont {Stephenson}}, \bibinfo {author} {\bibfnamefont {David~P.}\
  \bibnamefont {Nadlinger}}, \bibinfo {author} {\bibfnamefont {Bethan~C.}\
  \bibnamefont {Nichol}}, \bibinfo {author} {\bibfnamefont {Shuoming}\
  \bibnamefont {An}}, \bibinfo {author} {\bibfnamefont {Peter}\ \bibnamefont
  {Drmota}}, \bibinfo {author} {\bibfnamefont {Timothy~G.}\ \bibnamefont
  {Ballance}}, \bibinfo {author} {\bibfnamefont {Keshav}\ \bibnamefont
  {Thirumalai}}, \bibinfo {author} {\bibfnamefont {Joseph~F.}\ \bibnamefont
  {Goodwin}}, \bibinfo {author} {\bibfnamefont {David~M.}\ \bibnamefont
  {Lucas}}, \ and\ \bibinfo {author} {\bibfnamefont {Christopher~J.}\
  \bibnamefont {Ballance}},\ }\emph {\enquote {\bibinfo {title} {{High-Rate,
  High-Fidelity Entanglement of Qubits Across an Elementary Quantum
  Network}},}\ }\href {https://doi.org/10.1103/PhysRevLett.124.110501}
  {\bibfield  {journal} {\bibinfo  {journal} {Phys. Rev. Lett.}\ }\textbf
  {\bibinfo {volume} {124}},\ \bibinfo {pages} {110501} (\bibinfo {year}
  {2020})},\ \Eprint {http://arxiv.org/abs/1911.10841}
  {arXiv:1911.10841}\BibitemShut {NoStop}%
\bibitem [{\citenamefont {Nichol}\ \emph {et~al.}(2022)\citenamefont {Nichol},
  \citenamefont {Srinivas}, \citenamefont {Nadlinger}, \citenamefont {Drmota},
  \citenamefont {Main}, \citenamefont {Araneda}, \citenamefont {Ballance},\
  and\ \citenamefont {Lucas}}]{Nichol2022}%
  \BibitemOpen
  \bibfield  {author} {\bibinfo {author} {\bibfnamefont {Bethan~C.}\
  \bibnamefont {Nichol}}, \bibinfo {author} {\bibfnamefont {Raghavendra}\
  \bibnamefont {Srinivas}}, \bibinfo {author} {\bibfnamefont {David~P.}\
  \bibnamefont {Nadlinger}}, \bibinfo {author} {\bibfnamefont {Peter}\
  \bibnamefont {Drmota}}, \bibinfo {author} {\bibfnamefont {Dougal}\
  \bibnamefont {Main}}, \bibinfo {author} {\bibfnamefont {Gabriel}\
  \bibnamefont {Araneda}}, \bibinfo {author} {\bibfnamefont {Christopher~J.}\
  \bibnamefont {Ballance}}, \ and\ \bibinfo {author} {\bibfnamefont {David~M.}\
  \bibnamefont {Lucas}},\ }\emph {\enquote {\bibinfo {title} {An elementary
  quantum network of entangled optical atomic clocks},}\ }\href
  {https://doi.org/10.1038/s41586-022-05088-z} {\bibfield  {journal} {\bibinfo
  {journal} {Nature}\ }\textbf {\bibinfo {volume} {609}},\ \bibinfo {pages}
  {689} (\bibinfo {year} {2022})},\ \Eprint {http://arxiv.org/abs/2111.10336}
  {arXiv:2111.10336}\BibitemShut {NoStop}%
\bibitem [{\citenamefont {Main}\ \emph
  {et~al.}(2025{\natexlab{a}})\citenamefont {Main}, \citenamefont {Drmota},
  \citenamefont {Nadlinger}, \citenamefont {Ainley}, \citenamefont {Agrawal},
  \citenamefont {Nichol}, \citenamefont {Srinivas}, \citenamefont {Araneda},\
  and\ \citenamefont {Lucas}}]{Main2025}%
  \BibitemOpen
  \bibfield  {author} {\bibinfo {author} {\bibfnamefont {Dougal}\ \bibnamefont
  {Main}}, \bibinfo {author} {\bibfnamefont {Peter}\ \bibnamefont {Drmota}},
  \bibinfo {author} {\bibfnamefont {David~P.}\ \bibnamefont {Nadlinger}},
  \bibinfo {author} {\bibfnamefont {Ellis~M.}\ \bibnamefont {Ainley}}, \bibinfo
  {author} {\bibfnamefont {Ayush}\ \bibnamefont {Agrawal}}, \bibinfo {author}
  {\bibfnamefont {Bethan~C.}\ \bibnamefont {Nichol}}, \bibinfo {author}
  {\bibfnamefont {Raghavendra}\ \bibnamefont {Srinivas}}, \bibinfo {author}
  {\bibfnamefont {Gabriel}\ \bibnamefont {Araneda}}, \ and\ \bibinfo {author}
  {\bibfnamefont {David~M.}\ \bibnamefont {Lucas}},\ }\emph {\enquote {\bibinfo
  {title} {Distributed quantum computing across an optical network link},}\
  }\href {https://doi.org/10.1038/s41586-024-08404-x} {\bibfield  {journal}
  {\bibinfo  {journal} {Nature}\ }\textbf {\bibinfo {volume} {638}},\ \bibinfo
  {pages} {383} (\bibinfo {year} {2025}{\natexlab{a}})},\ \Eprint
  {http://arxiv.org/abs/2407.00835} {arXiv:2407.00835}\BibitemShut {NoStop}%
\bibitem [{\citenamefont {Main}\ \emph
  {et~al.}(2025{\natexlab{b}})\citenamefont {Main}, \citenamefont {Drmota},
  \citenamefont {Ainley}, \citenamefont {Agrawal}, \citenamefont {Webb},
  \citenamefont {Saner}, \citenamefont {Bazavan}, \citenamefont {Nichol},
  \citenamefont {Srinivas}, \citenamefont {Nadlinger}, \citenamefont
  {Araneda},\ and\ \citenamefont
  {Lucas}}]{main2025multipartitemixedspeciesentanglementquantum}%
  \BibitemOpen
  \bibfield  {author} {\bibinfo {author} {\bibfnamefont {Dougal}\ \bibnamefont
  {Main}}, \bibinfo {author} {\bibfnamefont {Peter}\ \bibnamefont {Drmota}},
  \bibinfo {author} {\bibfnamefont {Ellis~M.}\ \bibnamefont {Ainley}}, \bibinfo
  {author} {\bibfnamefont {Ayush}\ \bibnamefont {Agrawal}}, \bibinfo {author}
  {\bibfnamefont {Donovan}\ \bibnamefont {Webb}}, \bibinfo {author}
  {\bibfnamefont {Sebastian}\ \bibnamefont {Saner}}, \bibinfo {author}
  {\bibfnamefont {Oana}\ \bibnamefont {Bazavan}}, \bibinfo {author}
  {\bibfnamefont {Bethan~C.}\ \bibnamefont {Nichol}}, \bibinfo {author}
  {\bibfnamefont {Raghavendra}\ \bibnamefont {Srinivas}}, \bibinfo {author}
  {\bibfnamefont {David~P.}\ \bibnamefont {Nadlinger}}, \bibinfo {author}
  {\bibfnamefont {Gabriel}\ \bibnamefont {Araneda}}, \ and\ \bibinfo {author}
  {\bibfnamefont {David~M.}\ \bibnamefont {Lucas}},\ }\href@noop {} {\emph
  {\enquote {\bibinfo {title} {{Multipartite Mixed-Species Entanglement over a
  Quantum Network}},}\ }}\Eprint {http://arxiv.org/abs/2506.14334}
  {arXiv:2506.14334} [quant-ph] (\bibinfo {year}
  {2025}{\natexlab{b}})\BibitemShut {NoStop}%
\bibitem [{\citenamefont {Krutyanskiy}\ \emph
  {et~al.}(2023{\natexlab{a}})\citenamefont {Krutyanskiy}, \citenamefont
  {Galli}, \citenamefont {Krcmarsky}, \citenamefont {Baier}, \citenamefont
  {Fioretto}, \citenamefont {Pu}, \citenamefont {Mazloom}, \citenamefont
  {Sekatski}, \citenamefont {Canteri}, \citenamefont {Teller}, \citenamefont
  {Schupp}, \citenamefont {Bate}, \citenamefont {Meraner}, \citenamefont
  {Sangouard}, \citenamefont {Lanyon},\ and\ \citenamefont
  {Northup}}]{Krut2022}%
  \BibitemOpen
  \bibfield  {author} {\bibinfo {author} {\bibfnamefont {Victor}\ \bibnamefont
  {Krutyanskiy}}, \bibinfo {author} {\bibfnamefont {Maria}\ \bibnamefont
  {Galli}}, \bibinfo {author} {\bibfnamefont {Vojtech}\ \bibnamefont
  {Krcmarsky}}, \bibinfo {author} {\bibfnamefont {Simon}\ \bibnamefont
  {Baier}}, \bibinfo {author} {\bibfnamefont {Dario~Alessandro}\ \bibnamefont
  {Fioretto}}, \bibinfo {author} {\bibfnamefont {Yunfei}\ \bibnamefont {Pu}},
  \bibinfo {author} {\bibfnamefont {Azadeh}\ \bibnamefont {Mazloom}}, \bibinfo
  {author} {\bibfnamefont {Pavel}\ \bibnamefont {Sekatski}}, \bibinfo {author}
  {\bibfnamefont {Marco}\ \bibnamefont {Canteri}}, \bibinfo {author}
  {\bibfnamefont {Markus}\ \bibnamefont {Teller}}, \bibinfo {author}
  {\bibfnamefont {Josef}\ \bibnamefont {Schupp}}, \bibinfo {author}
  {\bibfnamefont {James}\ \bibnamefont {Bate}}, \bibinfo {author}
  {\bibfnamefont {Martin}\ \bibnamefont {Meraner}}, \bibinfo {author}
  {\bibfnamefont {Nicolas}\ \bibnamefont {Sangouard}}, \bibinfo {author}
  {\bibfnamefont {Ben~P.}\ \bibnamefont {Lanyon}}, \ and\ \bibinfo {author}
  {\bibfnamefont {Tracy~E.}\ \bibnamefont {Northup}},\ }\emph {\enquote
  {\bibinfo {title} {{Entanglement of Trapped-Ion Qubits Separated by 230
  Meters}},}\ }\href {https://doi.org/10.1103/PhysRevLett.130.050803}
  {\bibfield  {journal} {\bibinfo  {journal} {Phys. Rev. Lett.}\ }\textbf
  {\bibinfo {volume} {130}},\ \bibinfo {pages} {050803} (\bibinfo {year}
  {2023}{\natexlab{a}})},\ \Eprint {http://arxiv.org/abs/2208.14907}
  {arXiv:2208.14907}\BibitemShut {NoStop}%
\bibitem [{\citenamefont {Bock}\ \emph {et~al.}(2018)\citenamefont {Bock},
  \citenamefont {Eich}, \citenamefont {Kucera}, \citenamefont {Kreis},
  \citenamefont {Lenhard}, \citenamefont {Becher},\ and\ \citenamefont
  {Eschner}}]{Bock2018}%
  \BibitemOpen
  \bibfield  {author} {\bibinfo {author} {\bibfnamefont {Matthias}\
  \bibnamefont {Bock}}, \bibinfo {author} {\bibfnamefont {Pascal}\ \bibnamefont
  {Eich}}, \bibinfo {author} {\bibfnamefont {Stephan}\ \bibnamefont {Kucera}},
  \bibinfo {author} {\bibfnamefont {Matthias}\ \bibnamefont {Kreis}}, \bibinfo
  {author} {\bibfnamefont {Andreas}\ \bibnamefont {Lenhard}}, \bibinfo {author}
  {\bibfnamefont {Christoph}\ \bibnamefont {Becher}}, \ and\ \bibinfo {author}
  {\bibfnamefont {J{\"u}rgen}\ \bibnamefont {Eschner}},\ }\emph {\enquote
  {\bibinfo {title} {High-fidelity entanglement between a trapped ion and a
  telecom photon via quantum frequency conversion},}\ }\href
  {https://doi.org/10.1038/s41467-018-04341-2} {\bibfield  {journal} {\bibinfo
  {journal} {Nat. Commun.}\ }\textbf {\bibinfo {volume} {9}},\ \bibinfo {pages}
  {1998} (\bibinfo {year} {2018})},\ \Eprint {http://arxiv.org/abs/1710.04866}
  {arXiv:1710.04866}\BibitemShut {NoStop}%
\bibitem [{\citenamefont {Walker}\ \emph {et~al.}(2018)\citenamefont {Walker},
  \citenamefont {Miyanishi}, \citenamefont {Ikuta}, \citenamefont {Takahashi},
  \citenamefont {Vartabi~Kashanian}, \citenamefont {Tsujimoto}, \citenamefont
  {Hayasaka}, \citenamefont {Yamamoto}, \citenamefont {Imoto},\ and\
  \citenamefont {Keller}}]{Walker2018}%
  \BibitemOpen
  \bibfield  {author} {\bibinfo {author} {\bibfnamefont {Thomas}\ \bibnamefont
  {Walker}}, \bibinfo {author} {\bibfnamefont {Koichiro}\ \bibnamefont
  {Miyanishi}}, \bibinfo {author} {\bibfnamefont {Rikizo}\ \bibnamefont
  {Ikuta}}, \bibinfo {author} {\bibfnamefont {Hiroki}\ \bibnamefont
  {Takahashi}}, \bibinfo {author} {\bibfnamefont {Samir}\ \bibnamefont
  {Vartabi~Kashanian}}, \bibinfo {author} {\bibfnamefont {Yoshiaki}\
  \bibnamefont {Tsujimoto}}, \bibinfo {author} {\bibfnamefont {Kazuhiro}\
  \bibnamefont {Hayasaka}}, \bibinfo {author} {\bibfnamefont {Takashi}\
  \bibnamefont {Yamamoto}}, \bibinfo {author} {\bibfnamefont {Nobuyuki}\
  \bibnamefont {Imoto}}, \ and\ \bibinfo {author} {\bibfnamefont {Matthias}\
  \bibnamefont {Keller}},\ }\emph {\enquote {\bibinfo {title} {{Long-Distance
  Single Photon Transmission from a Trapped Ion via Quantum Frequency
  Conversion}},}\ }\href {https://doi.org/10.1103/PhysRevLett.120.203601}
  {\bibfield  {journal} {\bibinfo  {journal} {Phys. Rev. Lett.}\ }\textbf
  {\bibinfo {volume} {120}},\ \bibinfo {pages} {203601} (\bibinfo {year}
  {2018})},\ \Eprint {http://arxiv.org/abs/1711.09644}
  {arXiv:1711.09644}\BibitemShut {NoStop}%
\bibitem [{\citenamefont {Krutyanskiy}\ \emph {et~al.}(2019)\citenamefont
  {Krutyanskiy}, \citenamefont {Meraner}, \citenamefont {Schupp}, \citenamefont
  {Krcmarsky}, \citenamefont {Hainzer},\ and\ \citenamefont
  {Lanyon}}]{Krutyanskiy2019}%
  \BibitemOpen
  \bibfield  {author} {\bibinfo {author} {\bibfnamefont {Victor}\ \bibnamefont
  {Krutyanskiy}}, \bibinfo {author} {\bibfnamefont {Martin}\ \bibnamefont
  {Meraner}}, \bibinfo {author} {\bibfnamefont {Josef}\ \bibnamefont {Schupp}},
  \bibinfo {author} {\bibfnamefont {Vojtech}\ \bibnamefont {Krcmarsky}},
  \bibinfo {author} {\bibfnamefont {H.}~\bibnamefont {Hainzer}}, \ and\
  \bibinfo {author} {\bibfnamefont {Ben~P.}\ \bibnamefont {Lanyon}},\ }\emph
  {\enquote {\bibinfo {title} {Light-matter entanglement over 50{\thinspace}km
  of optical fibre},}\ }\href {https://doi.org/10.1038/s41534-019-0186-3}
  {\bibfield  {journal} {\bibinfo  {journal} {npj Quantum Inf.}\ }\textbf
  {\bibinfo {volume} {5}},\ \bibinfo {pages} {72} (\bibinfo {year} {2019})},\
  \Eprint {http://arxiv.org/abs/1901.06317} {arXiv:1901.06317}\BibitemShut
  {NoStop}%
\bibitem [{\citenamefont {Krutyanskiy}\ \emph {et~al.}(2024)\citenamefont
  {Krutyanskiy}, \citenamefont {Canteri}, \citenamefont {Meraner},
  \citenamefont {Krcmarsky},\ and\ \citenamefont
  {Lanyon}}]{KrutyanskiyCanteriMeranerKrcmarskyLanyon2024}%
  \BibitemOpen
  \bibfield  {author} {\bibinfo {author} {\bibfnamefont {Victor}\ \bibnamefont
  {Krutyanskiy}}, \bibinfo {author} {\bibfnamefont {Marco}\ \bibnamefont
  {Canteri}}, \bibinfo {author} {\bibfnamefont {Martin}\ \bibnamefont
  {Meraner}}, \bibinfo {author} {\bibfnamefont {Vojtech}\ \bibnamefont
  {Krcmarsky}}, \ and\ \bibinfo {author} {\bibfnamefont {Ben~P.}\ \bibnamefont
  {Lanyon}},\ }\emph {\enquote {\bibinfo {title} {{Multimode Ion-Photon
  Entanglement over 101 Kilometers}},}\ }\href
  {https://doi.org/10.1103/PRXQuantum.5.020308} {\bibfield  {journal} {\bibinfo
   {journal} {PRX Quantum}\ }\textbf {\bibinfo {volume} {5}},\ \bibinfo {pages}
  {020308} (\bibinfo {year} {2024})},\ \Eprint
  {http://arxiv.org/abs/2308.08891} {arXiv:2308.08891}\BibitemShut {NoStop}%
\bibitem [{\citenamefont {Krutyanskiy}\ \emph
  {et~al.}(2023{\natexlab{b}})\citenamefont {Krutyanskiy}, \citenamefont
  {Canteri}, \citenamefont {Meraner}, \citenamefont {Bate}, \citenamefont
  {Krcmarsky}, \citenamefont {Schupp}, \citenamefont {Sangouard},\ and\
  \citenamefont {Lanyon}}]{Krut2023}%
  \BibitemOpen
  \bibfield  {author} {\bibinfo {author} {\bibfnamefont {Victor}\ \bibnamefont
  {Krutyanskiy}}, \bibinfo {author} {\bibfnamefont {Marco}\ \bibnamefont
  {Canteri}}, \bibinfo {author} {\bibfnamefont {Martin}\ \bibnamefont
  {Meraner}}, \bibinfo {author} {\bibfnamefont {James}\ \bibnamefont {Bate}},
  \bibinfo {author} {\bibfnamefont {Vojtech}\ \bibnamefont {Krcmarsky}},
  \bibinfo {author} {\bibfnamefont {Josef}\ \bibnamefont {Schupp}}, \bibinfo
  {author} {\bibfnamefont {Nicolas}\ \bibnamefont {Sangouard}}, \ and\ \bibinfo
  {author} {\bibfnamefont {Ben~P.}\ \bibnamefont {Lanyon}},\ }\emph {\enquote
  {\bibinfo {title} {{Telecom-Wavelength Quantum Repeater Node Based on a
  Trapped-Ion Processor}},}\ }\href
  {https://doi.org/10.1103/PhysRevLett.130.213601} {\bibfield  {journal}
  {\bibinfo  {journal} {Phys. Rev. Lett.}\ }\textbf {\bibinfo {volume} {130}},\
  \bibinfo {pages} {213601} (\bibinfo {year} {2023}{\natexlab{b}})},\ \Eprint
  {http://arxiv.org/abs/2210.05418} {arXiv:2210.05418}\BibitemShut {NoStop}%
\bibitem [{\citenamefont {Bergerhoff}\ \emph {et~al.}(2024)\citenamefont
  {Bergerhoff}, \citenamefont {Elshehy}, \citenamefont {Kucera}, \citenamefont
  {Kreis},\ and\ \citenamefont {Eschner}}]{PhysRevA.110.032603}%
  \BibitemOpen
  \bibfield  {author} {\bibinfo {author} {\bibfnamefont {Max}\ \bibnamefont
  {Bergerhoff}}, \bibinfo {author} {\bibfnamefont {Omar}\ \bibnamefont
  {Elshehy}}, \bibinfo {author} {\bibfnamefont {Stephan}\ \bibnamefont
  {Kucera}}, \bibinfo {author} {\bibfnamefont {Matthias}\ \bibnamefont
  {Kreis}}, \ and\ \bibinfo {author} {\bibfnamefont {J\"urgen}\ \bibnamefont
  {Eschner}},\ }\emph {\enquote {\bibinfo {title} {Quantum repeater node with
  free-space coupled trapped ions},}\ }\href
  {https://doi.org/10.1103/PhysRevA.110.032603} {\bibfield  {journal} {\bibinfo
   {journal} {Phys. Rev. A}\ }\textbf {\bibinfo {volume} {110}},\ \bibinfo
  {pages} {032603} (\bibinfo {year} {2024})},\ \Eprint
  {http://arxiv.org/abs/2312.14805} {arXiv:2312.14805}\BibitemShut {NoStop}%
\bibitem [{\citenamefont {Schupp}\ \emph {et~al.}(2021)\citenamefont {Schupp},
  \citenamefont {Krcmarsky}, \citenamefont {Krutyanskiy}, \citenamefont
  {Meraner}, \citenamefont {Northup},\ and\ \citenamefont
  {Lanyon}}]{SchuppKrcmarskyKrutyanskiyMeranerNorthupLanyon2021}%
  \BibitemOpen
  \bibfield  {author} {\bibinfo {author} {\bibfnamefont {Josef}\ \bibnamefont
  {Schupp}}, \bibinfo {author} {\bibfnamefont {Vojtech}\ \bibnamefont
  {Krcmarsky}}, \bibinfo {author} {\bibfnamefont {Victor}\ \bibnamefont
  {Krutyanskiy}}, \bibinfo {author} {\bibfnamefont {Martin}\ \bibnamefont
  {Meraner}}, \bibinfo {author} {\bibfnamefont {Tracy~E.}\ \bibnamefont
  {Northup}}, \ and\ \bibinfo {author} {\bibfnamefont {Ben~P.}\ \bibnamefont
  {Lanyon}},\ }\emph {\enquote {\bibinfo {title} {{Interface between
  Trapped-Ion Qubits and Traveling Photons with Close-to-Optimal
  Efficiency}},}\ }\href {https://doi.org/10.1103/PRXQuantum.2.020331}
  {\bibfield  {journal} {\bibinfo  {journal} {PRX Quantum}\ }\textbf {\bibinfo
  {volume} {2}},\ \bibinfo {pages} {020331} (\bibinfo {year} {2021})},\ \Eprint
  {http://arxiv.org/abs/2105.02121} {arXiv:2105.02121}\BibitemShut {NoStop}%
\bibitem [{\citenamefont {Schupp}(2021)}]{SchuppPhDthesis2021}%
  \BibitemOpen
  \bibfield  {author} {\bibinfo {author} {\bibfnamefont {Josef}\ \bibnamefont
  {Schupp}},\ }\emph {\bibinfo {title} {Interface between trapped-ion qubits
  and travelling photons with close-to-optimal efficiency}},\ \href
  {https://quantumoptics.at/images/publications/dissertation/Schupp_thesis.pdf}
  {Ph.D. thesis},\ \bibinfo  {school} {University of Innsbruck} (\bibinfo
  {year} {2021})\BibitemShut {NoStop}%
\bibitem [{\citenamefont {Stute}\ \emph {et~al.}(2012)\citenamefont {Stute},
  \citenamefont {Casabone}, \citenamefont {Schindler}, \citenamefont {Monz},
  \citenamefont {Schmidt}, \citenamefont {Brandst{\"a}tter}, \citenamefont
  {Northup},\ and\ \citenamefont {Blatt}}]{StuteEtAl2012}%
  \BibitemOpen
  \bibfield  {author} {\bibinfo {author} {\bibfnamefont {Andreas}\ \bibnamefont
  {Stute}}, \bibinfo {author} {\bibfnamefont {Bernardo}\ \bibnamefont
  {Casabone}}, \bibinfo {author} {\bibfnamefont {Philipp}\ \bibnamefont
  {Schindler}}, \bibinfo {author} {\bibfnamefont {Thomas}\ \bibnamefont
  {Monz}}, \bibinfo {author} {\bibfnamefont {Piet~O.}\ \bibnamefont {Schmidt}},
  \bibinfo {author} {\bibfnamefont {Birgit}\ \bibnamefont {Brandst{\"a}tter}},
  \bibinfo {author} {\bibfnamefont {Tracy~E.}\ \bibnamefont {Northup}}, \ and\
  \bibinfo {author} {\bibfnamefont {Rainer}\ \bibnamefont {Blatt}},\ }\emph
  {\enquote {\bibinfo {title} {Tunable ion-photon entanglement in an optical
  cavity},}\ }\href {https://doi.org/10.1038/nature11120} {\bibfield  {journal}
  {\bibinfo  {journal} {Nature}\ }\textbf {\bibinfo {volume} {485}},\ \bibinfo
  {pages} {482} (\bibinfo {year} {2012})},\ \Eprint
  {http://arxiv.org/abs/1301.0275} {arXiv:1301.0275}\BibitemShut {NoStop}%
\bibitem [{\citenamefont {S{\o}rensen}\ and\ \citenamefont
  {M{\o}lmer}(1999)}]{SoerensenMoelmer1999}%
  \BibitemOpen
  \bibfield  {author} {\bibinfo {author} {\bibfnamefont {Anders}\ \bibnamefont
  {S{\o}rensen}}\ and\ \bibinfo {author} {\bibfnamefont {Klaus}\ \bibnamefont
  {M{\o}lmer}},\ }\emph {\enquote {\bibinfo {title} {{Quantum Computation with
  Ions in Thermal Motion}},}\ }\href {\doibase 10.1103/PhysRevLett.82.1971}
  {\bibfield  {journal} {\bibinfo  {journal} {Phys. Rev. Lett.}\ }\textbf
  {\bibinfo {volume} {82}},\ \bibinfo {pages} {1971{\textendash}1974} (\bibinfo
  {year} {1999})},\ \Eprint {http://arxiv.org/abs/quant-ph/9810039}
  {arXiv:quant-ph/9810039}\BibitemShut {NoStop}%
\bibitem [{\citenamefont {S{\o}rensen}\ and\ \citenamefont
  {M{\o}lmer}(2000)}]{SoerensenMoelmer2000}%
  \BibitemOpen
  \bibfield  {author} {\bibinfo {author} {\bibfnamefont {Anders}\ \bibnamefont
  {S{\o}rensen}}\ and\ \bibinfo {author} {\bibfnamefont {Klaus}\ \bibnamefont
  {M{\o}lmer}},\ }\emph {\enquote {\bibinfo {title} {Entanglement and quantum
  computation with ions in thermal motion},}\ }\href {\doibase
  10.1103/PhysRevA.62.022311} {\bibfield  {journal} {\bibinfo  {journal} {Phys.
  Rev. A}\ }\textbf {\bibinfo {volume} {62}},\ \bibinfo {pages} {022311}
  (\bibinfo {year} {2000})},\ \Eprint {http://arxiv.org/abs/quant-ph/0002024}
  {arXiv:quant-ph/0002024}\BibitemShut {NoStop}%
\bibitem [{\citenamefont {Benhelm}\ \emph {et~al.}(2008)\citenamefont
  {Benhelm}, \citenamefont {Kirchmair}, \citenamefont {Roos},\ and\
  \citenamefont {Blatt}}]{BenhelmKirchmairRoosBlatt2008}%
  \BibitemOpen
  \bibfield  {author} {\bibinfo {author} {\bibfnamefont {Jan}\ \bibnamefont
  {Benhelm}}, \bibinfo {author} {\bibfnamefont {Gerhard}\ \bibnamefont
  {Kirchmair}}, \bibinfo {author} {\bibfnamefont {Christian~F.}\ \bibnamefont
  {Roos}}, \ and\ \bibinfo {author} {\bibfnamefont {Rainer}\ \bibnamefont
  {Blatt}},\ }\emph {\enquote {\bibinfo {title} {Towards fault-tolerant quantum
  computing with trapped ions},}\ }\href {https://doi.org/10.1038/nphys961}
  {\bibfield  {journal} {\bibinfo  {journal} {Nat. Phys.}\ }\textbf {\bibinfo
  {volume} {4}},\ \bibinfo {pages} {463} (\bibinfo {year} {2008})},\ \Eprint
  {http://arxiv.org/abs/0803.2798} {arXiv:0803.2798}\BibitemShut {NoStop}%
\bibitem [{\citenamefont {Sackett}\ \emph {et~al.}(2000)\citenamefont
  {Sackett}, \citenamefont {Kielpinski}, \citenamefont {King}, \citenamefont
  {Langer}, \citenamefont {Meyer}, \citenamefont {Myatt}, \citenamefont {Rowe},
  \citenamefont {Turchette}, \citenamefont {Itano}, \citenamefont {Wineland},\
  and\ \citenamefont {Monroe}}]{Sackett2000}%
  \BibitemOpen
  \bibfield  {author} {\bibinfo {author} {\bibfnamefont {Charles~A.}\
  \bibnamefont {Sackett}}, \bibinfo {author} {\bibfnamefont {Dave}\
  \bibnamefont {Kielpinski}}, \bibinfo {author} {\bibfnamefont {B.~E.}\
  \bibnamefont {King}}, \bibinfo {author} {\bibfnamefont {C.}~\bibnamefont
  {Langer}}, \bibinfo {author} {\bibfnamefont {V.}~\bibnamefont {Meyer}},
  \bibinfo {author} {\bibfnamefont {Christopher~J.}\ \bibnamefont {Myatt}},
  \bibinfo {author} {\bibfnamefont {Mary}\ \bibnamefont {Rowe}}, \bibinfo
  {author} {\bibfnamefont {Quentin~A.}\ \bibnamefont {Turchette}}, \bibinfo
  {author} {\bibfnamefont {Wayne~M.}\ \bibnamefont {Itano}}, \bibinfo {author}
  {\bibfnamefont {David~J.}\ \bibnamefont {Wineland}}, \ and\ \bibinfo {author}
  {\bibfnamefont {Christopher}\ \bibnamefont {Monroe}},\ }\emph {\enquote
  {\bibinfo {title} {Experimental entanglement of four particles},}\ }\href
  {https://doi.org/10.1038/35005011} {\bibfield  {journal} {\bibinfo  {journal}
  {Nature}\ }\textbf {\bibinfo {volume} {404}},\ \bibinfo {pages} {256}
  (\bibinfo {year} {2000})}\BibitemShut {NoStop}%
\bibitem [{\citenamefont {Leibfried}\ \emph {et~al.}(2005)\citenamefont
  {Leibfried}, \citenamefont {Knill}, \citenamefont {Seidelin}, \citenamefont
  {Britton}, \citenamefont {Blakestad}, \citenamefont {Chiaverini},
  \citenamefont {Hume}, \citenamefont {Itano}, \citenamefont {Jost},
  \citenamefont {Langer}, \citenamefont {Ozeri}, \citenamefont {Reichle},\ and\
  \citenamefont {Wineland}}]{Leibfried2005}%
  \BibitemOpen
  \bibfield  {author} {\bibinfo {author} {\bibfnamefont {Dietrich}\
  \bibnamefont {Leibfried}}, \bibinfo {author} {\bibfnamefont {Emanuel}\
  \bibnamefont {Knill}}, \bibinfo {author} {\bibfnamefont {Signe}\ \bibnamefont
  {Seidelin}}, \bibinfo {author} {\bibfnamefont {Joseph}\ \bibnamefont
  {Britton}}, \bibinfo {author} {\bibfnamefont {R.~Bradford}\ \bibnamefont
  {Blakestad}}, \bibinfo {author} {\bibfnamefont {John}\ \bibnamefont
  {Chiaverini}}, \bibinfo {author} {\bibfnamefont {David~B.}\ \bibnamefont
  {Hume}}, \bibinfo {author} {\bibfnamefont {Wayne~M.}\ \bibnamefont {Itano}},
  \bibinfo {author} {\bibfnamefont {John~D.}\ \bibnamefont {Jost}}, \bibinfo
  {author} {\bibfnamefont {C.}~\bibnamefont {Langer}}, \bibinfo {author}
  {\bibfnamefont {Roee}\ \bibnamefont {Ozeri}}, \bibinfo {author}
  {\bibfnamefont {Rainer}\ \bibnamefont {Reichle}}, \ and\ \bibinfo {author}
  {\bibfnamefont {David~J.}\ \bibnamefont {Wineland}},\ }\emph {\enquote
  {\bibinfo {title} {{Creation of a six-atom `Schr{\"o}dinger cat' state}},}\
  }\href {https://doi.org/10.1038/nature04251} {\bibfield  {journal} {\bibinfo
  {journal} {Nature}\ }\textbf {\bibinfo {volume} {438}},\ \bibinfo {pages}
  {639} (\bibinfo {year} {2005})}\BibitemShut {NoStop}%
\bibitem [{\citenamefont {Monz}\ \emph {et~al.}(2011)\citenamefont {Monz},
  \citenamefont {Schindler}, \citenamefont {Barreiro}, \citenamefont {Chwalla},
  \citenamefont {Nigg}, \citenamefont {Coish}, \citenamefont {Harlander},
  \citenamefont {H\"ansel}, \citenamefont {Hennrich},\ and\ \citenamefont
  {Blatt}}]{MonzEtAl2011}%
  \BibitemOpen
  \bibfield  {author} {\bibinfo {author} {\bibfnamefont {Thomas}\ \bibnamefont
  {Monz}}, \bibinfo {author} {\bibfnamefont {Philipp}\ \bibnamefont
  {Schindler}}, \bibinfo {author} {\bibfnamefont {Julio~T.}\ \bibnamefont
  {Barreiro}}, \bibinfo {author} {\bibfnamefont {Michael}\ \bibnamefont
  {Chwalla}}, \bibinfo {author} {\bibfnamefont {Daniel}\ \bibnamefont {Nigg}},
  \bibinfo {author} {\bibfnamefont {William~A.}\ \bibnamefont {Coish}},
  \bibinfo {author} {\bibfnamefont {Maximilian}\ \bibnamefont {Harlander}},
  \bibinfo {author} {\bibfnamefont {Wolfgang}\ \bibnamefont {H\"ansel}},
  \bibinfo {author} {\bibfnamefont {Markus}\ \bibnamefont {Hennrich}}, \ and\
  \bibinfo {author} {\bibfnamefont {Rainer}\ \bibnamefont {Blatt}},\ }\emph
  {\enquote {\bibinfo {title} {{14-Qubit Entanglement: Creation and
  Coherence}},}\ }\href {https://doi.org/10.1103/PhysRevLett.106.130506}
  {\bibfield  {journal} {\bibinfo  {journal} {Phys. Rev. Lett.}\ }\textbf
  {\bibinfo {volume} {106}},\ \bibinfo {pages} {130506} (\bibinfo {year}
  {2011})},\ \Eprint {http://arxiv.org/abs/1009.6126}
  {arXiv:1009.6126}\BibitemShut {NoStop}%
\bibitem [{\citenamefont {G{\"u}hne}\ and\ \citenamefont
  {T{\'o}th}(2009)}]{GuehneToth2009}%
  \BibitemOpen
  \bibfield  {author} {\bibinfo {author} {\bibfnamefont {Otfried}\ \bibnamefont
  {G{\"u}hne}}\ and\ \bibinfo {author} {\bibfnamefont {G{\'e}za}\ \bibnamefont
  {T{\'o}th}},\ }\emph {\enquote {\bibinfo {title} {Entanglement detection},}\
  }\href {https://doi.org/10.1016/j.physrep.2009.02.004} {\bibfield  {journal}
  {\bibinfo  {journal} {Phys. Rep.}\ }\textbf {\bibinfo {volume} {474}},\
  \bibinfo {pages} {1{\textendash}75} (\bibinfo {year} {2009})},\ \Eprint
  {http://arxiv.org/abs/0811.2803} {arXiv:0811.2803}\BibitemShut {NoStop}%
\bibitem [{\citenamefont {Friis}\ \emph {et~al.}(2019)\citenamefont {Friis},
  \citenamefont {Vitagliano}, \citenamefont {Malik},\ and\ \citenamefont
  {Huber}}]{FriisVitaglianoMalikHuber2019}%
  \BibitemOpen
  \bibfield  {author} {\bibinfo {author} {\bibfnamefont {Nicolai}\ \bibnamefont
  {Friis}}, \bibinfo {author} {\bibfnamefont {Giuseppe}\ \bibnamefont
  {Vitagliano}}, \bibinfo {author} {\bibfnamefont {Mehul}\ \bibnamefont
  {Malik}}, \ and\ \bibinfo {author} {\bibfnamefont {Marcus}\ \bibnamefont
  {Huber}},\ }\emph {\enquote {\bibinfo {title} {{Entanglement Certification
  From Theory to Experiment}},}\ }\href
  {https://doi.org/10.1038/s42254-018-0003-5} {\bibfield  {journal} {\bibinfo
  {journal} {Nat. Rev. Phys.}\ }\textbf {\bibinfo {volume} {1}},\ \bibinfo
  {pages} {72{\textendash}87} (\bibinfo {year} {2019})},\ \Eprint
  {http://arxiv.org/abs/1906.10929} {arXiv:1906.10929}\BibitemShut {NoStop}%
\bibitem [{\citenamefont {Bavaresco}\ \emph {et~al.}(2018)\citenamefont
  {Bavaresco}, \citenamefont {Herrera~Valencia}, \citenamefont {Kl{\"o}ckl},
  \citenamefont {Pivoluska}, \citenamefont {Erker}, \citenamefont {Friis},
  \citenamefont {Malik},\ and\ \citenamefont {Huber}}]{BavarescoEtAl2018}%
  \BibitemOpen
  \bibfield  {author} {\bibinfo {author} {\bibfnamefont {Jessica}\ \bibnamefont
  {Bavaresco}}, \bibinfo {author} {\bibfnamefont {Natalia}\ \bibnamefont
  {Herrera~Valencia}}, \bibinfo {author} {\bibfnamefont {Claude}\ \bibnamefont
  {Kl{\"o}ckl}}, \bibinfo {author} {\bibfnamefont {Matej}\ \bibnamefont
  {Pivoluska}}, \bibinfo {author} {\bibfnamefont {Paul}\ \bibnamefont {Erker}},
  \bibinfo {author} {\bibfnamefont {Nicolai}\ \bibnamefont {Friis}}, \bibinfo
  {author} {\bibfnamefont {Mehul}\ \bibnamefont {Malik}}, \ and\ \bibinfo
  {author} {\bibfnamefont {Marcus}\ \bibnamefont {Huber}},\ }\emph {\enquote
  {\bibinfo {title} {Measurements in two bases are sufficient for certifying
  high-dimensional entanglement},}\ }\href
  {https://doi.org/10.1038/s41567-018-0203-z} {\bibfield  {journal} {\bibinfo
  {journal} {Nat. Phys.}\ }\textbf {\bibinfo {volume} {14}},\ \bibinfo {pages}
  {1032{\textendash}1037} (\bibinfo {year} {2018})},\ \Eprint
  {http://arxiv.org/abs/1709.07344} {arXiv:1709.07344}\BibitemShut {NoStop}%
\bibitem [{\citenamefont {Nielsen}\ and\ \citenamefont
  {Chuang}(2011)}]{NielsenChuang2011}%
  \BibitemOpen
  \bibfield  {author} {\bibinfo {author} {\bibfnamefont {Michael~A.}\
  \bibnamefont {Nielsen}}\ and\ \bibinfo {author} {\bibfnamefont {Isaac~L.}\
  \bibnamefont {Chuang}},\ }\href
  {https://www.amazon.com/Quantum-Computation-Information-10th-Anniversary/dp/1107002176?SubscriptionId=AKIAIOBINVZYXZQZ2U3A&tag=chimbori05-20&linkCode=xm2&camp=2025&creative=165953&creativeASIN=1107002176}
  {\emph {\bibinfo {title} {Quantum Computation and Quantum Information: 10th
  Anniversary Edition}}}\ (\bibinfo  {publisher} {Cambridge University Press},\
  \bibinfo {year} {2011})\BibitemShut {NoStop}%
\bibitem [{\citenamefont {Bertlmann}\ and\ \citenamefont
  {Friis}(2023)}]{BertlmannFriis2023}%
  \BibitemOpen
  \bibfield  {author} {\bibinfo {author} {\bibfnamefont {Reinhold~A.}\
  \bibnamefont {Bertlmann}}\ and\ \bibinfo {author} {\bibfnamefont {Nicolai}\
  \bibnamefont {Friis}},\ }\href
  {https://doi.org/10.1093/oso/9780199683338.001.0001} {\emph {\bibinfo {title}
  {Modern Quantum Theory {\textendash} From Quantum Mechanics to Entanglement
  and Quantum Information}}}\ (\bibinfo  {publisher} {Oxford University
  Press},\ \bibinfo {address} {Oxford, U.K.},\ \bibinfo {year}
  {2023})\BibitemShut {NoStop}%
\bibitem [{\citenamefont {Zhou}\ \emph {et~al.}(2024)\citenamefont {Zhou},
  \citenamefont {Malik}, \citenamefont {Fertig}, \citenamefont {Bock},
  \citenamefont {Bauer}, \citenamefont {van Leent}, \citenamefont {Zhang},
  \citenamefont {Becher},\ and\ \citenamefont {Weinfurter}}]{Zhou2024}%
  \BibitemOpen
  \bibfield  {author} {\bibinfo {author} {\bibfnamefont {Yiru}\ \bibnamefont
  {Zhou}}, \bibinfo {author} {\bibfnamefont {Pooja}\ \bibnamefont {Malik}},
  \bibinfo {author} {\bibfnamefont {Florian}\ \bibnamefont {Fertig}}, \bibinfo
  {author} {\bibfnamefont {Matthias}\ \bibnamefont {Bock}}, \bibinfo {author}
  {\bibfnamefont {Tobias}\ \bibnamefont {Bauer}}, \bibinfo {author}
  {\bibfnamefont {Tim}\ \bibnamefont {van Leent}}, \bibinfo {author}
  {\bibfnamefont {Wei}\ \bibnamefont {Zhang}}, \bibinfo {author} {\bibfnamefont
  {Christoph}\ \bibnamefont {Becher}}, \ and\ \bibinfo {author} {\bibfnamefont
  {Harald}\ \bibnamefont {Weinfurter}},\ }\emph {\enquote {\bibinfo {title}
  {{Long-Lived Quantum Memory Enabling Atom-Photon Entanglement over 101 km of
  Telecom Fiber}},}\ }\href {https://doi.org/10.1103/PRXQuantum.5.020307}
  {\bibfield  {journal} {\bibinfo  {journal} {PRX Quantum}\ }\textbf {\bibinfo
  {volume} {5}},\ \bibinfo {pages} {020307} (\bibinfo {year} {2024})},\ \Eprint
  {http://arxiv.org/abs/2308.08892} {arXiv:2308.08892}\BibitemShut {NoStop}%
\bibitem [{\citenamefont {van Leent}\ \emph {et~al.}(2022)\citenamefont {van
  Leent}, \citenamefont {Bock}, \citenamefont {Fertig}, \citenamefont
  {Garthoff}, \citenamefont {Eppelt}, \citenamefont {Zhou}, \citenamefont
  {Malik}, \citenamefont {Seubert}, \citenamefont {Bauer}, \citenamefont
  {Rosenfeld}, \citenamefont {Zhang}, \citenamefont {Becher},\ and\
  \citenamefont {Weinfurter}}]{vanLeent2022}%
  \BibitemOpen
  \bibfield  {author} {\bibinfo {author} {\bibfnamefont {Tim}\ \bibnamefont
  {van Leent}}, \bibinfo {author} {\bibfnamefont {Matthias}\ \bibnamefont
  {Bock}}, \bibinfo {author} {\bibfnamefont {Florian}\ \bibnamefont {Fertig}},
  \bibinfo {author} {\bibfnamefont {Robert}\ \bibnamefont {Garthoff}}, \bibinfo
  {author} {\bibfnamefont {Sebastian}\ \bibnamefont {Eppelt}}, \bibinfo
  {author} {\bibfnamefont {Yiru}\ \bibnamefont {Zhou}}, \bibinfo {author}
  {\bibfnamefont {Pooja}\ \bibnamefont {Malik}}, \bibinfo {author}
  {\bibfnamefont {Matthias}\ \bibnamefont {Seubert}}, \bibinfo {author}
  {\bibfnamefont {Tobias}\ \bibnamefont {Bauer}}, \bibinfo {author}
  {\bibfnamefont {Wenjamin}\ \bibnamefont {Rosenfeld}}, \bibinfo {author}
  {\bibfnamefont {Wei}\ \bibnamefont {Zhang}}, \bibinfo {author} {\bibfnamefont
  {Christoph}\ \bibnamefont {Becher}}, \ and\ \bibinfo {author} {\bibfnamefont
  {Harald}\ \bibnamefont {Weinfurter}},\ }\emph {\enquote {\bibinfo {title}
  {Entangling single atoms over 33{\thinspace}km telecom fibre},}\ }\href
  {\doibase 10.1038/s41586-022-04764-4} {\bibfield  {journal} {\bibinfo
  {journal} {Nature}\ }\textbf {\bibinfo {volume} {607}},\ \bibinfo {pages}
  {69} (\bibinfo {year} {2022})}\BibitemShut {NoStop}%
\bibitem [{\citenamefont {Hannegan}\ \emph {et~al.}(2022)\citenamefont
  {Hannegan}, \citenamefont {Siverns},\ and\ \citenamefont
  {Quraishi}}]{Hannegan2022}%
  \BibitemOpen
  \bibfield  {author} {\bibinfo {author} {\bibfnamefont {John}\ \bibnamefont
  {Hannegan}}, \bibinfo {author} {\bibfnamefont {James~D.}\ \bibnamefont
  {Siverns}}, \ and\ \bibinfo {author} {\bibfnamefont {Qudsia}\ \bibnamefont
  {Quraishi}},\ }\emph {\enquote {\bibinfo {title} {Entanglement between a
  trapped-ion qubit and a 780-nm photon via quantum frequency conversion},}\
  }\href {https://doi.org/10.1103/PhysRevA.106.042441} {\bibfield  {journal}
  {\bibinfo  {journal} {Phys. Rev. A}\ }\textbf {\bibinfo {volume} {106}},\
  \bibinfo {pages} {042441} (\bibinfo {year} {2022})},\ \Eprint
  {http://arxiv.org/abs/2207.13680} {arXiv:2207.13680}\BibitemShut {NoStop}%
\bibitem [{\citenamefont {Bate}\ \emph {et~al.}(2025)\citenamefont {Bate},
  \citenamefont {Hamann}, \citenamefont {Canteri}, \citenamefont {Winkler},
  \citenamefont {Koong}, \citenamefont {Krutyanskiy}, \citenamefont {D{\"u}r},\
  and\ \citenamefont
  {Lanyon}}]{BateHamannCanteriWinklerKoongKrutyanskiyDuerLanyon2025}%
  \BibitemOpen
  \bibfield  {author} {\bibinfo {author} {\bibfnamefont {James}\ \bibnamefont
  {Bate}}, \bibinfo {author} {\bibfnamefont {Arne}\ \bibnamefont {Hamann}},
  \bibinfo {author} {\bibfnamefont {Marco}\ \bibnamefont {Canteri}}, \bibinfo
  {author} {\bibfnamefont {Armin}\ \bibnamefont {Winkler}}, \bibinfo {author}
  {\bibfnamefont {Zhe~Xian}\ \bibnamefont {Koong}}, \bibinfo {author}
  {\bibfnamefont {Victor}\ \bibnamefont {Krutyanskiy}}, \bibinfo {author}
  {\bibfnamefont {Wolfgang}\ \bibnamefont {D{\"u}r}}, \ and\ \bibinfo {author}
  {\bibfnamefont {Benjamin~Peter}\ \bibnamefont {Lanyon}},\ }\href@noop {}
  {\emph {\enquote {\bibinfo {title} {{Experimental distributed quantum sensing
  in a noisy environment}},}\ }}\Eprint {http://arxiv.org/abs/2501.08940}
  {arXiv:2501.08940} [quant-ph] (\bibinfo {year} {2025})\BibitemShut {NoStop}%
\bibitem [{\citenamefont {Sekatski}\ \emph {et~al.}(2020)\citenamefont
  {Sekatski}, \citenamefont {W\"olk},\ and\ \citenamefont
  {D\"ur}}]{PhysRevResearch.2.023052}%
  \BibitemOpen
  \bibfield  {author} {\bibinfo {author} {\bibfnamefont {Pavel}\ \bibnamefont
  {Sekatski}}, \bibinfo {author} {\bibfnamefont {Sabine}\ \bibnamefont
  {W\"olk}}, \ and\ \bibinfo {author} {\bibfnamefont {Wolfgang}\ \bibnamefont
  {D\"ur}},\ }\emph {\enquote {\bibinfo {title} {Optimal distributed sensing in
  noisy environments},}\ }\href
  {https://doi.org/10.1103/PhysRevResearch.2.023052} {\bibfield  {journal}
  {\bibinfo  {journal} {Phys. Rev. Res.}\ }\textbf {\bibinfo {volume} {2}},\
  \bibinfo {pages} {023052} (\bibinfo {year} {2020})},\ \Eprint
  {http://arxiv.org/abs/1905.06765} {arXiv:1905.06765}\BibitemShut {NoStop}%
\bibitem [{\citenamefont {Canteri}\ \emph {et~al.}(2025)\citenamefont
  {Canteri}, \citenamefont {Koong}, \citenamefont {Bate}, \citenamefont
  {Winkler}, \citenamefont {Krutyanskiy},\ and\ \citenamefont
  {Lanyon}}]{Canteri2025}%
  \BibitemOpen
  \bibfield  {author} {\bibinfo {author} {\bibfnamefont {Marco}\ \bibnamefont
  {Canteri}}, \bibinfo {author} {\bibfnamefont {Zhe~Xian}\ \bibnamefont
  {Koong}}, \bibinfo {author} {\bibfnamefont {James}\ \bibnamefont {Bate}},
  \bibinfo {author} {\bibfnamefont {Armin}\ \bibnamefont {Winkler}}, \bibinfo
  {author} {\bibfnamefont {Victor}\ \bibnamefont {Krutyanskiy}}, \ and\
  \bibinfo {author} {\bibfnamefont {Ben~P.}\ \bibnamefont {Lanyon}},\ }\emph
  {\enquote {\bibinfo {title} {{Photon-Interfaced Ten-Qubit Register of Trapped
  Ions}},}\ }\href {https://doi.org/doi/10.1103/v5k1-whwz} {\bibfield
  {journal} {\bibinfo  {journal} {Phys. Rev. Lett.}\ }\textbf {\bibinfo
  {volume} {135}},\ \bibinfo {pages} {080801} (\bibinfo {year} {2025})},\
  \Eprint {http://arxiv.org/abs/2406.09480} {arXiv:2406.09480}\BibitemShut
  {NoStop}%
\bibitem [{\citenamefont {You}\ \emph {et~al.}(2024)\citenamefont {You},
  \citenamefont {Wu}, \citenamefont {Miron}, \citenamefont {Ke}, \citenamefont
  {Monga}, \citenamefont {Saglamyurek},\ and\ \citenamefont
  {Haeffner}}]{you2024}%
  \BibitemOpen
  \bibfield  {author} {\bibinfo {author} {\bibfnamefont {Bingran}\ \bibnamefont
  {You}}, \bibinfo {author} {\bibfnamefont {Qiming}\ \bibnamefont {Wu}},
  \bibinfo {author} {\bibfnamefont {David}\ \bibnamefont {Miron}}, \bibinfo
  {author} {\bibfnamefont {Wenjun}\ \bibnamefont {Ke}}, \bibinfo {author}
  {\bibfnamefont {Inder}\ \bibnamefont {Monga}}, \bibinfo {author}
  {\bibfnamefont {Erhan}\ \bibnamefont {Saglamyurek}}, \ and\ \bibinfo {author}
  {\bibfnamefont {Hartmut}\ \bibnamefont {Haeffner}},\ }\href@noop {} {\emph
  {\enquote {\bibinfo {title} {Temporally multiplexed ion-photon quantum
  interface via fast ion-chain transport},}\ }}\Eprint
  {http://arxiv.org/abs/2405.10501} {arXiv:2405.10501} [quant-ph] (\bibinfo
  {year} {2024})\BibitemShut {NoStop}%
\bibitem [{\citenamefont {Drmota}\ \emph {et~al.}(2023)\citenamefont {Drmota},
  \citenamefont {Main}, \citenamefont {Nadlinger}, \citenamefont {Nichol},
  \citenamefont {Weber}, \citenamefont {Ainley}, \citenamefont {Agrawal},
  \citenamefont {Srinivas}, \citenamefont {Araneda}, \citenamefont {Ballance},\
  and\ \citenamefont {Lucas}}]{Drmota2023}%
  \BibitemOpen
  \bibfield  {author} {\bibinfo {author} {\bibfnamefont {Peter}\ \bibnamefont
  {Drmota}}, \bibinfo {author} {\bibfnamefont {Dougal}\ \bibnamefont {Main}},
  \bibinfo {author} {\bibfnamefont {David~P.}\ \bibnamefont {Nadlinger}},
  \bibinfo {author} {\bibfnamefont {Bethan~C.}\ \bibnamefont {Nichol}},
  \bibinfo {author} {\bibfnamefont {M.~A.}\ \bibnamefont {Weber}}, \bibinfo
  {author} {\bibfnamefont {Ellis~M.}\ \bibnamefont {Ainley}}, \bibinfo {author}
  {\bibfnamefont {Ayush}\ \bibnamefont {Agrawal}}, \bibinfo {author}
  {\bibfnamefont {Raghavendra}\ \bibnamefont {Srinivas}}, \bibinfo {author}
  {\bibfnamefont {Gabriel}\ \bibnamefont {Araneda}}, \bibinfo {author}
  {\bibfnamefont {Christopher~J.}\ \bibnamefont {Ballance}}, \ and\ \bibinfo
  {author} {\bibfnamefont {David~M.}\ \bibnamefont {Lucas}},\ }\emph {\enquote
  {\bibinfo {title} {{Robust Quantum Memory in a Trapped-Ion Quantum Network
  Node}},}\ }\href {https://doi.org/10.1103/PhysRevLett.130.090803} {\bibfield
  {journal} {\bibinfo  {journal} {Phys. Rev. Lett.}\ }\textbf {\bibinfo
  {volume} {130}},\ \bibinfo {pages} {090803} (\bibinfo {year} {2023})},\
  \Eprint {http://arxiv.org/abs/2210.11447} {arXiv:2210.11447}\BibitemShut
  {NoStop}%
\end{thebibliography}%

\end{document}